%% file: ms.tex
\documentclass[12pt]{article}

\usepackage{nature}

\usepackage{times}
\usepackage{graphicx}
\usepackage{amsmath,amssymb}
\usepackage{url}
\usepackage{xcolor, soul}
\usepackage{pdfpages}

\usepackage[tableposition=top]{caption}
\usepackage{caption}

\usepackage{bm}                                                                       
\usepackage{enumerate}
\usepackage{subfig}
\usepackage{pict2e}

\usepackage{times}

\renewcommand{\aap}{{Astron. Astrophys.}}
\renewcommand{\araa}{{Annual Review of Astron and Astrophys}}
\newcommand{\lrsp}{Living Rev. Sol. Phys.}
\newcommand{\rmpp}{Rev. Mod. Plasma Phys.}
\renewcommand{\aa}{Astron. \& Astrophys.}
\renewcommand{\apj}{Astrophys. J.}   
\renewcommand{\apjl}{Astrophys. J. Lett.}   
\newcommand{\sph}{Sol. Phys.}
\newcommand{\nastro}{Nat. Astron.} 
\renewcommand{\ssr}{Space Sci. Rev.}
\renewcommand{\jcp}{J. Comput. Phys.}
\newcommand{\cpc}{Comput. Phys. Commun.}
\renewcommand{\mnras}{Mon. Not. R. Astron. Soc.}
\newcommand{\asr}{Adv. Space Res.}

\newcommand{\ion}[2]{\textup{#1\,\textsc{\uppercase{#2}}}}

\usepackage[colorlinks=true,linkcolor=blue,urlcolor=blue,citecolor=blue,bookmarksopenlevel=2,bookmarksopen=true]{hyperref}

\title{Self-consistent numerical simulations for the formation and dynamics of solar prominences}

\author
{{ Lisa-Marie Zessner*$^1$}, {Robert H. Cameron$^1$}, {Sami K. Solanki$^1$}, {Damien Przybylski$^1$}
\\
\normalsize{$^{1}$Max-Planck-Institut f\"ur Sonnensystemforschung,
   G{\"o}ttingen 37077, Germany}
\\
\normalsize{To whom correspondence should be addressed: Lisa-Marie Zessner;
            E-mail: zessner@mps.mpg.de}
}

\date{}

\begin{document} 
\baselineskip18pt
\maketitle

\begin{abstract}
Solar prominences are cool and dense plasma structures floating in the hot solar corona. They are ubiquitous features in the solar atmosphere, but their formation mechanism is still unclear.
Here we perform comprehensive fully three-dimensional numerical simulations of prominence formation including the physics necessary to describe all atmospheric layers of the sun.  With appropriate initial conditions for the magnetic field, solar prominences form self-consistently in the simulations. The formation starts by the random ejection of a dense plasma seed from the chromosphere into the corona. Subsequently, the prominence is built up by a combination of plasma injections from the chromosphere and condensation of inflowing coronal plasma. The prominence properties qualitatively match those of observed prominences. Our findings demonstrate the importance of the dynamics at and below the solar surface in the formation and evolution of solar prominences. This suggests that subsurface dynamics should also be considered in the study of prominence eruptions, which can be associated with coronal mass ejections.
\end{abstract}

\section{Introduction}

Solar prominences are beautiful and complex objects in the solar atmosphere. As gas clouds that are two orders of magnitude cooler and denser than the surrounding hot solar corona, they appear in emission when observed at the solar limb and as dark filaments on the solar disk. Solar prominences are as diverse as they are common - they can have very different sizes, fine structures, and underlying magnetic field configurations \cite{parenti2014solar}. Even though the prominence plasma is very dynamic, prominences can form long-lived, stable structures, with lifetimes ranging from days to months. At the end of their lifetimes, they can disappear quietly or erupt. Erupting prominences are often associated with Coronal Mass Ejections {\cite{gopalswamy2003prominence}}, which can produce violent space weather effects when they reach Earth. When prominence material is present, the prominence mass can influence or trigger the eruption process{\cite{fan2025mhd}}. Realistic models of the formation and evolution of solar prominences should thus also contribute to long-term space weather forecasts. In this paper, we present self-consistent simulations that include the relevant physical processes in all atmospheric layers guiding the formation and dynamics of solar prominences. 

The solar atmosphere is filled with plasma with temperatures of about $5800$\,K at the solar surface (the photosphere), $6\,000$--$20\,000$\,K in the overlying chromosphere and over a million degrees in the upper solar atmosphere (the corona) \cite{sturrock2014atmosphere}. Turbulent convective motions continually generate and rearrange the magnetic field below the surface. These changes rearrange the magnetic field in the upper chromosphere and corona where it dominates the energetics.  
Such sub-photospheric turbulent driving also strongly affects the dynamics and evolution of prominences, but has so far only been included heuristically in prominence models{\cite{jercic2024prominence}} and not in a self-consistent manner.

The cool and dense plasma in prominences is held up against gravity by magnetic fields \cite{kippenhahnTheorieSolarenFilamente1957, kuperus1974support}. For small prominences, it is possible to accumulate the prominence mass by in-situ condensation from the coronal environment{\cite{donne2024mass}}, whereas massive prominences need an additional mass reservoir{\cite{pikelnerOriginQuiescentProminences1971}}. In this case, the plasma must be transported up from the lower-lying chromosphere.
Several mechanisms have been proposed for this mass supply{\cite{zhou2025formationsolarprominencesplasma}}. The most basic models are direct injection, levitation, and condensation. In the direct injection model \cite{chaeFormationProminenceNOAA2003,li2025response}, cool plasma is directly transported along the magnetic field lines from the chromosphere into the corona. In the levitation model \cite{rustHelicalMagneticFields1994a, galsgaardFormationSolarProminences1999,zhao2017formation}, emerging magnetic fields capture and lift the cool plasma on their way up to the corona. In the condensation model \cite{pikelnerOriginQuiescentProminences1971, antiochosDynamicFormationProminence1999, antiochosThermalNonequilibriumProminences2000a,keppens2025modelingmultiphaseplasmacorona}, chromospheric gas is first heated and evaporated into the corona, where it then condenses again to form the cool prominence plasma.  The evaporation-condensation model has been studied extensively in a number of numerical simulations of prominence formation \cite{jercic2024prominence,xiaSIMULATIONSPROMINENCEFORMATION2012,karpenOriginHighSpeedMotions2006b,fanMHDSimulationProminence2018}. Combinations of the three basic mechanisms have also been proposed \cite{ kaneko2015numerical,huangUnifiedModelSolar2021,jenkinsProminenceFormationLevitationcondensation2021}. Numerical simulations have also been able to reproduce some prominence properties like fine structures, instabilities, and oscillations \cite{donne2024mass,gunar20153d,jenkins2022resolving, zhou2020simulations,braileanu2021two,terradas2015morphology,arregui2018prominence}. Despite the large variety of existing prominence models, there is so far no simulation of a solar prominence that includes a self-consistent treatment of the upper part of the convection zone and the solar surface. The photospheric dynamics are expected to have an important influence on especially low-lying active region prominences. We present the formation and properties of an active-region like prominence in the 3D radiative magnetohydrodynamic (MHD) code MURaM, which captures the most important physical processes to describe the photosphere, chromosphere and corona self-consistently. Our results show that the surface dynamics plays an important role in the formation and mass supply of the prominence. 

\section{Numerical methods and simulation setup}

In this paper we gain insights into the physics that drives the formation and the mass supply of solar prominences by performing numerical simulations. These simulations, carried out with the MURaM code  \cite{voglerSimulationsMagnetoconvectionSolar2005a, przybylskiChromosphericExtensionMURaM2022, rempelEXTENSIONMURAMRADIATIVE2016}, capture the dominant physical processes in each layer of the solar atmosphere and are therefore self-consistent.
MURaM is a box-in-a-star code that explicitly treats the energy transport in the different solar layers, including convection and small-scale dynamo action \cite{borreroSolarMagnetoconvectionSmallScale2017} in the photosphere \cite{voglerSimulationsMagnetoconvectionSolar2005a}, a non-equilibrium treatment of hydrogen and a multi-group scattering scheme in the chromosphere \cite{przybylskiChromosphericExtensionMURaM2022}, as well as optically thin radiative losses and field aligned heat conduction in the corona \cite{rempelEXTENSIONMURAMRADIATIVE2016}. The MURaM code has been used to simulate sunspot umbrae, plage \cite{voglerSimulationsMagnetoconvectionSolar2005a}, pores \cite{cameronRadiativeMagnetohydrodynamicSimulations2007}, sunspots \cite{rempelRADIATIVEMAGNETOHYDRODYNAMICSIMULATION2009}, flux emergence and flares \cite{cheungComprehensiveThreedimensionalRadiative2019}.

The most commonly proposed magnetic field configurations for prominences are either a magnetic flux rope or a dipped arcade \cite{kippenhahnTheorieSolarenFilamente1957, kuperus1974support, gibson2018solar}. In either case, a region is present above the solar surface where the magnetic field lines are concave upwards, forming a dip. Since plasma is free to flow along field lines, such a dip is essential to support the prominence material against gravity. We chose a dipped arcade structure for the magnetic field configuration in our simulations. Starting with a potential field based on this (see Extended Data Figure \ref{fig:ed-fig1} and \ref{fig:ed-fig2}), the resulting quadrupolar magnetic field structure is initially not sheared. We present three different simulations that are based on this initial condition, which we call Run I, Run II and Shear in the following. Run I and Run II are both based on the non-sheared, potential setup from the Extended Data Figures \ref{fig:ed-fig1} and \ref{fig:ed-fig2}. Run II has a higher magnetic field strength, a deeper convection zone and a slightly different bottom boundary condition compared to Run I (see the Methods Section for details on the implementation). To account for the high degree of shear that is normally observed for real prominences \cite{parenti2014solar}, the third simulation is a sheared configuration based on the setup of Run I with an already built-up prominence (see the description in the Methods Section and Extended Data Figure \ref{fig:ed-fig3}). We chose the presented setups because the resulting prominences are governed by the same physical processes, but have different appearances. 

All runs were simulated with the local thermodynamic equilibrium (LTE) version of the MURaM code {\cite{rempelEXTENSIONMURAMRADIATIVE2016}}. Run I and II were additionally run with the non-LTE (NLTE) version{\cite{przybylskiChromosphericExtensionMURaM2022}}, which includes a time-dependent treatment of hydrogen ionisation and a scattering radiation transfer scheme in addition to the physics captured by the LTE version of the code (see the Methods Section). Results in the main text concentrate on the LTE results. The only exception to this is the presented approximated H$\mathrm{\alpha}$ emission, which uses the non-equilibrium hydrogen populations that are calculated in the NLTE simulations. Further NLTE results are shown in the Supplementary  Information in Section \ref{sec:sup-sec1}.

The lower boundary condition on the magnetic field pins the footpoints of the magnetic arcade. Boundary conditions are periodic in the horizontal directions and matched to a potential field at the upper boundary. 

\begin{figure}
    \centering
     \includegraphics[width = 1 \textwidth]{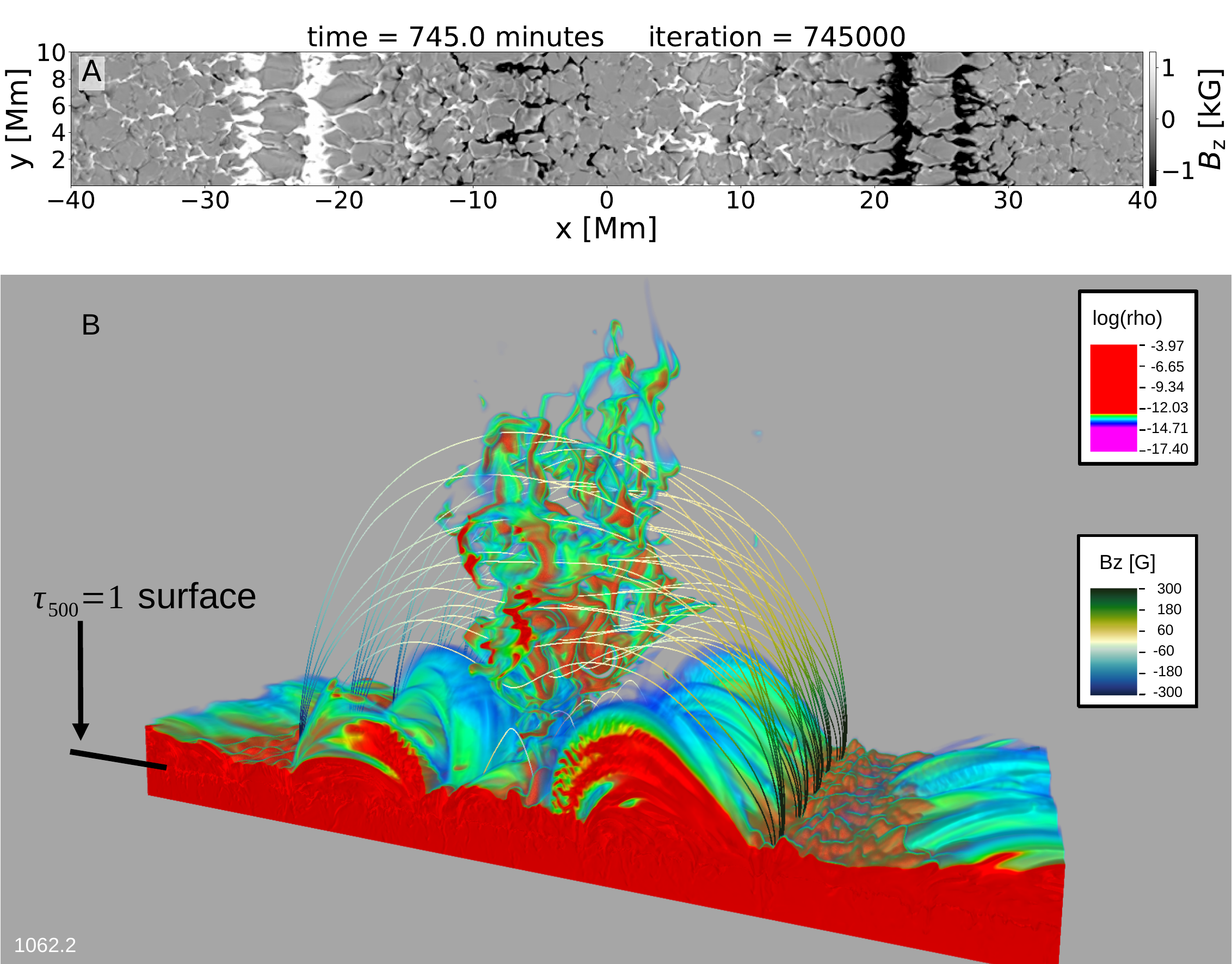}

    \caption{\textbf{Visualisation of the prominence and magnetic field structure for Run I}. A: Vertical magnetic field at the $\tau_{500} = 1$ surface at around 60 minutes after prominence formation starts. Supplementary Video 1 shows an animation of this panel. B: 3D rendering of the prominence density with a set of field lines to demonstrate the quadrupolar structure of the magnetic field. This snapshot shows the prominence 300 minutes after the formation starts. The coloring of the plasma (rainbow colormap) shows the logarithmic gas density (in $\mathrm{g/cm^3}$). For plasma with a density below $10^{-14}\,\mathrm{g/cm^3}$, the opacity is set to zero for this image, such that the surrounding corona is not visible. The coloring of the magnetic field lines (blue-green colormap) shows the value of the vertical magnetic field component. The number in the bottom left shows the time in minutes since the start of the simulation. The thick black line indicates the average height where $\tau_{500} = 1$, roughly corresponding to the location of the photosphere. For the corresponding images of Run II, see Extended Data Figure \ref{fig:ed-fig5}.}
   
    \label{fig:fig1}
\end{figure}

Except at the lower boundary, the magnetic field evolves in accordance with the induction equation (see Equation~\ref{eq:induction} in the Methods).
Panel A of Figure \ref{fig:fig1} shows the photospheric magnetic field for Run I after the initial magnetic field and interior convection have been evolved to a statistically stationary state. The imposed quadrupolar structure is still clear (positive polarity on the left hand side, a ridge of negative polarity at about $-8$\,Mm, a ridge of positive polarity at $+8$\,Mm, and negative polarity field to the right).
The effect of the turbulent convection is seen in the fine structure. Supplementary Video\footnote{Supplementary Videos can accessed here \url{https://doi.org/10.1038/s41550-026-02840-7}.} 1 shows that the magnetic structure is constantly evolving. Panel B of Figure \ref{fig:fig1} shows a 3D rendering of the prominence density and magnetic field lines for Run I. We can see how the quadrupolar structure from the magnetogram translates to a dipped magnetic arcade in the corona. The simulated prominence is hanging in the magnetic dips of this arcade, above the polarity inversion line. Due to the magnetic nullpoint in our setup, the magnetic field strength in the dips first increases, then decreases with increasing height above the surface (see Extended Data Figure \ref{fig:ed-fig4}). Extended Data Figure \ref{fig:ed-fig5} equivalently shows the magnetic field configuration for Run II.

The magnetic field strength within the prominence is in the range of $\approx 15$--$37 \mathrm{G}$ for Run I, $\approx 30$--$80 \mathrm{G}$ for Run II, and $\approx 38$--$45 \mathrm{G}$ for the Shear simulation, as shown in Extended Data Figure \ref{fig:ed-fig4}. The average field at these heights is with $\approx 50$--$120 \mathrm{G}$ in Run I/the sheared setup and $\approx 160$-$360 \mathrm{G}$ in Run II much higher than the field in the prominence itself. The magnetic field values within the prominences are not as high as the very strongest measured values for active region prominences \cite{kuckein2009magnetic}, but significantly higher than the values for most quiescent prominences \cite{parenti2014solar}. Furthermore, the simulated prominences are low-lying ($4$--$7\,\mathrm{Mm}$ over the surface) and have heights of $6$--$20\,\mathrm{Mm}$ that are similar to prominences in active regions. Due to these dimensions and the surrounding magnetic field strength, we classify the simulated prominences as active region prominences. However, magnetic field values of 30\,G and higher are also reasonable for intermediate prominences \cite{berger2013solar} and have even been observed in some quiescent prominences \cite{casini2003magnetic}. We therefore compare the fine-structure of the simulated prominences with all prominence types. 

In the sheared setup, the shear angle of the magnetic field with respect to the long prominence axis is in the range of 15-60\textdegree\ within the prominence (see Extended Data Figure \ref{fig:ed-fig3}), which makes the shear angle in the central part of the prominence body comparable to observed values of around 35\textdegree \cite{parenti2014solar}.

\section{Prominence formation and mass supply}

\begin{figure}
    \centering

     \includegraphics[width =  0.85 \textwidth]{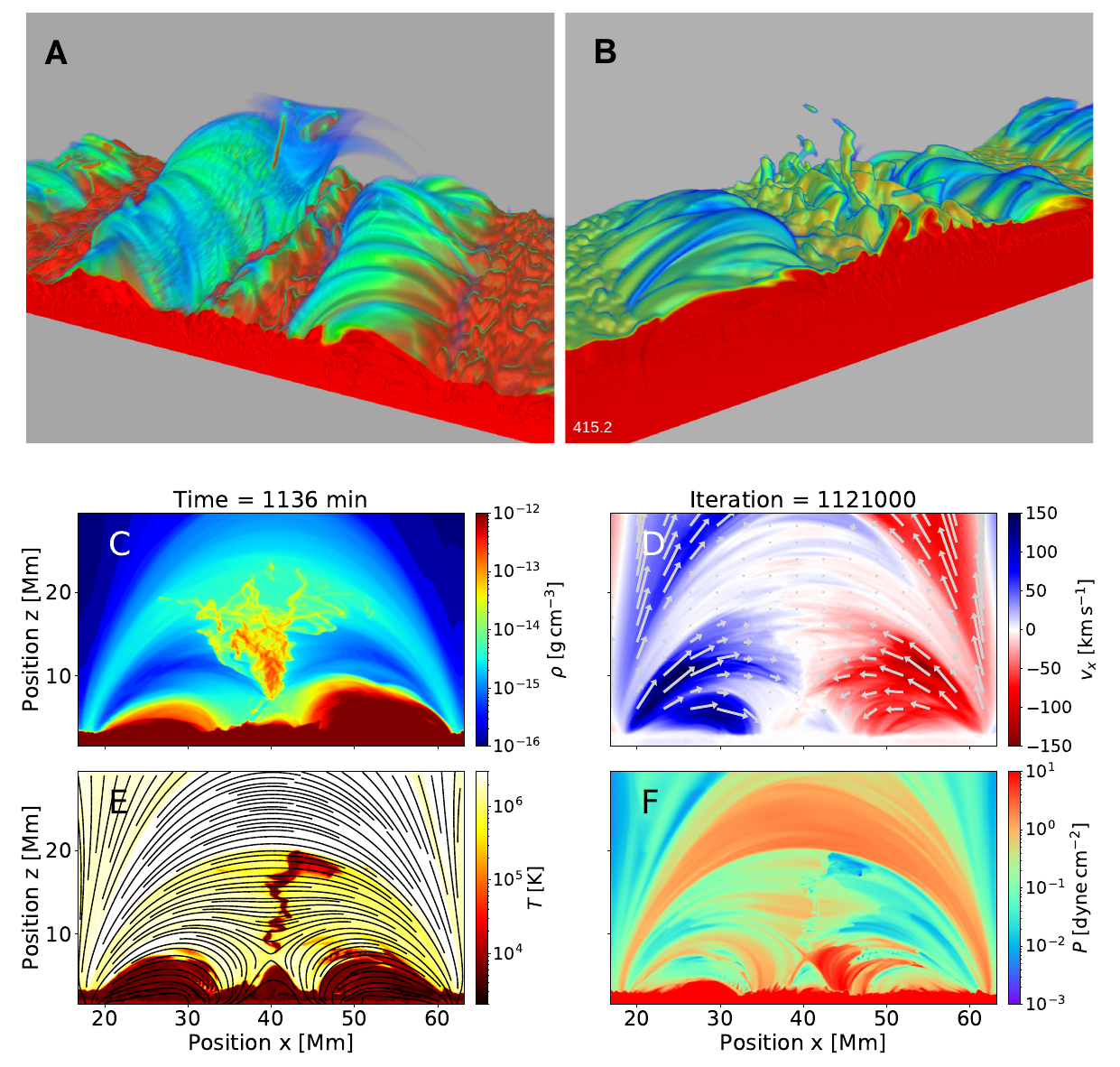}

    \caption{\textbf{Formation of the simulated prominences in two steps}. A: 3D rendering of the prominence density, showing a snapshot during the formation phase in Run I. A seed of cool plasma is transported along the loops on the left side into the magnetic dips. B: 3D rendering of the prominence density, showing a snapshot during the formation phase in Run II. The initial cool seed is here ejected into the dips from the polarity inversion line. The number in the bottom left shows the time in minutes since the start of the simulation. C-F: A vertical slice through one snapshot when the prominence in Run I is fully built up. It shows how the prominence is fed via condensation of hot plasma that is flowing along the magnetic field lines onto the cool prominence structure, driven by a pressure drop at the cool prominence material. The density (C) and horizontal velocity (D) are averaged over the current line of sight. For better visibility, the temperature (E) and pressure (F) are taken along one vertical slice of the box. The arrows in panel D show the line-of-sight averaged velocity field. Line-of-sight averaged magnetic field lines are added to panel E. Supplementary Videos 2 (panel A), 3 (panel B) and 4 (panel C-F) show an animation of this Figure. Supplementary Video 5 shows the equivalent of panels C-F for Run II.}
    \label{fig:fig2}
\end{figure}

Prominences form spontaneously in this simulation. The formation is initiated by random chromospheric motions (driven by the turbulent convection) that inject a small blob of cool chromospheric material into the magnetic dips. For Run I this happens at 670 minutes, for Run II at 410 minutes after the start of the 3D simulation. The ejection of this seed of dense material into the corona is caught on the Supplementary Videos 2 and 3. In Run I, this blob is supplied by a cool flow along the small magnetic loops on the left side (see Figure \ref{fig:fig2}A and Supplementary Video 2). In Run II, this blob is ejected from below the magnetic dips (see Figure \ref{fig:fig2}B and Supplementary Video 3). Once in the corona, two processes cause the blob to grow in mass.

One of these mass supply mechanisms is caused by a radiative instability that occurs as the cool prominence material (and its surrounding transition region) loses energy by radiation (see Supplementary Figure \ref{fig:sup-fig1} and \ref{fig:sup-fig2} in Supplementary Section \ref{sec:sup-sec1}). This causes the pressure to drop, and drives a siphon flow {\cite{pikelnerOriginQuiescentProminences1971}}, where plasma flows onto the prominence structure along the magnetic field lines passing through the corona (see Figure \ref{fig:fig2}C-F, as well as Supplementary Figure \ref{fig:sup-fig3}). The resulting siphon flow can be seen in panel D of Figure \ref{fig:fig2} as an inflow of plasma in the horizontal velocities as soon as the initial plasma blob settles into the magnetic dip (see Supplementary Videos 4 and 5). Hot plasma (Figure \ref{fig:fig2}E) flows up along the field lines on both sides and condenses onto the prominence structure. This mass supply of the simulated prominence therefore corresponds to the condensation model
{\cite{donne2024mass,xiaSIMULATIONSPROMINENCEFORMATION2012}}. Supplementary Figure~\ref{fig:sup-fig4} and \ref{fig:sup-fig5} in Supplementary Section~\ref{sec:sup-sec2} show that thermal instabilities play a role in the transition region around the prominence and the cool injected seeds. This aligns with what has been found in previous work\cite{keppens2025modelingmultiphaseplasmacorona}. In contrast to models where the prominence forms solely by condensation, the condensation process in this simulation does not start the prominence formation process, as it sets in only after the first dense seed has been injected from the chromosphere into the corona. 

The second process that supplies mass is the continued ejection of dense plasma blobs from the chromosphere, which happens from time to time throughout the whole simulation. This mechanism is the same as that which is responsible for supplying the initial prominence seed. The ejections mainly happen from below the prominence for Run II, and from the magnetic side loops next to the prominence for Run I (see Supplementary Videos 2 and 3). Supplementary Section \ref{sec:sup-sec3} takes a closer look on the dynamics of an injection event in Run II. For the chosen event as shown in Supplementary Figure~\ref{fig:sup-fig6} (Supplementary Video 10), the chromospheric plasma is injected into the corona from below the Nullpoint. Around the time of the injection, flux cancellation occurs around the footpoints of the loops that lie under the Nullpoint (see Supplementary Figure \ref{fig:sup-fig9} and Supplementary Video 14). Above the surface, reconfigurations of the magnetic field are visible while the chromospheric plasma is accelerated upwards from the chromosphere. Both the Lorentz force and the pressure gradient force contribute to the dynamics of the accelerated plasma (see Supplementary Figure \ref{fig:sup-fig7} and Supplementary Video 11). Through rearrangements of the magnetic field at the Nullpoint, the chromospheric plasma reaches the corona and contributes to the prominence mass (see Supplementary Figure \ref{fig:sup-fig8} and Supplementary Videos 12 and 13). Once in the corona, the material is supported against gravity by the Lorentz force (see Supplementary Figure \ref{fig:sup-fig10}). Further examples of ejected blobs can be seen in the Supplementary Videos 6 and 7 (left and top right panel). This mass supply mechanism is ubiquitous in the simulation. In Run II, for example, at $t \thicksim 560$\,min, $y \thicksim 6$--$8$\,Mm, $x \thicksim 40$\,Mm we see a large ejection from the bottom which supplies part of its mass to the prominence, and a small ejection supplying mass at $t \thicksim 624$\,min, $y \thicksim 4$--$6$\,Mm, $x \thicksim 40$\,Mm. However, not all of these events supply mass to the prominence: a third example event occurs at $t \thicksim 523$\,min, $y \thicksim 1$\,Mm, $x \thicksim 40$\,Mm that only disturbs the prominence before falling back to the chromosphere. Supplementary Figure \ref{fig:sup-fig11} and \ref{fig:sup-fig12} in Supplementary Section \ref{sec:sup-sec3} demonstrate that only sufficiently slow blobs are successful in building up the prominence.

The siphon flow (condensation) and turbulent injections both contribute mass to the prominence throughout the simulation. For both simulations, the turbulent injection has a higher contribution to the mass supply (${\sim}\,58$--$82\,$\%) than the condensation mechanism (${\sim}\,18$--$42\,$\%) (see Supplementary Figures \ref{fig:sup-fig13} and \ref{fig:sup-fig14} in Supplementary Section \ref{sec:sup-sec4} for more details).

\begin{figure}
    \centering

    \includegraphics[width = 0.95 \textwidth]{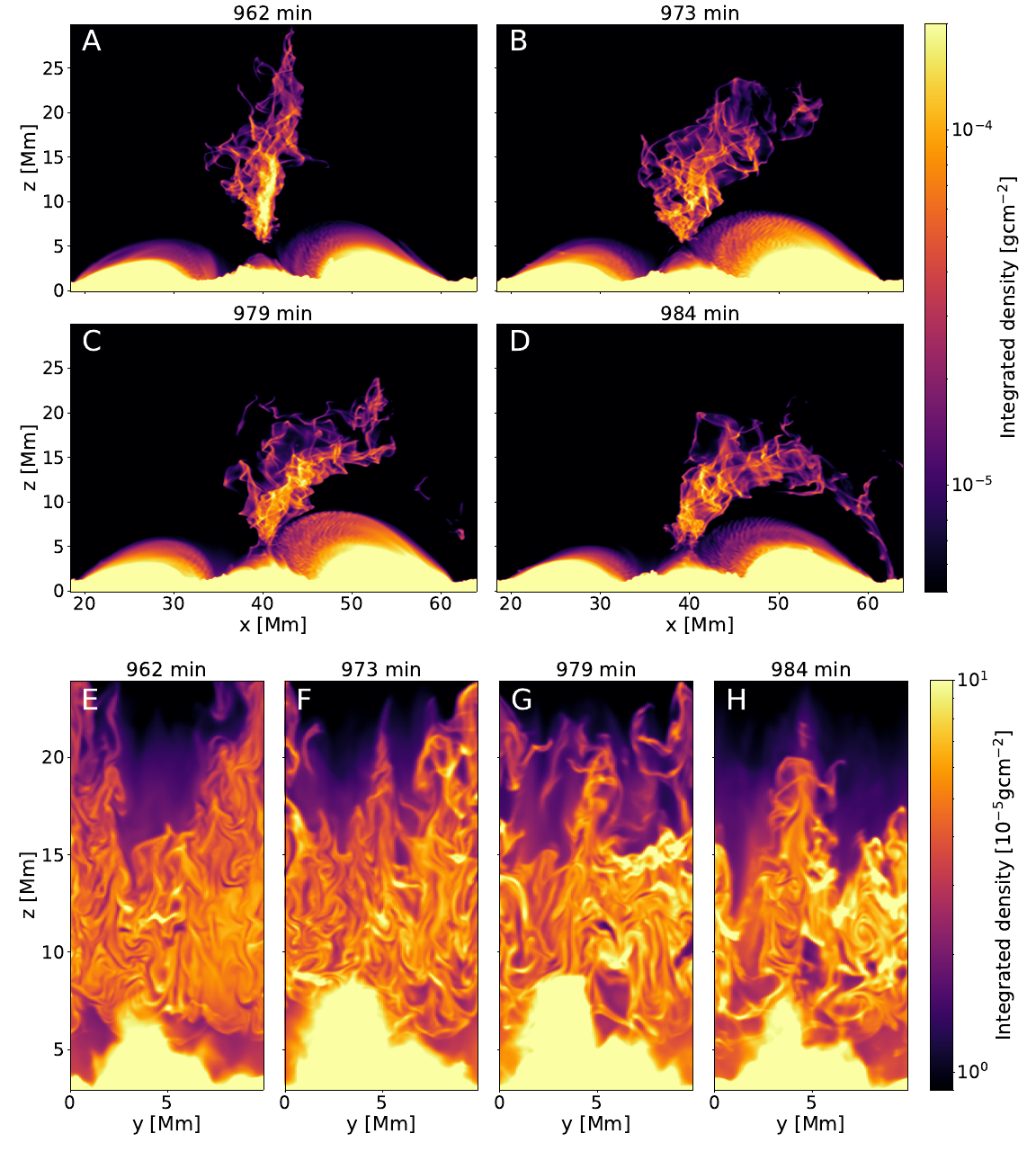}

    \caption{\textbf{Prominence dynamics seen from two directions}. A-D: Integrated density through the prominence from the front (integration along the y-axis) for a series of four snapshots. A: a straight state of the prominence, followed by a draining event to the right in the next three snapshots (B-D). B-D: The prominence in three separate time steps during the draining event. E-H:  Integrated density through the prominence from the side (integration along the x-axis) for the same time series as in A-D. The Supplementary  Videos 6, 7 and 8 show an animation of this Figure for Run I, Run II and the Shear run.}  
    \label{fig:fig3}
\end{figure}

\section{Prominence properties and dynamics}

The mass supplied to the prominence by injection and condensation is partly drained away again. The Supplementary Videos 6, 7 and 8 show that the prominence structure is continuously swaying from side to side along the x-axis. During this motion, from time to time, cool plasma rains down along the magnetic field lines from the top of the prominence to the solar surface. This is possible because the prominence gas in our dipped arcade setup is not confined at the sides, such that the mass can in principle drain down at every point along the long prominence axis. Observations often show 
prominence mass draining in the context of eruptions \cite{gilbert2000active}, but draining events during oscillatory motions have also been observed, in general agreement to what we find \cite{zhang2017large}. The prominence is slightly tilted during the draining process. An example of a draining event is shown in Figure \ref{fig:fig3}. As soon as the draining stops, the prominence straightens and the siphon flow sets in again. Alternating between mass supply by condensation and injection, and mass loss by draining, the prominence is dynamically stable over a timescale of hours. For Run I, the prominence structure is stable over the whole simulation time, which is 12 hours from the formation time. The prominence mass increases for the first 200 minutes after the start of the formation, then stays approximately constant for the next 200 minutes and starts to increase slightly towards the end of the simulated time frame (see Supplementary Figure \ref{fig:sup-fig14}). For Run II, the prominence structure is not stable. After the formation, the prominence grows in mass for the next 230 minutes. After this time, the prominence material drains down completely in a big draining event. The formation of a new prominence starts again 134 minutes after complete disappearance. This new prominence disappears and reforms again before we stop the simulation (see Supplementary Figure \ref{fig:sup-fig14}). The formation and mass supply procedure is the same for all three formations. As observed in real prominences \cite{liuFIRSTSDOAIA2012}, the total mass that circulates through the prominence is several times the prominence mass for both runs, which shows how dynamic the structure is (see Supplementary Section \ref{sec:sup-sec4}).  

In the prominence cores, we get temperatures of 6000-8000\,K and densities of $10^{-13}$--$10^{-12}\,\mathrm{g\,cm^{-3}}$, which are comparable to observations \cite{parenti2014solar, mackay2021solar}. Also the pressure values (see Supplementary Figure \ref{fig:sup-fig3}) are in line with the values cited in ref. \cite{parenti2014solar}. 

 \begin{figure}
    \centering
    \includegraphics[width = 1 \textwidth]{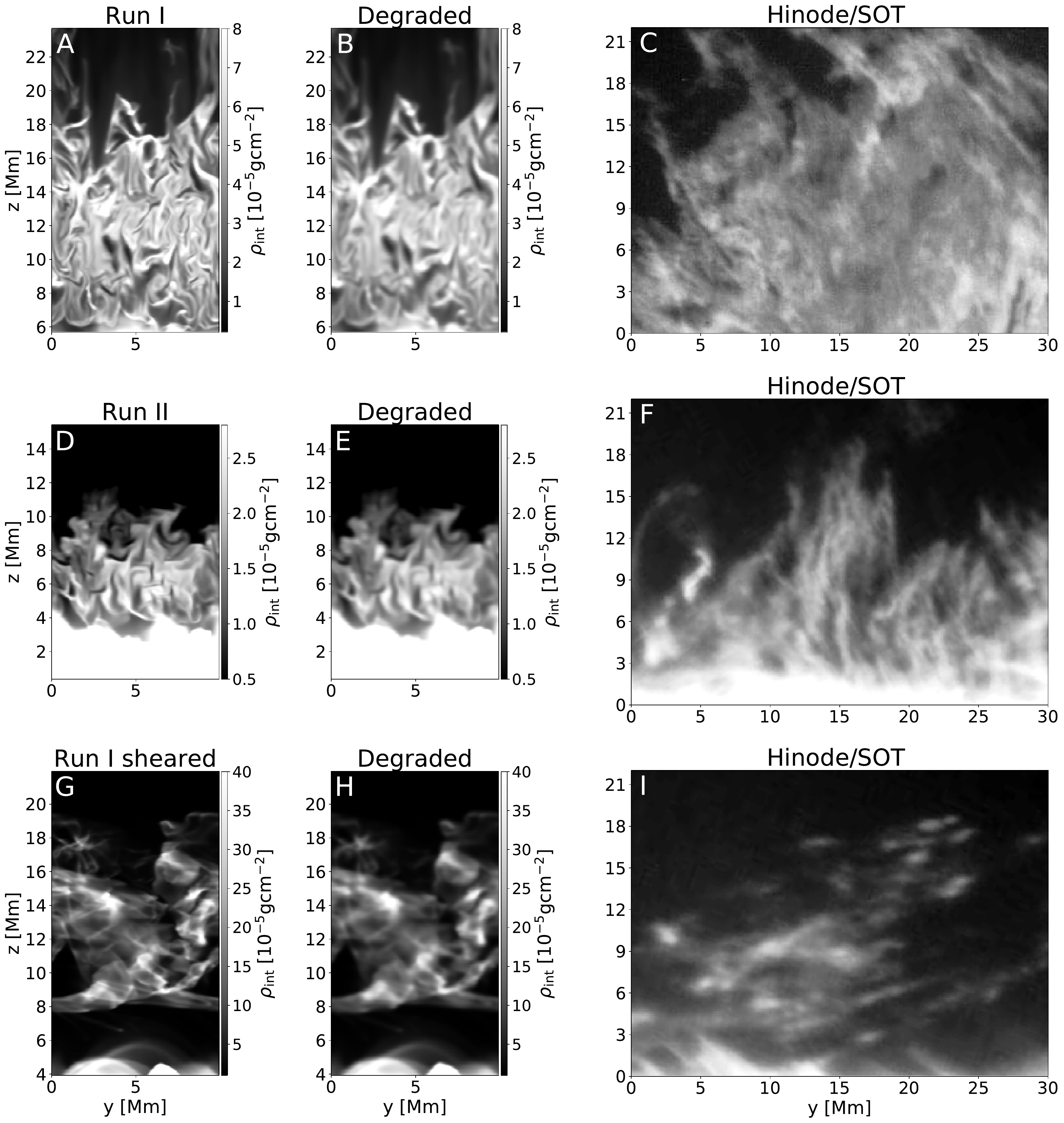}
   
    \caption{\textbf{Fine structure of the three simulated prominences in comparison to Hinode/SOT observations}. A,D,G: Integrated density in the simulation from the side for Run I (A), Run II (D) and the Shear simulation (G). B,E,H: Panels A,D and G degraded to Hinode/SOT resolution. C,F,I: Hinode/SOT observations of individual prominences above the limb. The observations were taken on 26th July 2013 (C),  25th June 2010 (F) and 16th October 2014 (I). The shown length scales were obtained by using a constant factor of 725\,km/arcsecond for all observations. The observed images were rotated such that the local plane of the solar surface is approximately parallel to the y-axis of the simulations.} 
    \label{fig:fig4}
\end{figure}

Figure \ref{fig:fig4} shows the general prominence appearance in integrated density for the three simulation runs in comparison with three observations from the Hinode Solar Optical Telescope (SOT) \cite{tsuneta2008solar} that resemble the structure of the simulated prominences. For a better comparison, the simulated images were degraded using a Gaussian filter with a FWHM of 0.3 arcseconds. The fine-structure of Run I (panels A and B) is more turbulent and resembles the diffuse structure of the observed prominence in panel C. Especially the left part of the observation shows a similar mixture of vertical and horizontal structures, which can often be observed in intermediate prominences \cite{berger2013solar}. In addition to this turbulent pattern, Run II (panels D and E) shows a few elongated, more vertical structures that resemble the observation in panel F. These vertical structures are generally more typical for quiescent prominences \cite{berger2013solar}. Both for Run I and for Run II the magnetic arcade is not sheared, which is the reason they do not feature the pronounced horizontal pattern that is typical for active-region prominences on the limb. Run I has an overall lower magnetic field, which allows the plasma to spread more freely across the magnetic field lines (in y-direction) and produces a more turbulent looking structure compared to Run II. The sheared setup of Run I (panels G and H) starts to show more horizontal structures compared to the original fine-structure of Run I, which is more in line with what is expected from an active region prominence \cite{okamotoCoronalTransverseMagnetohydrodynamic2008}. Along the polarity inversion line, the magnetic field configuration in this run is strongly sheared, which makes the magnetic field orientation mostly horizontal in the yz-plane, leading to horizontal structures. The shear strength decreases with height and is in the range of 15-60\,\textdegree\ (see Extended Data Figure \ref{fig:ed-fig3}). As the magnetic field value within our prominences is still on the lower side for an active region prominence, we consider it reasonable to compare the simulated fine structures with all prominence types.

\begin{figure}
    \centering
    \includegraphics[width = 0.95 \textwidth]{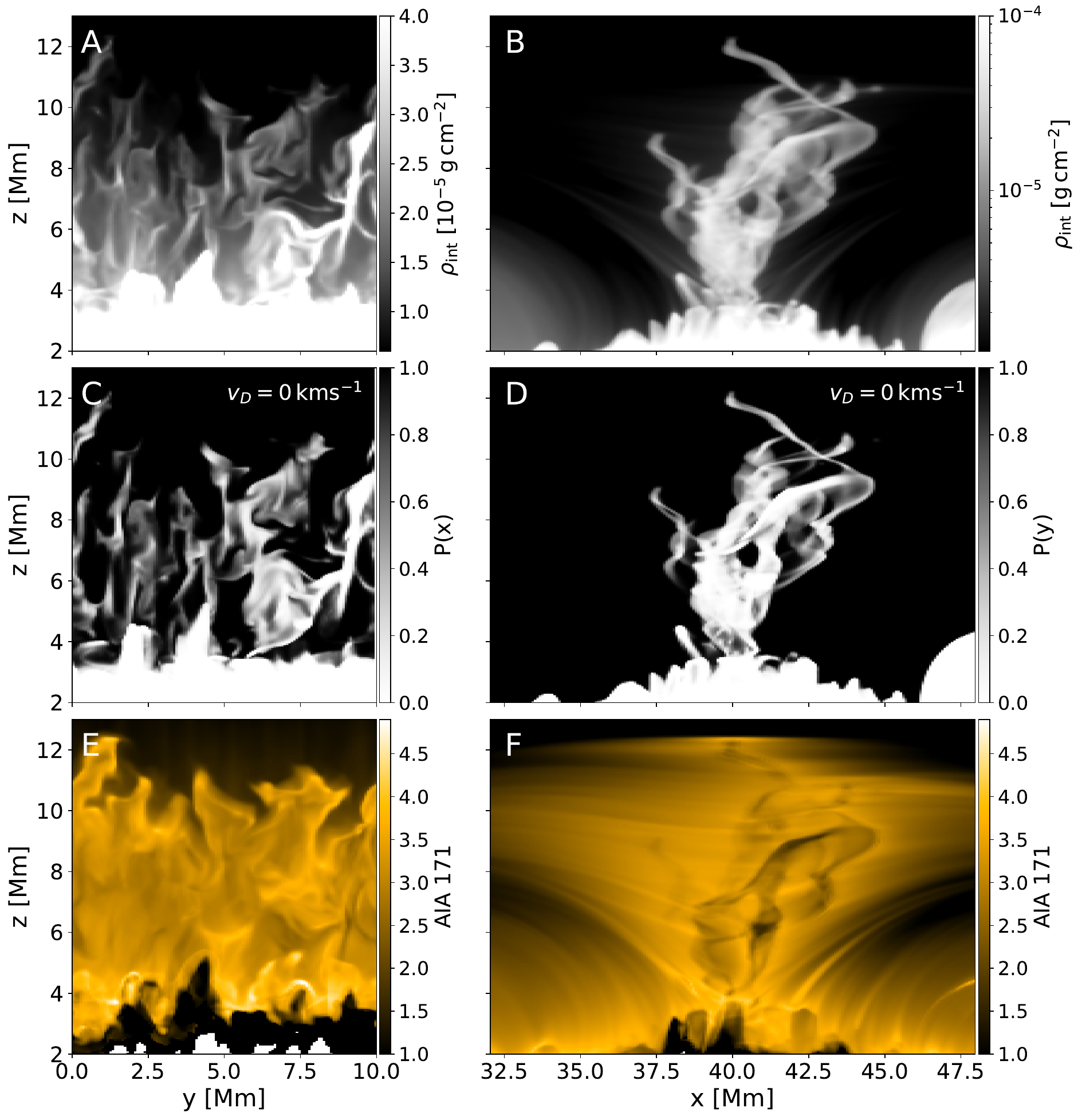}
   
    \caption{\textbf{Appearance of the prominence in Run II NLTE in H$\mathrm{\alpha}$ and AIA 171}. Comparison of integrated density (A,B), H$\mathrm{\alpha}$ emission in the line core (C,D) and AIA 171 emission (E,F) for the prominence view in Run II. The left panels (A,C,E) show the prominence from the side (line-of-sight along the x-axis) and the right panels (B,D,F) show the prominence from the front (line-of-sight along the y-axis). Following ref. \cite{chandra2025probing}, the H$\mathrm{\alpha}$ emission is approximated by calculating an integrated hydrogen opacity that uses the non-equilibrium hydrogen populations from the NLTE version of the MURaM code. The shown emission is taken in the H$\mathrm{\alpha}$ line core (that is, at a Dopper velocity of $v_D = 0$). The AIA 171 emission is calculated following ref. \cite{chen2021transient} and takes absorption into account. Extended Data Figure \ref{fig:ed-fig6} equivalently shows the appearance of the prominence in Run I NLTE.} 
    \label{fig:fig5}
\end{figure}

In addition to the appearance of the prominence in integrated density as shown in Figure \ref{fig:fig4}, Figure \ref{fig:fig5} and Extended Data Figures \ref{fig:ed-fig6} and \ref{fig:ed-fig7} show the appearance of the prominences for Run I and Run II in H$\mathrm{\alpha}$ and AIA 171 emission for the prominence view. We calculated the H$\mathrm{\alpha}$ images following the approximative approach developed in {\cite{chandra2025probing}}. To calculate the synthesised H$\mathrm{\alpha}$ emission, these authors used an integrated hydrogen opacity that takes the non-equilibrium populations from the MURaM NLTE simulations into account. We follow this method and use the NLTE simulations of Run I and Run II with the corresponding non-equilibrium hydrogen populations. Figure \ref{fig:fig5} and Extended Data Figure \ref{fig:ed-fig6} show the resulting appearance for the prominence view in the line core of the H$\mathrm{\alpha}$ proxy ($v_D = 0\,\mathrm{km\,s^{-1}}$) in comparison to the appearance in integrated density. The structures visible in H$\mathrm{\alpha}$ (panels C and D) are similar to the structures seen in integrated density (panels A and B). Due to the saturation of the H$\mathrm{\alpha}$ proxy in the line core, some of the substructures visible in the integrated density are washed out in the H$\mathrm{\alpha}$ images. In the side view (panel 5C) the fine structure looks more pronounced in H$\mathrm{\alpha}$ compared to the smoother appearance in integrated density (panel 5A). For comparison to the appearance in the H$\mathrm{\alpha}$ line core, we show an image of Run I in the blue wing of H$\mathrm{\alpha}$ at a Doppler velocity of $v_D = 12 \,\mathrm{km\,s^{-1}}$ in Extended Data Figure \ref{fig:ed-fig7}. The AIA 171
emission is calculated as described in {\cite{chen2021transient}}. The method used here also includes absorption. From the front (panel 5F), the prominence structure that is visible in integrated density and H$\mathrm{\alpha}$ can be seen in absorption in AIA 171. From the side (panel 5E), the prominence structure looks relatively smooth, with some visible fine structure that appears slightly darker. No cavity structure is visible in AIA 171, in contrast to what is often seen in observations{\cite{gibson2014coronal}}. We plan to study if this changes in the future when moving to other magnetic field configurations.

\begin{figure}
    \centering
    \includegraphics[width = 1 \textwidth]{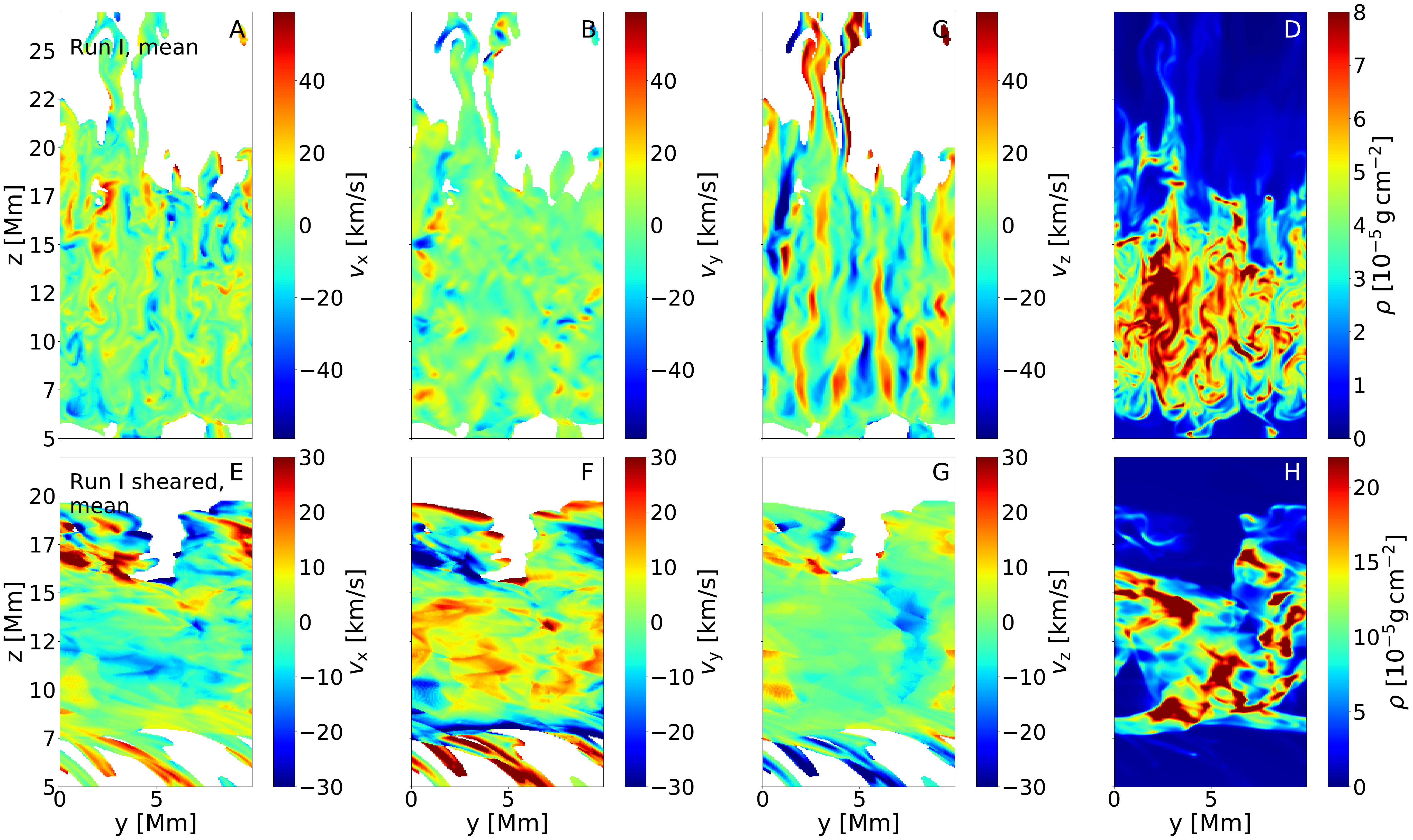}
    \caption{\textbf{Flow structures for one snapshot of the non-sheared and the sheared setup of Run I.} The top row shows the non-sheared and the bottom row the sheared setup. A-D: velocities in x (A), y (B), and z-direction (C), and integrated plasma density (D) along a vertical slice in the yz-plane for Run I. For all quantities apart from the density, only prominence plasma with $\rho > 10^{-14}\,\mathrm{g\,cm^{-3}}$ is shown. All velocities are averaged along the line-of-sight (here, the x-axis). E-H: The same for the sheared setup of Run I. The density in panel D and H is integrated along the x-axis.}
    \label{fig:fig6}
\end{figure}

Figure \ref{fig:fig6} shows flow patterns of the prominence fine structure for one snapshot of Run I and the Shear simulation. The velocities are averaged over the line of sight. For both simulations, counterstreaming flows are visible. In Run I (panel A-D), we see vertically oriented up-and-downflows side by side (panel C). Also along the x-direction (panel A), flows in both directions are visible. Like the fine-structure in Figure \ref{fig:fig4}, these counterstreaming flows develop a strong component along the y-axis in the Shear setup (panel E-H) due to the build-up of a y-component in the magnetic field during the shearing process. In observations, a variety of flows is seen, often including these counterstreaming flows as a striking feature \cite{zirker1998counter}. Quiescent prominences mostly show vertical \cite{berger2013solar} and active region prominences horizontal flows \cite{okamotoCoronalTransverseMagnetohydrodynamic2008}.

\section{Summary}

We have simulated the formation and properties of a solar prominence in MURaM for three different configurations of our setup. We find that prominence formation happens self-consistently by the random injection of a dense seed into the magnetic dips. The subsequent mass build-up of the prominence happens via a mixture of the injection and the condensation mechanism.  In contrast to other simulations that show the formation of prominences via condensation \cite{xiaSIMULATIONSPROMINENCEFORMATION2012, karpenOriginHighSpeedMotions2006b, fanMHDSimulationProminence2018}, condensation does not start the formation process in our simulation, but only sets in after the first dense seed settles in the magnetic dips via injection (see Supplementary Section \ref{sec:sup-sec5} for a discussion). This shows the importance of a self-consistent connection of the solar corona with the underlying photosphere and chromosphere for the formation of low-lying solar prominences. Even for larger prominences, the injection process may initiate prominence formation if the underlying magnetic field configuration rises from low altitudes into the corona during the prominence formation process. In this case, the presented model of injection followed by condensation would not be limited to low-lying prominences. Instead, it could also feed larger prominences by providing the initial dense material via injection before the magnetic structure rises higher into the atmosphere. After the initial seed has been provided, the prominence can grow via condensation at larger heights.

The formation and dynamics of the simulated prominences are similar for all runs, whereas the sizes and appearances in the fine structure differ. This shows that different-looking prominences can form in our simulations despite the simple underlying magnetic field configuration. We see similarities to observations of different prominence types in the fine-structure, whereas our sheared setup is more in line with the fine structures observed in active-region prominences than the non-sheared setups. In the future, we plan to set up more realistic magnetic field configurations, such as a magnetic flux rope, to study the differences in the structure and dynamics compared to our current setups, and to do more detailed comparisons with observations.

\section{Methods}
\subsection{The MURaM code}

MURaM is a 3D box-in-a-star radiative MHD code incorporating the physics to simulate the solar photosphere \cite{voglerSimulationsMagnetoconvectionSolar2005a}, chromosphere \cite{przybylskiChromosphericExtensionMURaM2022} and corona \cite{rempelEXTENSIONMURAMRADIATIVE2016}. MURaM solves the non-ideal MHD equations in 3D Cartesian coordinates

\begin{equation}
     \frac{\partial \rho}{\partial t} = - \nabla \cdot (\rho \mathbf{v}),
     \label{eq:mass}
\end{equation}

\begin{equation}
    \frac{\partial \rho \mathbf{v}}{\partial t} = -\nabla \cdot \left(\rho \mathbf{v} \mathbf{v}\right) - \nabla p + \rho \mathbf{g} + \mathbf{F}_{\mathrm{L}} + \mathbf{F}_{\mathrm{SR}},
    \label{eq:momentum}
\end{equation}

\begin{equation}
    \frac{\partial E_{\mathrm{HD}}}{\partial t}  =  -\nabla \cdot \left[ \mathbf{v} \left(E_{\mathrm{HD}} + p\right) + q \frac{\mathbf{B}}{|\mathbf{B}|} \right]+\rho \mathbf{v} \cdot \mathbf{g} + \mathbf{v} \cdot \mathbf{F_L}
+ \mathbf{v} \cdot \mathbf{F_{\mathrm{SR}}} + Q_\mathrm{rad} + Q_{\mathrm{res}}, 
\label{eq:energy}
\end{equation}

\begin{equation}
     \frac{\partial \mathbf{B}}{\partial t} = \nabla \times (\mathbf{v} \times \mathbf{B}),
     \label{eq:induction}
\end{equation}

with density $\rho$, velocity $\mathbf{v}$, pressure $p$, gravitational acceleration $\mathbf{g}$, magnetic field $\mathbf{B}$, the numerically computed Lorentz force $\mathbf{F_L}$ (see equation~\ref{eq:FL}), the hydrodynamic energy $E_{\mathrm{HD}} = E_{\mathrm{int}} + \frac{1}{2} \rho \mathbf{v}^2$, the radiative cooling/heating $Q_{\mathrm{rad}}$, the resistive heating $Q_{\mathrm{res}}$ from the diffusion scheme, the Spitzer heat flux $q$ and a semi-relativistic (Boris) correction $\mathbf{F_{\mathrm{SR}}}$ to prevent severe time step constraints (see \cite{przybylskiChromosphericExtensionMURaM2022,rempelEXTENSIONMURAMRADIATIVE2016} for details). The term $q \mathbf{B}/|\mathbf{B}|$ describes the conductive heat flux along the magnetic field lines, whereas the Spitzer heat flux $q$ is calculated via
 \begin{equation}
    \frac{\partial q}{\partial t} = \frac{1}{\tau}\left(-f_{\mathrm{\mathrm{sat}}}\sigma T^{\frac{5}{2}}\left(\frac{\mathbf{B}}{|\mathbf{B}|}\cdot\nabla\right)T-q\right) .
\end{equation}
The prefactor
\begin{equation}
    f_{\mathrm{sat}} = \left( 1 + \frac{ |\sigma T^{\frac{5}{2}} (\hat{\mathbf{b}} \cdot \nabla) T|}{1.5 \rho c_s^3}\right)^{-1} 
\end{equation}
takes the saturation of the conductive heat flux into account \cite{fisher1985flare,meyer2012second} (see \cite{rempelEXTENSIONMURAMRADIATIVE2016} for details). $\sigma = 10^{-6} \mathrm{erg\,cm^{-1}\,s^{-1}\,K^{-7/2}}$ is the Spitzer heat conductivity and $c_s$ is the sound speed.
 
The equations~\ref{eq:mass} to \ref{eq:induction} describe the conservation of mass (\ref{eq:mass}), momentum (\ref{eq:momentum}), and energy (\ref{eq:energy}) as well as the evolution of the magnetic field, following the induction equation \ref{eq:induction}. A hyperbolic divergence cleaner is applied to fulfill $\nabla \cdot \mathbf{B} = 0 $ \cite{dednerHyperbolicDivergenceCleaning2002}. There is no explicit treatment for viscosity and magnetic resistivity. Only numerical diffusivities are used as described in \cite{przybylskiChromosphericExtensionMURaM2022}, \cite{rempelEXTENSIONMURAMRADIATIVE2016} and \cite{rempelNUMERICALSIMULATIONSACTIVE2014}. Numerical diffusive fluxes are calculated from a slope-limited diffusion scheme \cite{rempelNUMERICALSIMULATIONSACTIVE2014}. Fourth order hyperdiffusion terms are added in the vertical direction to $\log \rho$, internal energy density $\epsilon = E_{\mathrm{int}}/\rho$ as well as the vertical components of the magnetic field and the velocity field ($B_z$,$v_z$). In the energy conservation equation \ref{eq:energy} the hydrodynamic energy $E_{\mathrm{HD}} = E_{\mathrm{int}} + \frac{1}{2} \rho v^2$, the sum of internal and kinetic energy, is used instead of the total energy to prevent numerical errors in regions where the magnetic energy is much larger than the hydrodynamic energy. Heating resulting from the diffusion scheme is included as  $Q_{\mathrm{res}}$. The viscous heating is not directly included in the energy conservation because energy that is removed from the kinetic energy by viscous heating is added to the internal energy.

The Lorentz force is numerically computed by \begin{equation}
    \mathbf{F}_{\mathrm{L}} = \frac{f_{A}}{4 \pi} \nabla \cdot \left( \mathbf{B} \mathbf{B} - \frac{1}{2}\mathbf{I}\mathbf{B}^2\right)
+ \frac{1-f_A}{4 \pi}  \left(\nabla \times \mathbf{B} \right) \times \mathbf{B},
\label{eq:FL}
\end{equation}
with identity matrix $\mathbf{I}$. The factor  $f_A=\frac{1}{\sqrt{1+\left(v_a/c_{\mathrm{max}}\right)^4}} $ describes the limitation of the Alfv\'en velocity $v_a$ that is used to prevent time step constraints. For the reduction of the Alfven velocity a semi-relativistic treatment with reduced speed of light is used (Boris correction). It introduces a force term $\mathbf{F_{\mathrm{SR}}}$ in the momentum and energy equation (see \cite{rempelEXTENSIONMURAMRADIATIVE2016}).

There are currently two different versions of the MURaM code: the coronal extension \cite{rempelEXTENSIONMURAMRADIATIVE2016} (also called MURaM-CE) and the chromospheric extension \cite{przybylskiChromosphericExtensionMURaM2022} (also called MURaM-ChE). The coronal extension includes optically thin losses and thermal conduction along the field lines. It treats the radiation in local thermodynamic equilibrium (LTE). The chromospheric extension \cite{przybylskiChromosphericExtensionMURaM2022} covers all that the coronal extension includes, and additionally includes non-LTE (NLTE) effects in the chromosphere. In comparison to the coronal extension, the chromospheric extension additionally includes a time-dependent treatment of hydrogen ionisation, an improved treatment of radiative losses in the chromosphere and a scattering multi-group radiation transfer scheme. In this paper, we distinguish between LTE and NLTE runs. We call it LTE when using the coronal extension only, whereas the NLTE runs also use the chromospheric extension.

Pressure, temperature and electron number are calculated from density and internal energy $E_{\mathrm{int}}$ using a pre-tabulated equation of state (EoS). In the LTE runs, an LTE EoS is used for the
plasma in all regimes. The LTE EoS is tabulated using the FreeEoS code \cite{2012ascl.soft11002I} and the abundances of \cite{asplund2009}. The NLTE runs additionally include an EoS with
a non-equilibirum treatment for hydrogen and $H_2$ molecules. This non-equilibrium EoS is used for for the near photosphere and atmosphere, where $p < 2 \cdot 10^5 \mathrm{\,dyne\,cm^{-2}}$. The two methods are smoothly joint at this pressure \cite{przybylskiChromosphericExtensionMURaM2022}. All non-hydrogen atoms are treated in LTE, also for low pressures. The system of hydrogen rate equations includes the ground state, four excited states and the continuum.

MURaM includes different radiative transfer treatments for the different physical regimes of the solar atmosphere. For the photosphere, where the populations of atomic levels are dominated by collisions, the time-independent LTE radiative transfer equation is solved. A multigroup method is used to calculate the photospheric radiation field \cite{nordlund1982, nordlundstein1990}. In the chromospheric extension \cite{przybylskiChromosphericExtensionMURaM2022} a scattering term is added in the source function, following the prescription of \cite{skartlien2000multigroup} and \cite{hayek2010radiative}. For the optically thin gas in the corona, optically thin losses are calculated from tabulated losses by an overlap interval approach \cite{rempelEXTENSIONMURAMRADIATIVE2016}. The transition between the two approaches is defined by thresholds in optical depth and pressure (see \cite{przybylskiChromosphericExtensionMURaM2022} for further details). Two different recipes are implemented for the optically thin losses: a Chianti loss function \cite{landiCHIANTIATOMICDATABASE2011} and a Carlsson \& Leenaarts description \cite{carlssonApproximationsRadiativeCooling2012}. The latter includes NLTE models for \ion{H}{i}, \ion{Ca}{ii} and \ion{Mg}{ii} line losses as well as optically thin losses. The chromospheric extension furthermore includes EUV back-heating of the chromospheric plasma, following \cite{carlssonApproximationsRadiativeCooling2012}. The resulting radiative heating/cooling term in equation \ref{eq:energy} is therefore $Q_{\mathrm{rad}} = Q_{\mathrm{RT}} + Q_{\mathrm{thin}} + Q_{\mathrm{H}} + Q_{\mathrm{Mg}} + Q_{\mathrm{Ca}} + Q_{\mathrm{back}}$ in the NLTE simulations, with optically thin losses $Q_{\mathrm{thin}}$, back-heating $Q_{\mathrm{back}}$, line losses $Q_{\mathrm{H}}$, $Q_{\mathrm{Mg}}$, $Q_{\mathrm{Ca}}$ due to the elements \ion{H}{i}, \ion{Mg}{ii} and \ion{Ca}{ii}, and cooling/heating $Q_{\mathrm{RT}}$ from the multigroup radiation transfer scheme. 

\subsection{Simulation setup}

 In the following, we differentiate between two simulation runs that have the same general structure, but different magnetic field configurations and a slightly different bottom boundary condition. We refer to them as Run I and Run II, according to the setup presented in this section. We used a 3D box with a size of 80 Mm (x) x 10 Mm (y) in horizontal direction and a height of 32 Mm (z) for Run I, and a size of 80 Mm (x) x 10 Mm (y) in horizontal direction and a height of 34 Mm (z) for Run II. The resolution is 80 km x 80 km horizontally and 50 km vertically, corresponding to 1000 x 125 x 640 grid cells for Run I and 1000 x 125 x 680 grid cells for Run II. The difference in the vertical size comes from different convection zone depths: the convection zone is 2\,Mm deep for Run I and 4\,Mm deep for Run II. The atmosphere goes up to 30\,Mm above the surface for both runs. The third Run that we present is a sheared configuration based on the setup of Run I, which is explained below. Apart from the bottom boundary condition that is used to drive the shearing motions, all properties that are presented here for Run I are also valid for the sheared configuration. The resolution used for the presented simulations is with a maximum of 50 km relatively low. In future setups, a higher resolution will be needed to study the small-scale dynamics of the simulated prominences in detail.

The 3D run was started from an evolved state of a 2D run in both cases. The 2D run makes up the xz-plane and is later extended along the y-direction for the 3D run. The initial conditions for the 2D runs were chosen such that they provide a dipped magnetic field, as shown in Extended Data Figure \ref{fig:ed-fig1} and \ref{fig:ed-fig2}. This initial magnetic field configuration was created by fixing a set of vertical magnetic field columns below the surface (up to 100 km below the surface) and performing a potential-field extrapolation into the atmosphere. For Run I, we chose six narrow columns for the vertical magnetic field. To get an overall higher field strength for Run II, we instead used four thicker columns while keeping the amplitude the same as in Run I. The columns in Run I all have the same spatial extend of 1.6\,Mm in the x-direction. The inner/outer vertical magnetic field columns for Run II each span
3.2 Mm/6.4 Mm. In both runs, the vertical magnetic field in the columns follows a Gaussian distribution in the horizontal (x) direction, with a maximum amplitude of $B_z=4.8 \,\mathrm{kG}$ for the four (Run I)/two (Run II) outer columns and $B_z=2.88\,\mathrm{kG}$ for the two inner columns. From left to right, the columns have polarity  $+,+,-,+,-,-$ for Run I and $+,-,+,-$ for Run II, such that the magnetic field resembles a quadrupolar structure in both cases. Their initial field strength is constant in the vertical direction within the convection zone. The horizontal field $B_x$ is zero initially in the whole convection zone.

We aimed to create a stable dipped magnetic field configuration, such that mass can settle in the dips and stay there to form a prominence. To make the magnetic field shown in Extended Data Figure \ref{fig:ed-fig1} and \ref{fig:ed-fig2} stable for extended lengths of time, we tied the field lines of the six/four vertical columns to the bottom of the convection zone. This is done by fixing the value of $B_z$ in the first ten (Run I)/ twenty (Run II) grid cells of the domain to the magnetic field distribution in the columns as described above. Outside of the vertical columns, $B_z$ is set to 0 at the bottom boundary. Apart from $B_z$, the only difference in the boundary conditions between Run I and Run II is the treatment of flows within the vertical magnetic field columns: for Run I, the bottom boundary is open for flows everywhere, inside and outside of the columns. For Run II, flows within the vertical magnetic field columns are suppressed by setting an asymmetric boundary condition for all velocity components. Outside of the columns, the boundary is open for flows. The bottom boundary condition for $B_x$ (and $B_y$ in the 3D run) is symmetric everywhere. The boundary conditions are periodic in both horizontal directions. The top boundary is open for outflows and a potential field extrapolation is used for the magnetic field.

After running the 2D calculation for a few hours of solar time, we picked one snapshot and stacked it in the y-direction to create a 3D simulation. To break the symmetry in this direction, random fluctuations were introduced in the slices. The resulting magnetic field configuration is a 3D dipped magnetic arcade (see Figure \ref{fig:fig1} and Extended Data Figure \ref{fig:ed-fig5}). No shearing motion was applied for Run I and Run II. Therefore, the initial magnetic field configuration includes no $B_y$ component, in contrast to what is normally observed in prominences\cite{parenti2014solar}. The resulting photospheric magnetic field and its evolution is shown Figure \ref{fig:fig1} for Run I, in Extended Figure \ref{fig:ed-fig5} for Run II and in the Supplementary Videos 1 and 9. For the sheared setup of Run I, a shearing motion was applied to the y-component of the velocity at the bottom boundary of the middle vertical magnetic field columns (the second and the fifth column from the left in Extended Data Figure \ref{fig:ed-fig1}). Starting with a snapshot that already includes a formed prominence, the two columns were sheared with a velocity amplitude of $v_y = -2\,\mathrm{km/s}$ and $v_y = +2\,\mathrm{km/s}$. The shearing profile follows the same Gaussian distribution along the x-axis as the magnetic field. This led to a shearing of the magnetic loops between the two middle columns. This shear is decreasing with height, such that the lower part of the prominence is sheared the most. With increasing height, the shear becomes weaker, and the sheared arcade smoothly transitions to the overlying coronal loops that are only weakly sheared. When the setup is sufficiently sheared, we start to continuously decrease the shear amplitude until a value of $100\,\mathrm{m/s}$ is reached, which is then kept constant. For the sheared setup, our goal is to compare the prominence properties with the non-sheared setup. We do not simulate the formation from scratch for the sheared prominence. 
Extended Data Figure \ref{fig:ed-fig3} shows the shear angle of the sheared setup along the polarity inversion line in dependence of height above the surface. The shear angle is calculated via $\alpha = \arctan{(B_x/B_y)}$. The $B_x$ component changes orientation at the nullpoint, so that the shear angle changes sign at ${\sim}\,7\,\mathrm{Mm}$. 

Extended Data Figure \ref{fig:ed-fig4} shows the horizontally averaged magnetic field and the magnetic field in the prominence core as a function of height for all runs. The average magnetic field strength at the surface is ${\sim}\,350\,\mathrm{G}$ for Run I, ${\sim}\,500\,\mathrm{G}$  for Run II and ${\sim}\,400\,\mathrm{G}$ for the sheared setup. The magnetic field strength in the magnetic dips at roughly half the height of the prominence is  ${\sim}\,35\,\mathrm{G}$ for Run I, ${\sim}\,60\,\mathrm{G}$ for Run II and ${\sim}\,45\,\mathrm{G}$ for the sheared setup. The average field at these heights is larger than the field in the prominence itself, being ${\sim}\,80\,\mathrm{G}$ in Run I and in the sheared setup, and ${\sim}\,250\,\mathrm{G}$ in Run II. For the sheared case, we see that $B_y > B_x$ over most of the prominence height along the polarity inversion line, but not when averaged over the full box width. This shows that the shear is stronger around the polarity inversion line than in the rest of the box.

Additionally to the different used magnetic field configurations, we performed simulations in LTE and NLTE, as described in the previous subsection. Due to the long time it needs for the prominence seed to be ejected into the corona (7-10 hours from the start of the 3D simulation), the formation process is only simulated in our LTE runs. NLTE runs for Run I and Run II are started from an evolved state of the already existing prominence, because of the higher computational needs of such runs. The LTE treatment uses the Chianti loss function \cite{landiCHIANTIATOMICDATABASE2011}, the NLTE treatment the Carlsson \& Leenaarts loss function \cite{carlssonApproximationsRadiativeCooling2012}. The Shear setup is only simulated in LTE.

\vspace{0.75cm}
{\bf Data availability}:
The data for snapshots of both Run I and II are available for download via the Edmond Open Research Data Repository of the Max Planck Society at \url{https://doi.org/10.17617/3.8YEPQW} (ref. \cite{3.8YEPQW_2026}).

{\bf Code availability}:
The coronal extension of the MURaM Code \cite{rempelEXTENSIONMURAMRADIATIVE2016} is available for download at \url{https://github.com/NCAR/MURaM_main}.

{\bf Acknowledgements}: 
This work was supported by the International Max Planck Research School (IMPRS) for Solar System Science at the University of Göttingen and at TU Braunschweig (L.M.Z). This work furthermore received funding from the European Research Council (ERC) under the European Union's Horizon 2020 research and innovation programme (grant agreement No. 101097844) (S.K.S and L.M.Z). The work of D.P. was funded by the Federal Ministry for Economic Affairs and Climate Action (BMWK) through the German Space Agency at DLR based on a decision of the German Bundestag (Funding code: 50OU2201). We highly appreciate the computing resources provided by the HPC systems Raven, Viper and Cobra at the Max Planck Computing and Data Facility. Hinode is a Japanese mission developed and launched by ISAS/JAXA, with NAOJ as domestic partner and NASA and STFC (UK) as international partners. It is operated by these agencies in co-operation with ESA and NSC (Norway).
\\

{\bf Author contributions}: 
L.M.Z. performed the simulations and data analysis, and wrote the initial text of the study. R.H.C., S.K.S and D.P. provided physical ideas, helped with the data interpretation, and provided expertise for the MURaM code. D.P. started the NLTE simulations with the MURaM-ChE code. 
\\

{\bf Corresponding author}:
Correspondence and requests for materials
should be addressed to Lisa-Marie Zessner (zessner@mps.mpg.de).
\\

{\bf Competing interests}:
The authors declare no competing interests.
\\

{\bf Additional Information}:
Supplementary Information is available for this paper (see page 37 onwards). It includes Supplementary Text with more information on the the prominence energetics, the injection events and the prominence mass supply. The last section of the Supplementary Information provides a description of the available Supplementary Videos, which can be accessed here \url{https://doi.org/10.1038/s41550-026-02840-7}.

\bibliographystyle{sn-nature}

\renewcommand{\figurename}{Extended Data Figure}
\setcounter{figure}{0}

\newpage
\clearpage
\section{Extended Data Figures}
\begin{figure}
    \centering
    \includegraphics[width = \textwidth]{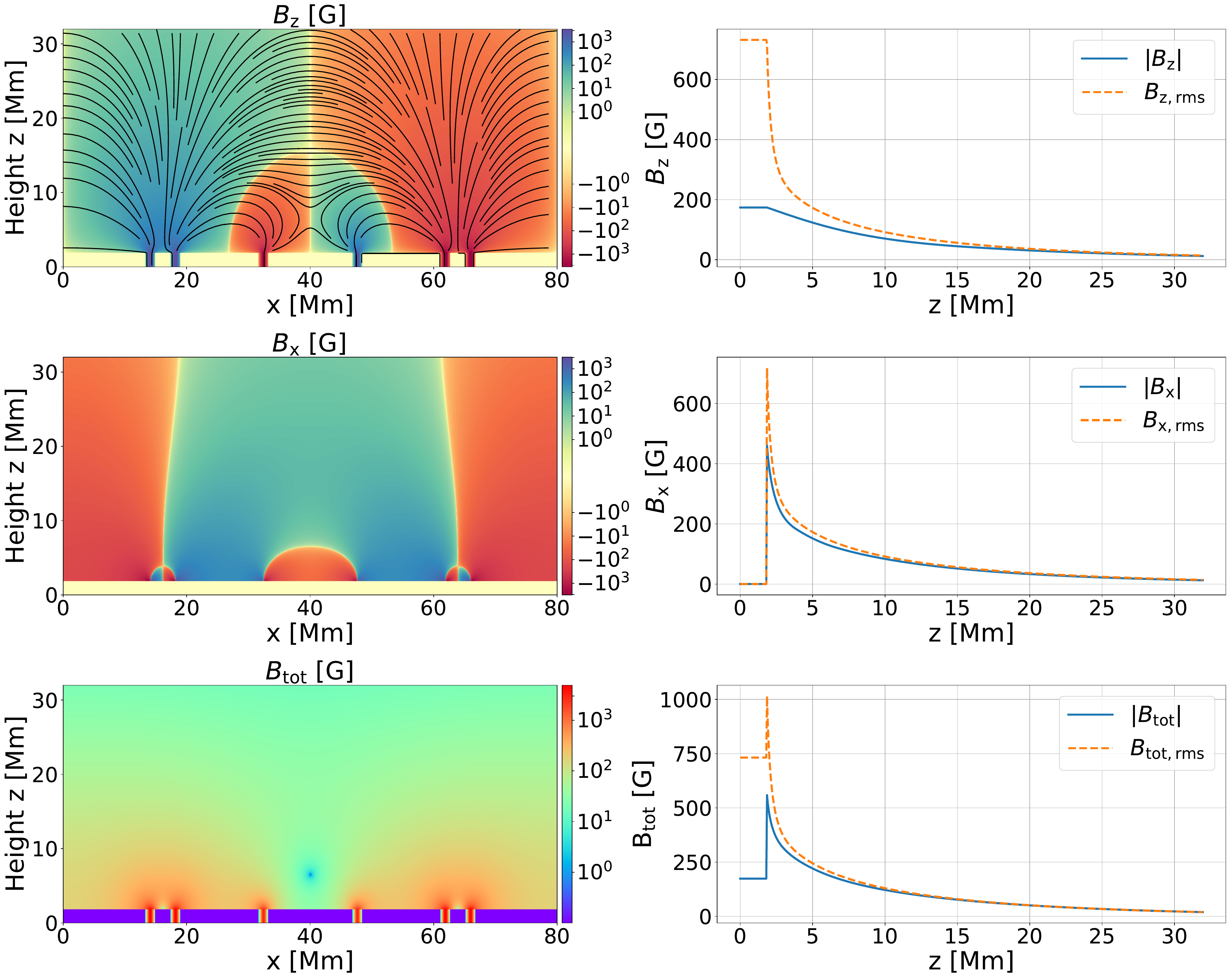}
   
    \caption{\textbf{2D initial condition for the magnetic field configuration in Run I}. The bottom boundary is here at $z = 0$, and the surface at $z=2\,\mathrm{Mm}$. The polarity inversion line over which the prominence forms later in the 3D simulation is at $x = 40\,\mathrm{Mm}$. Left panels: 2D Maps of the vertical magnetic field component (top), the horizontal magnetic field component (middle) and the total magnetic field (bottom). The top left panel additionally shows magnetic field lines of the initial condition in black. Right panels: Horizontally averaged magnetic field strength in dependence of height over the bottom boundary for the vertical magnetic field component (top), the horizontal magnetic field component (middle) and the total magnetic field (bottom). The blue lines show the unsigned magnetic field and the dashed orange lines show the root mean square value of the magnetic field. The horizontal averages are taken over the whole box.} 
    \label{fig:ed-fig1}
\end{figure}

\newpage
\clearpage
\begin{figure}
    \centering
    \includegraphics[width = \textwidth]{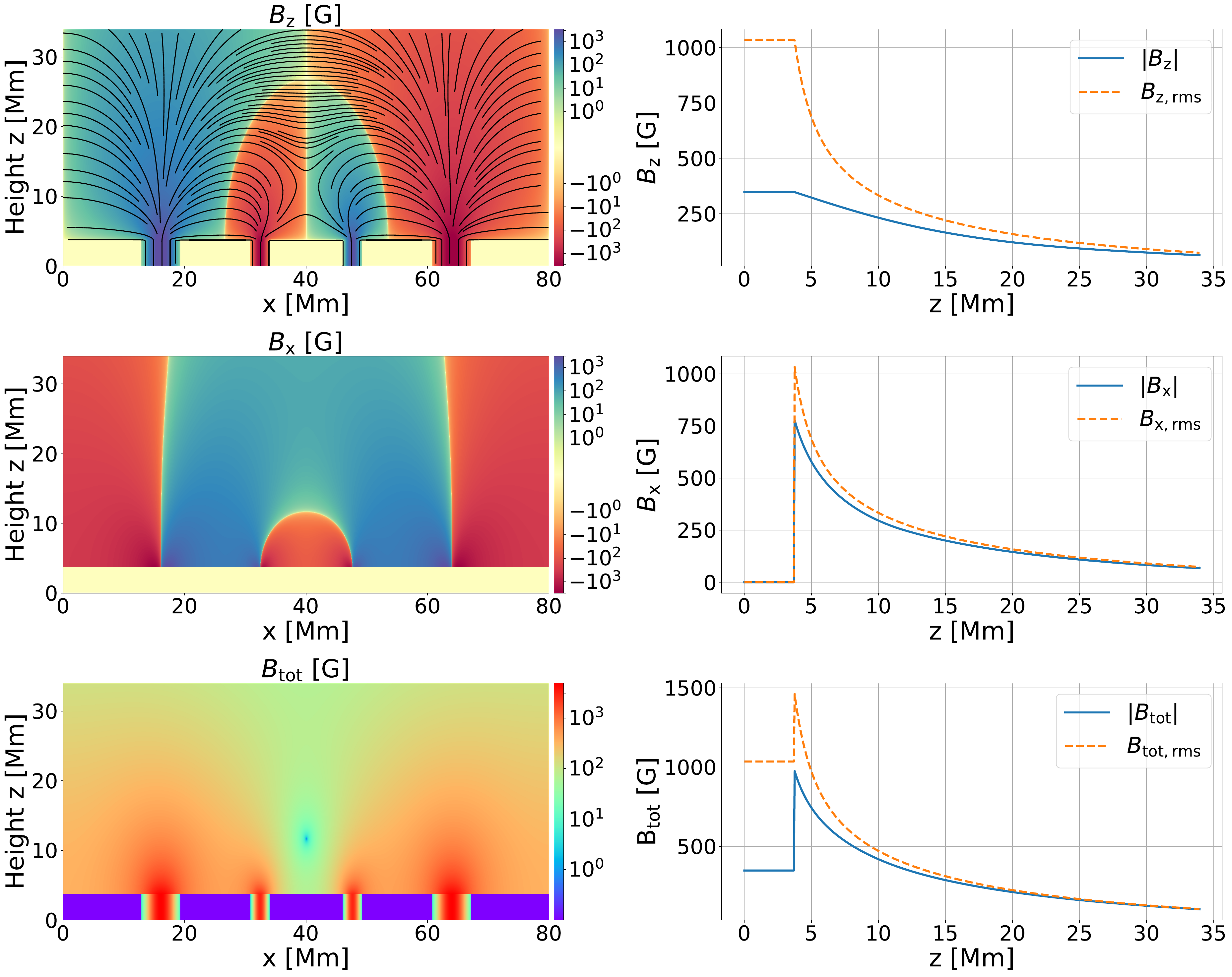}
   
    \caption{\textbf{2D initial condition for the magnetic field configuration in Run II}. The bottom boundary is here at $z = 0$, and the surface at $z=4\,\mathrm{Mm}$. The polarity inversion line over which the prominence forms later in the 3D simulation is at $x = 40\,\mathrm{Mm}$. Left panels: 2D Maps of the vertical magnetic field component (top), the horizontal magnetic field component (middle) and the total magnetic field (bottom). The top left panel additionally shows magnetic field lines of the initial condition in black. Right panels: Horizontally averaged magnetic field strength in dependence of height over the bottom boundary for the vertical magnetic field component (top), the horizontal magnetic field component (middle) and the total magnetic field (bottom). The blue lines show the unsigned magnetic field and the dashed orange lines show the root mean square value of the magnetic field. The horizontal averages are taken over the whole box.} 
    \label{fig:ed-fig2}
\end{figure}

\newpage
\begin{figure}
    \centering

     \includegraphics[width= 1 \textwidth]{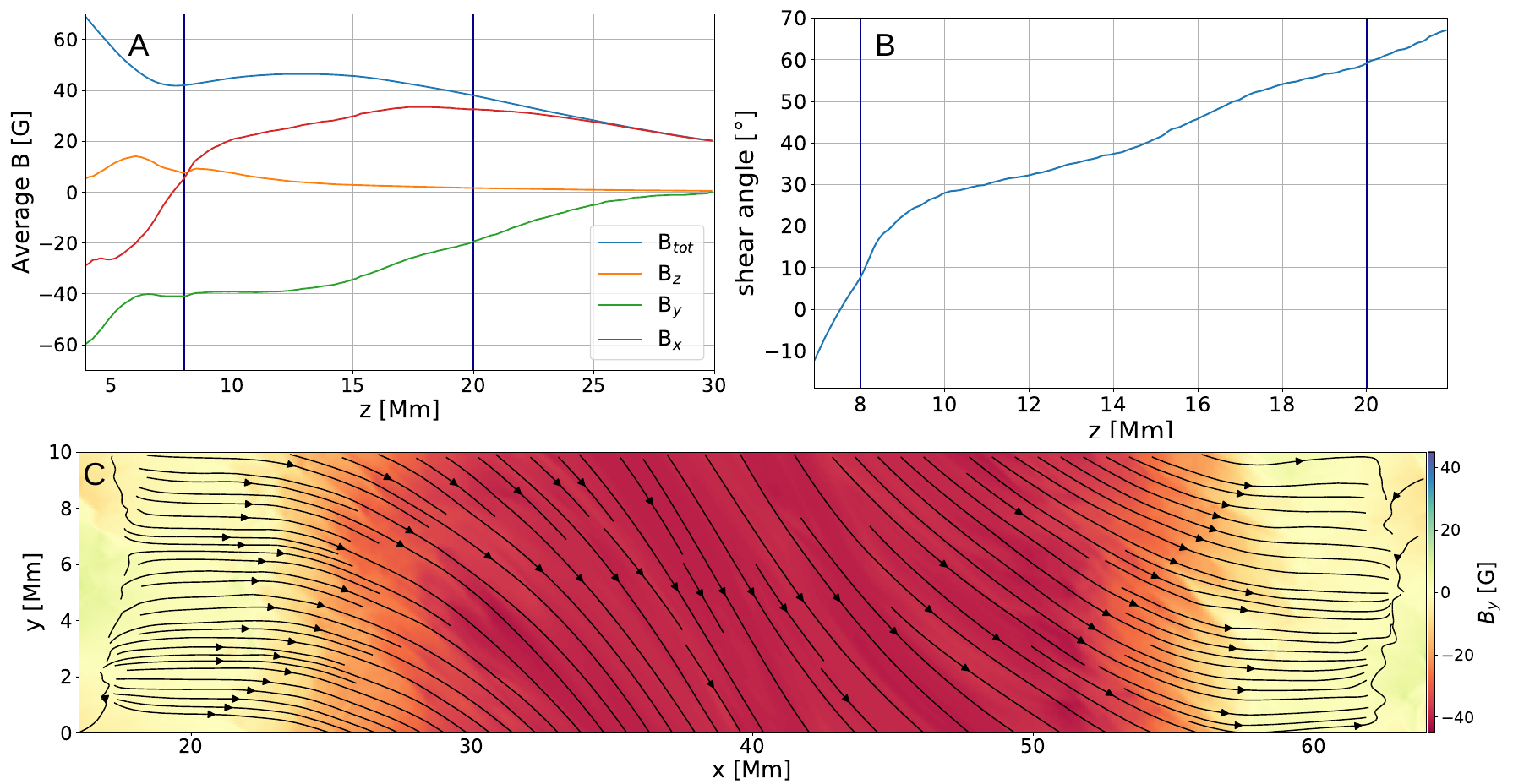}
    
    \caption{\textbf{Magnetic field strength and orientation for the Shear run.} A: Magnetic field components for the sheared setup of Run I, averaged over the polarity inversion line and a time frame of 10 minutes. The different curves show the total magnetic field ($B_{\mathrm{tot}}$, blue line), the vertical magnetic field component ($B_{\mathrm{z}}$, orange line) and the two horizontal magnetic field components ($B_{\mathrm{y}}$, green line, and $B_{\mathrm{x}}$, red line). The components $B_x$ and $B_y$ are used to determine the shear angle in panel B. B: Shear angle for the sheared setup of Run I in dependence of height above the surface, averaged over the polarity inversion line and a time frame of 10 minutes. The shear angle measures the angle between the y-direction (along which the long axis of the prominence is located) and the horizontal direction of the magnetic field. The blue vertical lines indicate the lower and upper height limit between which the prominence mass is mainly found. The prominences can occasionally be higher than the indicated sizes. C: Example for the magnetic field component in the y-direction for the sheared setup of Run I, taken along a horizontal slice. Magnetic field lines projected onto the plane are added in black. The height of the horizontal slice is $12\,\mathrm{Mm}$ above the surface.}
    \label{fig:ed-fig3}
\end{figure}

\newpage
\begin{figure}
    \centering

    \includegraphics[width= 1 \textwidth]{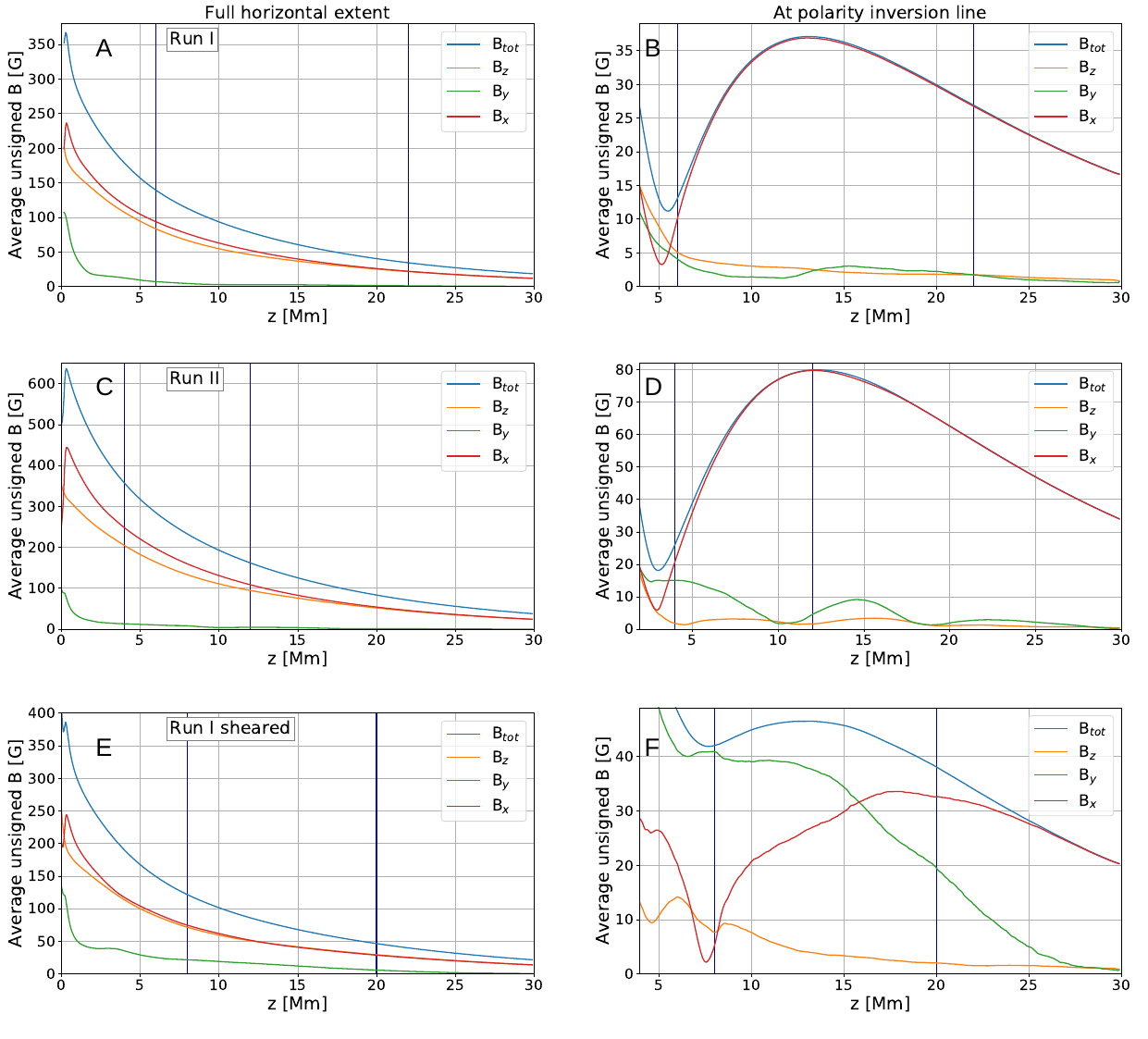}
    
    \caption{\textbf{Magnetic field strength within the whole simulation box and at the location of the prominence.} Magnetic field strength for Run I (A,B), Run II (C,D) and the sheared setup of Run I (E,F), averaged over a time frame of 50 minutes for Run I and Run II, and over 10 minutes for the sheared configuration. The chosen time frames start after the prominences are fully built up. The different curves show the total magnetic field ($B_{\mathrm{tot}}$, blue line), the vertical magnetic field component ($B_{\mathrm{z}}$, orange line) and the two horizontal magnetic field components ($B_{\mathrm{y}}$, green line, and $B_{\mathrm{x}}$, red line). The blue vertical lines indicate the lower and upper height limit between which the prominence mass is mainly found. The prominences can occasionally be higher than the indicated sizes. A,C,E: Horizontally averaged unsigned magnetic field in dependence of height over the surface. B,D,F: Unsigned magnetic field averaged along the polarity inversion line as a measurement for the magnetic field in the prominence core, in dependence of height over the surface. The height for the sheared setup is higher than for Run I because the shearing process lifts the height of the magnetic dips.}
    \label{fig:ed-fig4}
\end{figure}

\newpage
\begin{figure}
    \centering
    \includegraphics[width = 1 \textwidth]{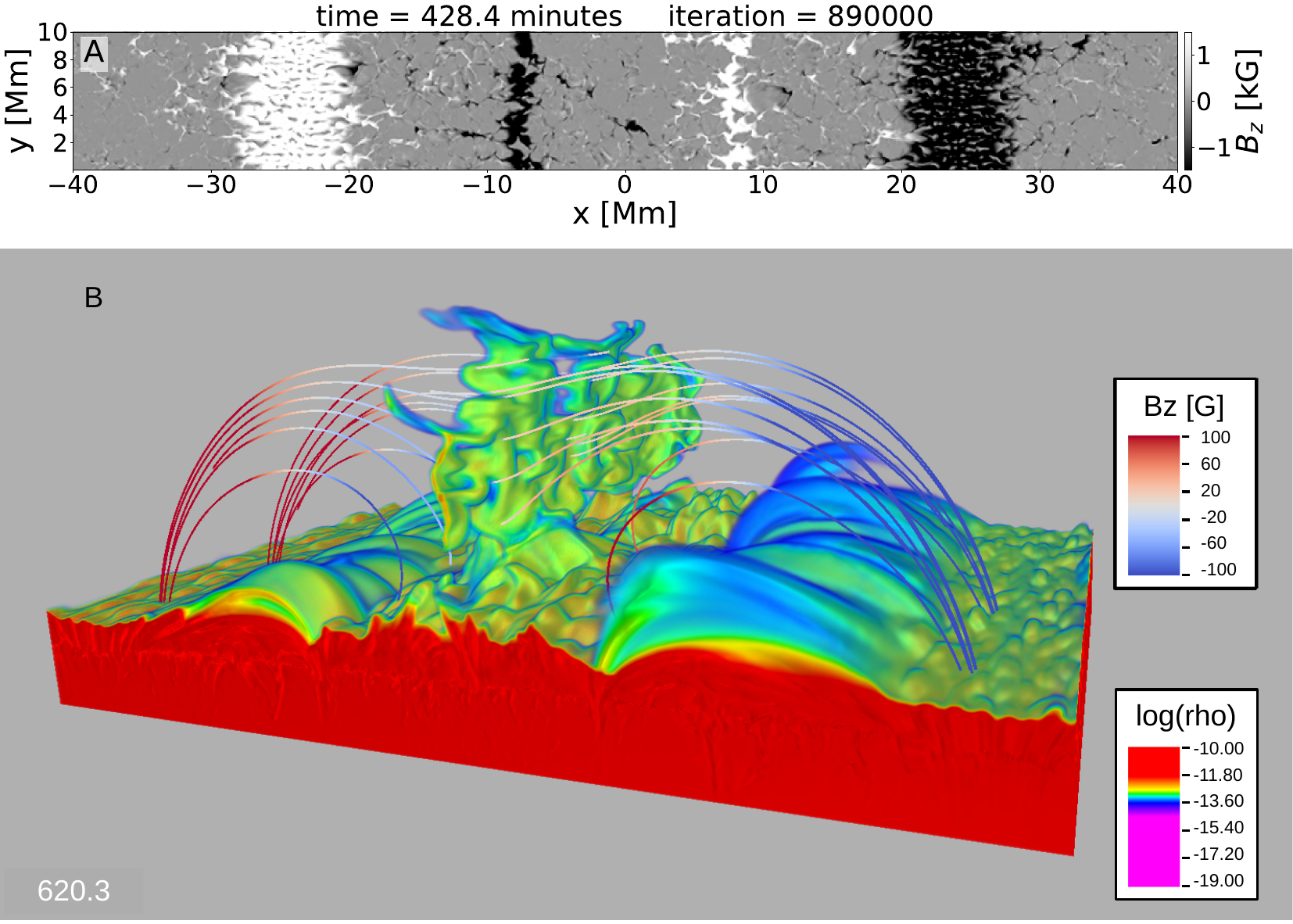}
    \caption{\textbf{Visualisation of the prominence and magnetic field structure for Run II.} Equivalent to Figure~\ref{fig:fig1} of the main text, showing the configuration of the magnetic field for Run II. A: Magnetic field at the $\tau_{500} = 1$ surface at around 30 minutes after prominence formation starts. Supplementary Video 9 shows an animation of this panel. B: 3D rendering of the prominence density (rainbow colored) with a set of field lines to demonstrate the quadrupolar structure of the magnetic field (blue-red field lines). This snapshot shows the prominence 200 minutes after the formation starts. The coloring of the plasma (rainbow colormap) shows the logarithmic gas density (in $\mathrm{g/cm^3}$). For plasma with a density below $10^{-14}\,\mathrm{g/cm^3}$, the opacity is set to zero for this image, such that the surrounding corona is not visible. The coloring of the magnetic field lines (blue-red colormap) shows the value of the vertical magnetic field component. The number at the bottom left shows the time in minutes from the start of the simulation.} 
   
    \label{fig:ed-fig5}
\end{figure}

\newpage
\begin{figure}
    \centering
    \includegraphics[width=0.9 \textwidth]{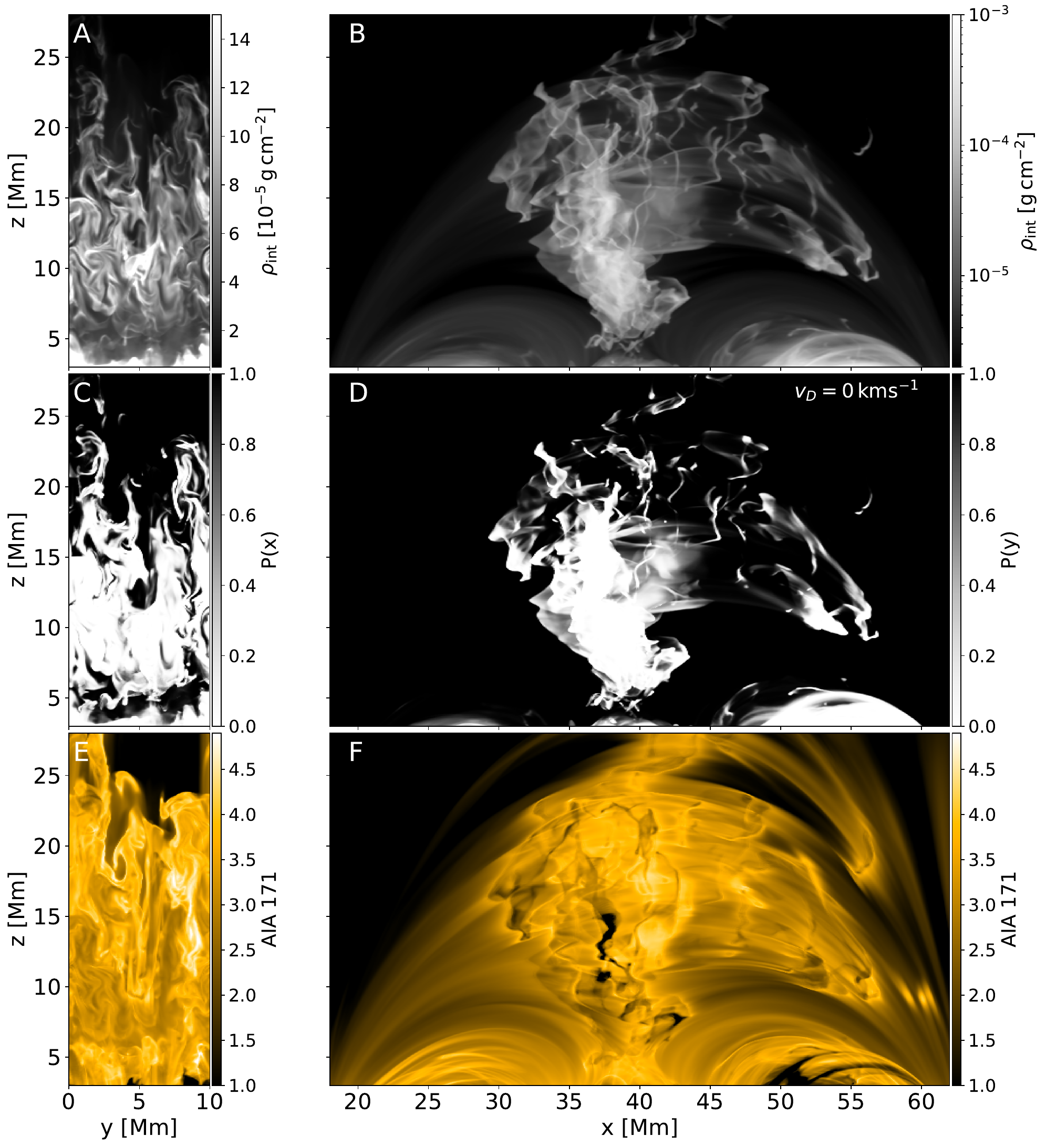}
    \caption{\textbf{Appearance of the prominence in Run I NLTE in H$\mathrm{\alpha}$ and AIA 171}. The same as Figure~\ref{fig:fig5} of the main text, but now for a snapshot from Run I NLTE. Due to the saturation of the H$\mathrm{\alpha}$ proxy in the line core, some of the substructures visible in the integrated density (A,B) are washed out in the H$\mathrm{\alpha}$ images (C,D). For Run I, the saturation in H$\mathrm{\alpha}$ is clearer, especially for the bottom and middle part of the prominence, because the prominence is thicker and more massive compared to the prominence in Run II. The less dense structures in the upper part of the prominence look more pronounced compared to the structures in integrated density. The prominence fine structure that is visible in integrated density and in H$\mathrm{\alpha}$ can be partly seen in absorption in AIA 171. For comparison to the appearance in the H$\mathrm{\alpha}$ line core, we show an image of Run I in the blue wing of H$\mathrm{\alpha}$ at a Doppler velocity of $v_D = 12 \,\mathrm{km\,s^{-1}}$ in Extended Data Figure \ref{fig:ed-fig7}.} 

    \label{fig:ed-fig6}
\end{figure}

\newpage
\begin{figure}
    \centering
    \includegraphics[width=\textwidth]{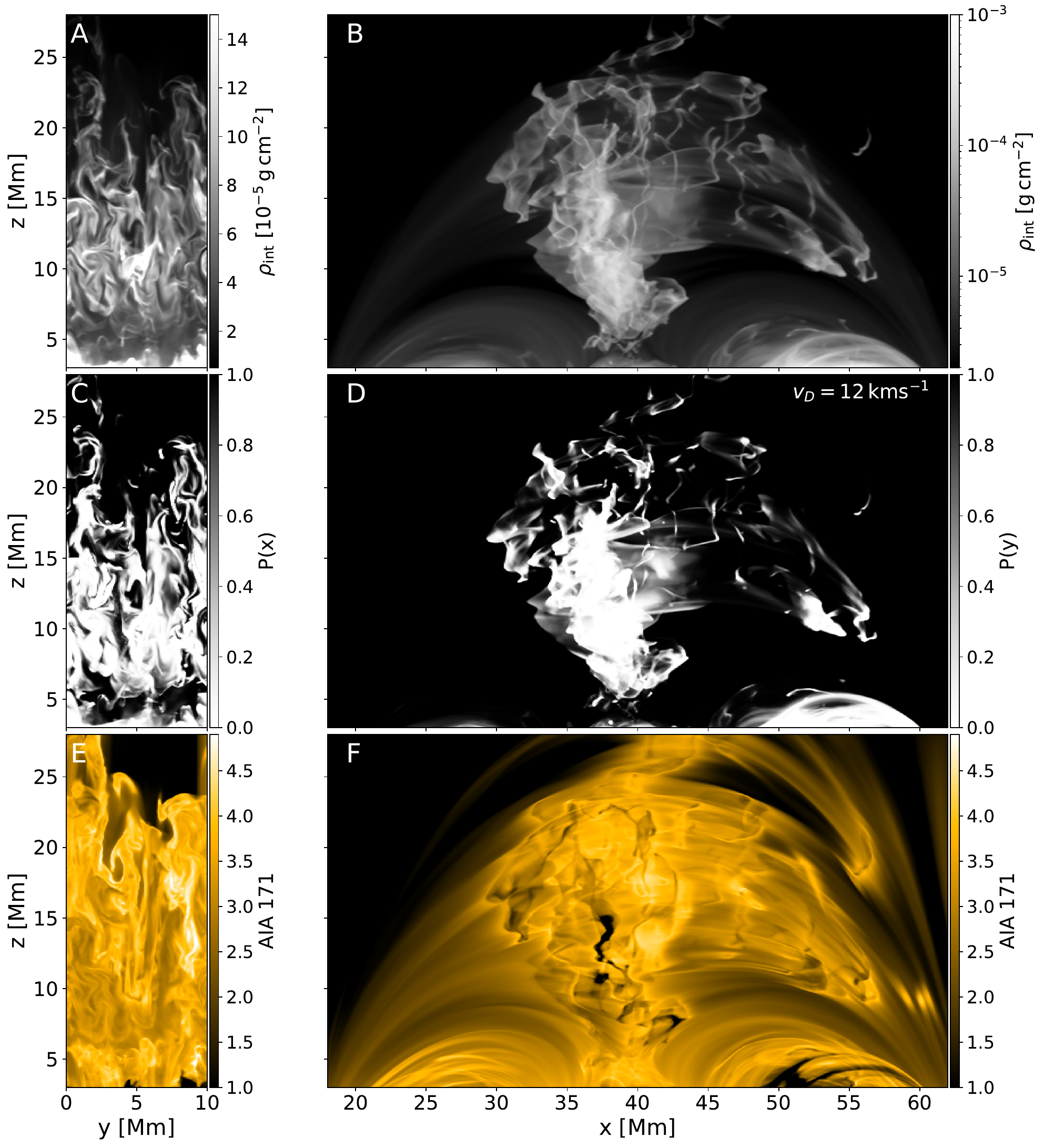}
    \caption{\textbf{Appearance of the prominence in Run I NLTE in the blue wing of H$\mathrm{\alpha}$ and in AIA 171}. Same as Extended Data Figure~\ref{fig:ed-fig6}, but the H$\mathrm{\alpha}$ emission (panels A and B) is here shown for the blue wing at a Dopper velocity of $v_D = 12\,\mathrm{km\,s^{-1}}$. In the side view (panel C), the fine structure is less saturated and thus more visible compared to the structure seen in the H$\mathrm{\alpha}$ line core (panel C in Extended Data Figure \ref{fig:ed-fig6}).}
    \label{fig:ed-fig7}
\end{figure}

\include{supplementary}

\end{document}

%% file: supplementary.tex
{\LARGE\noindent\textbf{Supplementary Text}}
\vspace{0.5cm}

\renewcommand{\figurename}{Supplementary Figure}
\renewcommand{\thetable}{S\arabic{table}}
\renewcommand{\thesection}{S\arabic{section}}
\renewcommand{\theequation}{S.\arabic{equation}}

\setcounter{figure}{0}
\setcounter{section}{0}
\setcounter{equation}{0}

{\noindent}The following sections include Supplementary Text, including Supplementary Figures 1 to 14 and the captions for the Supplementary Videos 1 to 14.

\setlength{\parindent}{0pt}

\section{Energetics, radiative cooling and pressure}
\label{sec:sup-sec1}

This section gives additional information on the energetics within the prominence and its surrounding transition region, as well as the siphon flows that are discussed in the main text (see Figure~\ref{fig:fig2}C-F). We present the heating and cooling rates, and show examples for the pressure and cooling distributions for LTE and NLTE runs. 

{\noindent}The basis of the thermodynamic balance within the prominence can be seen in Supplementary Figure~\ref{fig:sup-fig1} for the LTE and NLTE versions of Run I. The left panels show the horizontal profiles of the terms dominating the energy evolution equation (see Equation~3 in the Methods Section), and the right panels show the corresponding composition of the radiative cooling term. The spatial profiles in each y-slice of the simulation were first centered on their respective center of mass before averaging a snapshot over the y-axis. These profiles change throughout the simulation, depending on the state of the prominence. Supplementary Figure~\ref{fig:sup-fig1} shows an example for an approximately vertical state of the prominence. For all runs, the cooling of the structure is largely dominated by the radiative losses (blue curve in the left panels, labeled as '$Q_{rad}$'), which are divided into different contributions as shown in the right panels of the Figure: for the LTE runs, this includes the cooling and heating contribution from the radiative transfer scheme (labeled as RT losses '$Q_{RT}$', blue curve in the right panels) and the contributions from a tabulated optically thin loss function for the corona (therefore labeled as optically thin losses '$Q_{thin}$', orange curve). Additionally, the Carlsson \& Leenaarts treatment in the NLTE runs includes tabulated line losses for H~{\sc i} ('$Q_{H}$', green curve), Mg~{\sc ii} ('$Q_{Mg}$', red curve), and Ca~{\sc ii} ('$Q_{Ca}$', purple curve) that were calculated with a tabulated NLTE model (see the Methods Section for more information about the radiative losses). The prominence structure is heated by heat conduction ('$\nabla \cdot F_{con}$', orange curve) at the edges and by advection in the core ('$\nabla \cdot F_{adv}$', light blue curve). The NLTE simulation is run for a short time frame compared to the LTE simulation. During the NLTE runtime, the prominence partly drains to one side and does thus not reach a properly vertical state along the whole y-axis. Therefore, the advective term for the NLTE simulation looks more variable than in the LTE case. In contrast to the LTE simulation, the NLTE case includes a chromospheric backheating term ('$Q_{back}$', darkred curve) that additionally heats the prominence core.

{\noindent}While the NLTE runs are generally similar to the LTE runs, the distribution of temperatures in the prominence differs between the LTE and the NLTE simulations. We will address this in a follow-up paper about different radiative treatments for the setup of Run I. The prominence temperatures stated in the main text correspond to the temperatures found in the LTE runs.

\begin{figure}[!htp]
    \centering

    \includegraphics[width = 0.49 \textwidth]{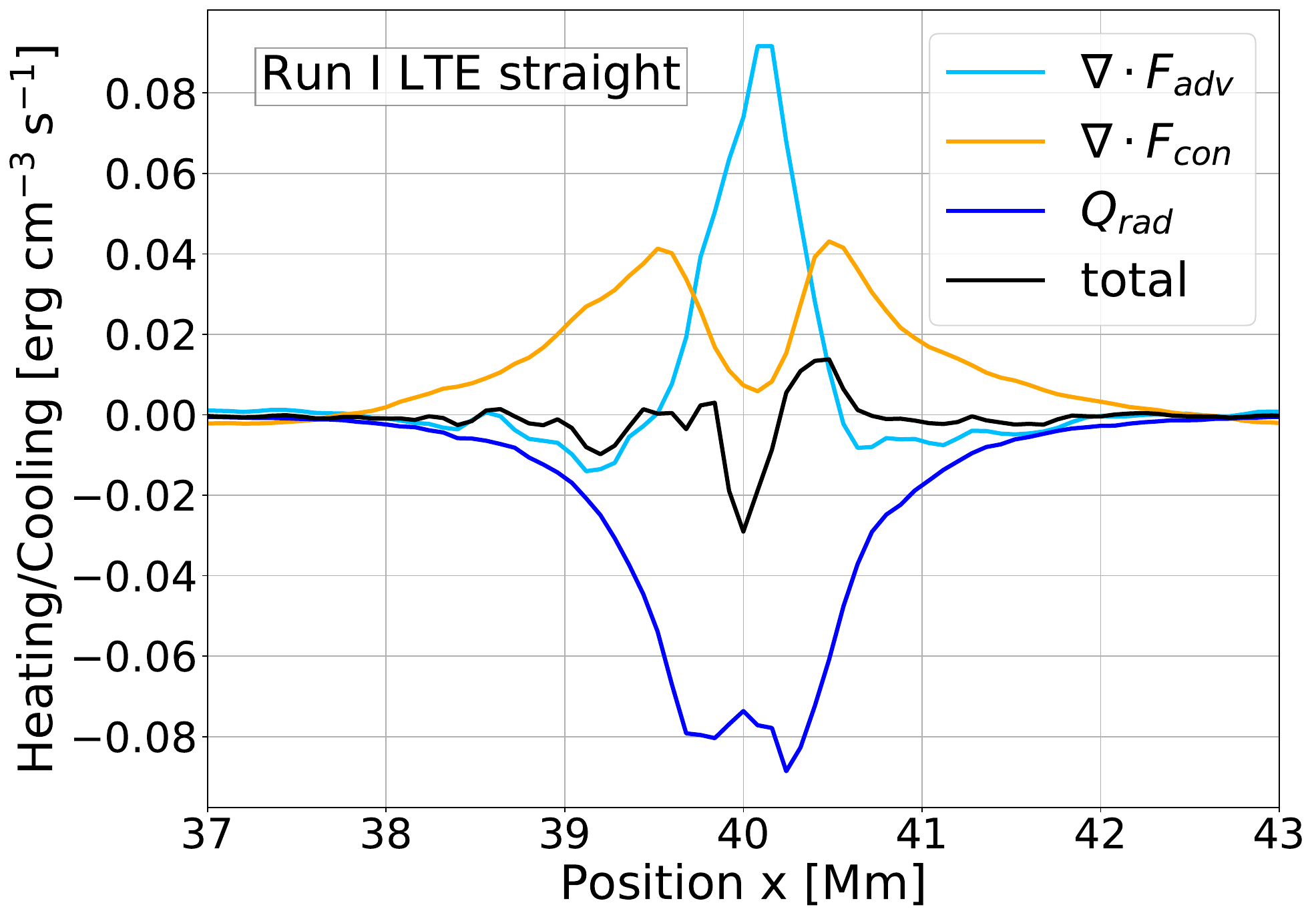}
    \includegraphics[width = 0.49 \textwidth]{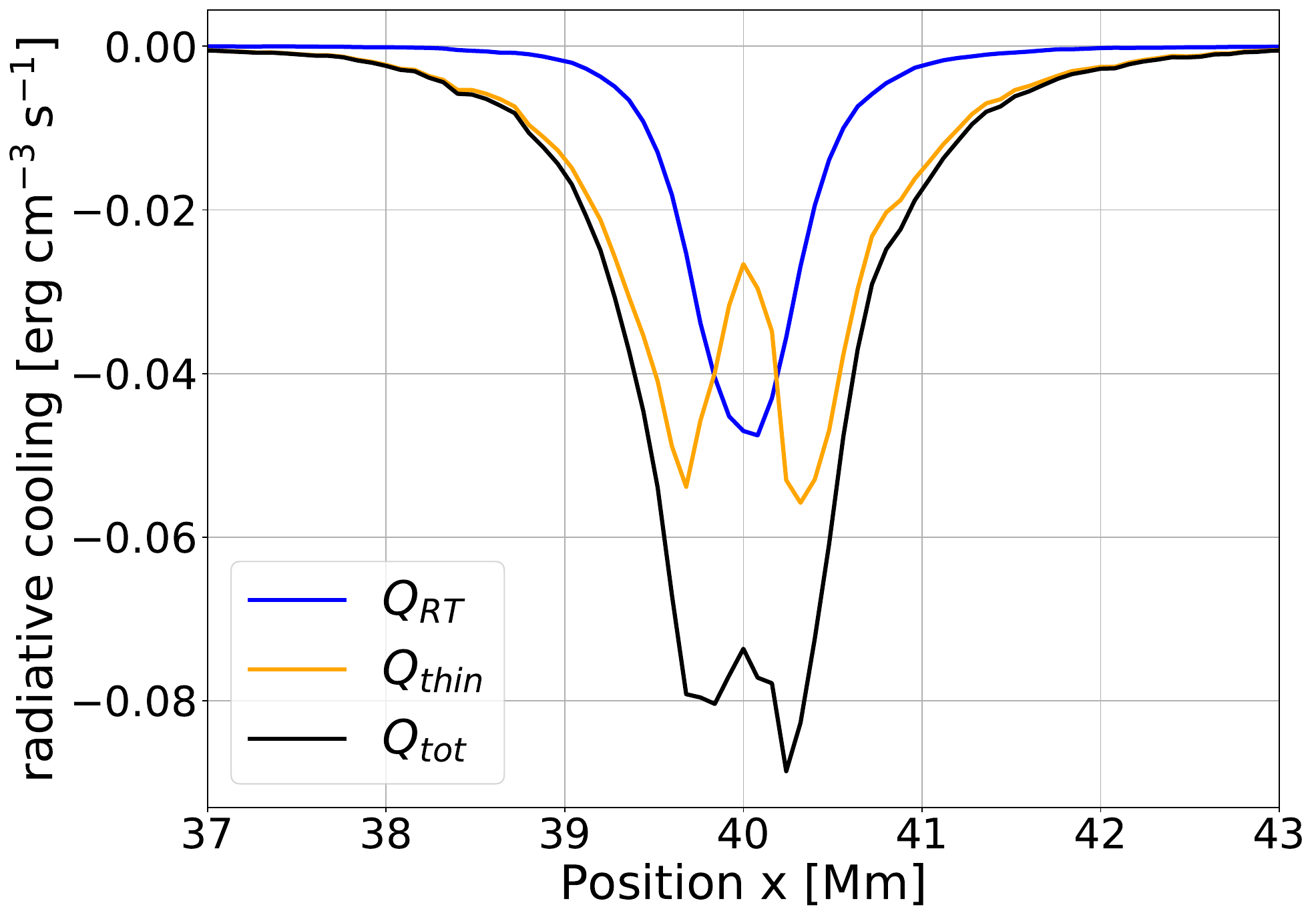}

    \includegraphics[width = 0.49 \textwidth]{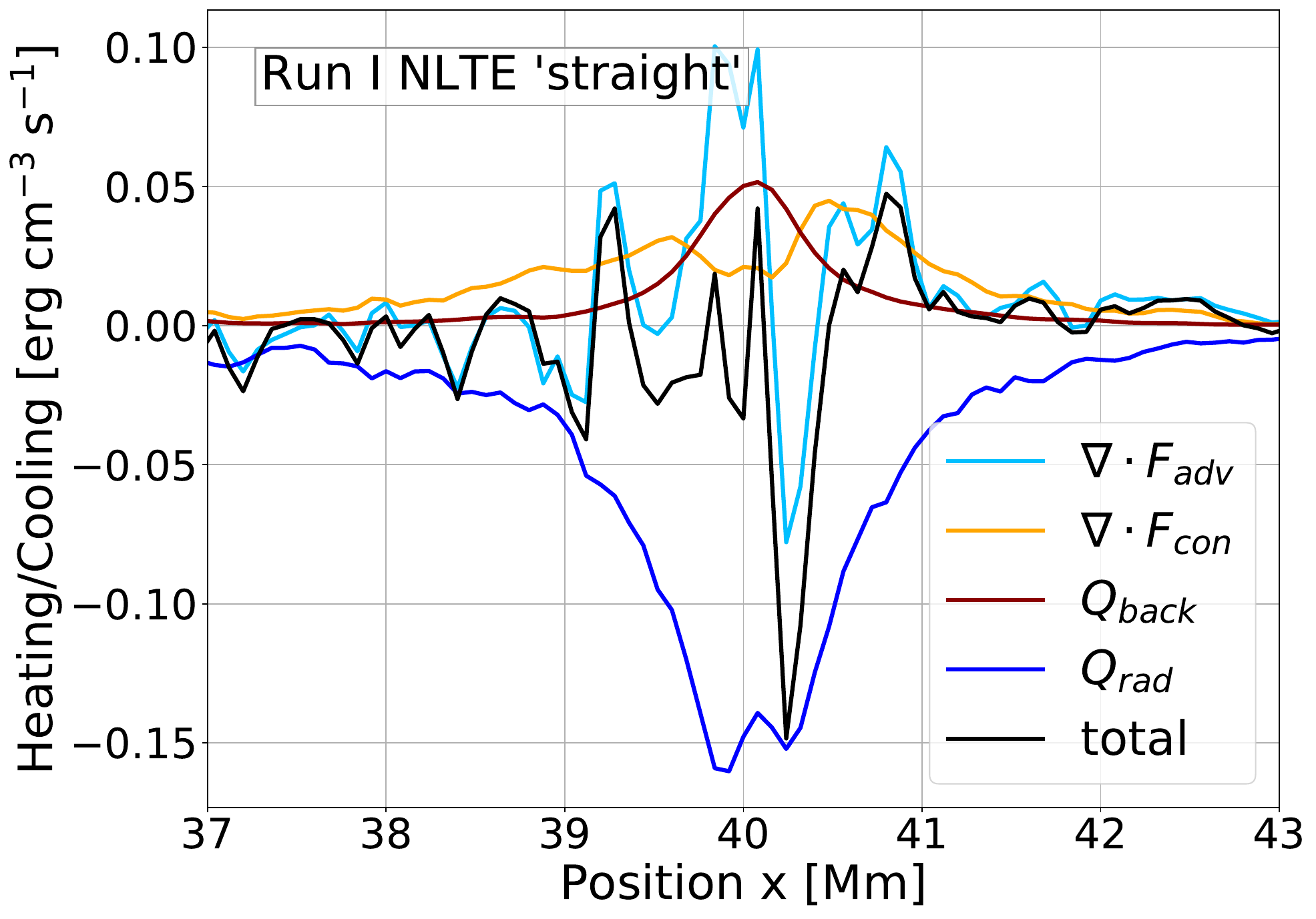}
    \includegraphics[width = 0.49 \textwidth]{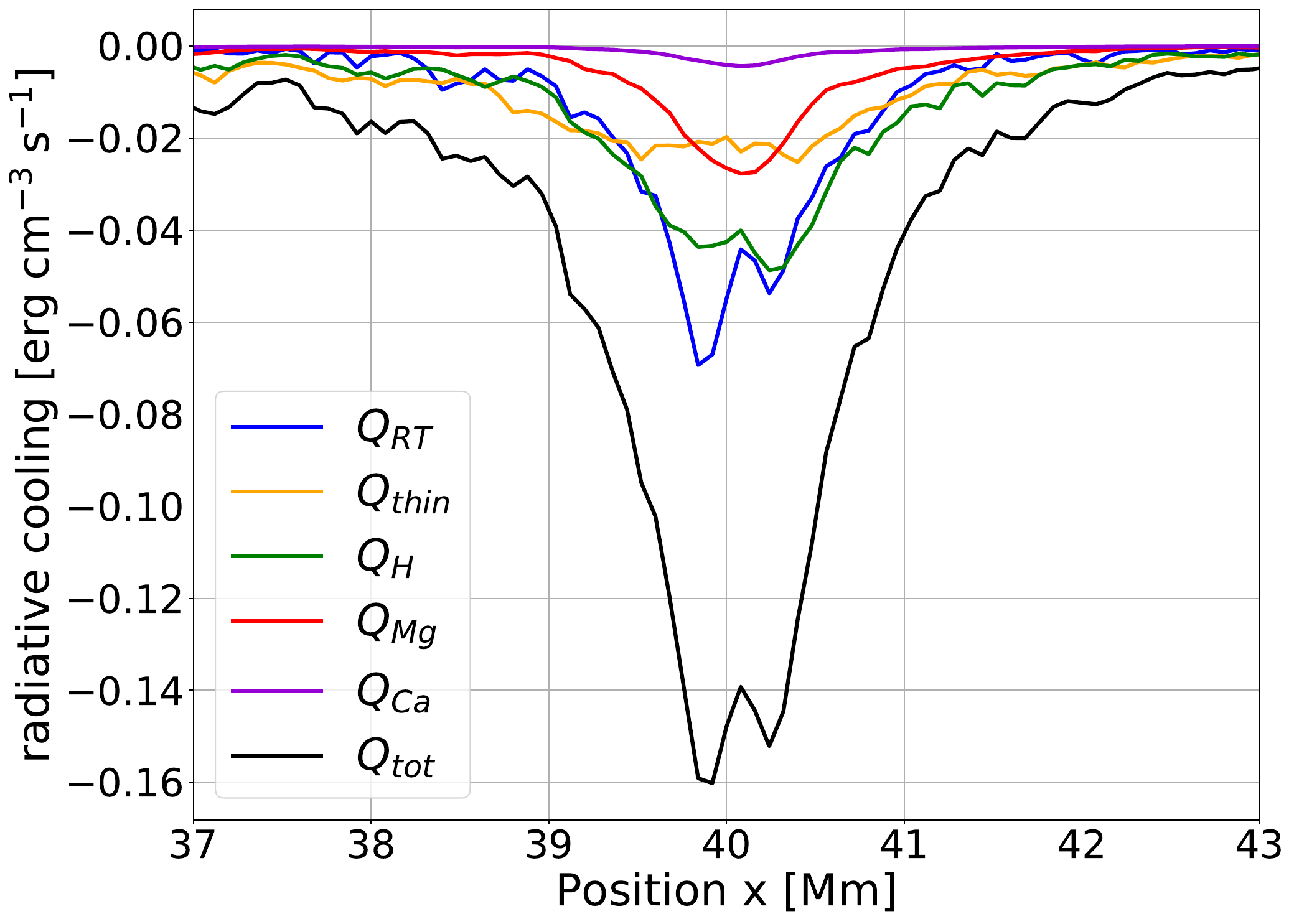}
    
    \caption{\textbf{Energetics of the prominence for Run I LTE and Run I NLTE.} The top row shows the LTE version and the bottom row the NLTE version of Run I. The energy terms are shown along a horizontal slice at a height of 10\,Mm above the surface, averaged over the y-axis and 60 minutes/5 minutes of simulation time for the LTE/NLTE run, respectively. Each vertical slice is centered on its respective center of mass before performing the y-average. The time frame is chosen such that the prominence is in an approximately straight vertical state. Due to the shorter runtime of the NLTE simulation, the simulated time frame does not include a prominence that is as vertical as for the LTE simulation. The advective term is therefore more variable. Left: The variously colored curves represent the dominating terms of the energy equation (Equation~3 in the Methods Section): the advective term ('$\nabla \cdot F_{adv}$', light blue), the conductive term ('$\nabla \cdot F_{con}$', orange), the total radiative cooling/heating ('$Q_{rad}$', blue) and the sum of all terms contributing to the energy equation (black).  Right: Contributions to the radiative cooling/heating term (here the black curve): losses from the RT scheme ('$Q_{RT}$', blue), optically thin losses ('$Q_{thin}$', orange), Hydrogen line losses ('$Q_{H}$, green), Magnesium line losses ('$Q_{Mg}$, red) and Calcium line losses ('$Q_{Ca}$, purple). The latter three are only applicable for NLTE computations.} 
    \label{fig:sup-fig1}
\end{figure}

\begin{figure}[h!]
    \centering
    \includegraphics[width=1\textwidth]{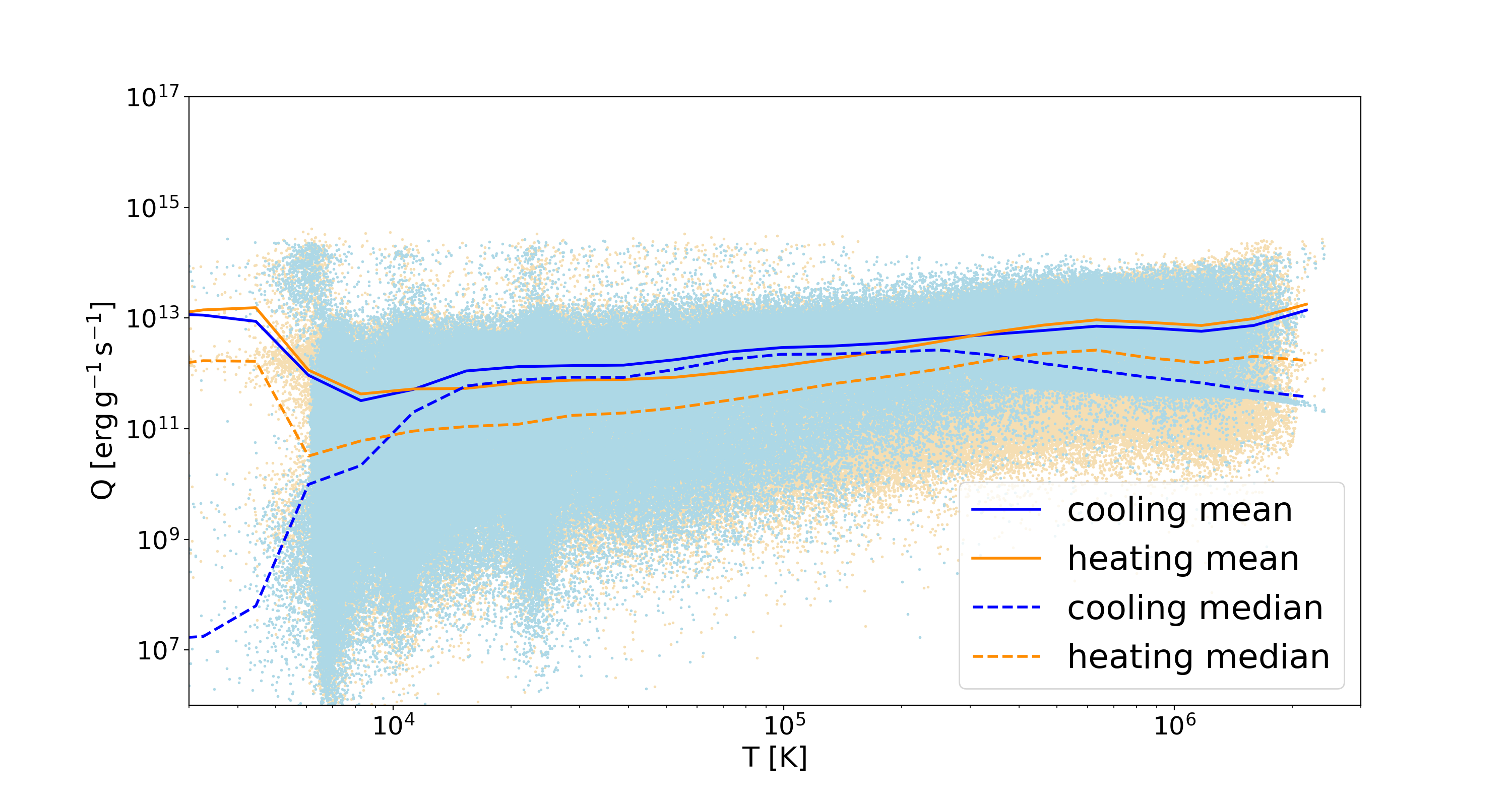}
    \caption{\textbf{Cooling and heating rates in the transition region around the prominence in Run I.} The scatter plot shows the cooling (blue) and heating (orange) contributions for individual pixels in one snapshot of Run I. The transition region pixels are selected by taking all pixels with $1.5 \cdot 10^{-14}\,\mathrm{g\,cm^{-3}} \leq \rho \leq 1.5 \cdot 10^{-13}\,\mathrm{g\,cm^{-3}}$ above a height of $5\,\mathrm{Mm}$ above the surface. Included in the cooling and heating rates are the optically thin losses, the radiative losses from the RT scheme, the advective term, the conductive term, the resistive heating and the viscous heating (see also Supplementary Figure~\ref{fig:sup-fig1}). The solid and dashed curves show the mean and median curves of the scattered points.}
    \label{fig:sup-fig2}
\end{figure}

To illustrate the cooling and heating at the edges of the prominence in this context, Supplementary Figure~{\ref{fig:sup-fig2}} shows the dependence of the cooling and heating rates on the temperature for all pixels in the transition region in one snapshot of Run I. To select the transition region around the prominence, we consider all pixels with $1.5 \cdot 10^{-14}\,\mathrm{g\,cm^{-3}} \leq \rho \leq 1.5 \cdot 10^{-13}\,\mathrm{g\,cm^{-3}}$ over a height of $5\,\mathrm{Mm}$ above the surface. The cooling and heating rates include the contributions from the optically thin losses, the radiative losses from the RT scheme, the advective term, the conductive term, the resistive heating and the viscous heating. The corresponding mean/median curves in Supplementary Figure~{\ref{fig:sup-fig2}} show that the heating is on average smaller than the cooling in the temperature range $10^4\,\mathrm{K} \leq T \leq 3 \cdot 10^5\,\mathrm{K}$. This leads to a pressure drop as shown below in Supplementary Figure~{\ref{fig:sup-fig3}} and drives the siphon flow. Also, the first injected cool blobs that start the prominence formation have a transition region like this associated with them (see Supplementary Figure~{\ref{fig:sup-fig5}} in the next section).

 To take a closer look at the radiative cooling and the Siphon flow described in the main text, Supplementary Figure~\ref{fig:sup-fig3} shows the radiative cooling and the pressure within the prominence. The maps on the left side show the distribution of the radiative cooling and those in the middle column the pressure for one vertical slice at one timestep for each run. The right panels of Supplementary Figure~\ref{fig:sup-fig3} show the corresponding distribution of the radiative cooling and the pressure along a horizontal slice, averaged over the y-axis as in Supplementary Figure~\ref{fig:sup-fig1}. The height of the horizontal slices is indicated in the respective pressure maps on the left. The times of the four snapshots are chosen such that they represent an approximately vertical state of the prominence. The profiles on the right side show that the strong cooling at the edges of the prominence coincides with a pressure drop. This indicates that the radiative cooling drives this pressure drop, which then leads to a siphon flow from the corona onto the prominence. As the mass has to follow the magnetic field lines, this pressure gradient creates an inflow from the chromosphere into the corona and then onto the prominence. As soon as the first dense plasma seed has settled in the magnetic dips, we can see the converging flows of hot plasma towards the prominence structure in the velocities, as shown in the Supplementary Videos 4 and 5, corresponding to Figure \ref{fig:fig2}C-F in the main text.

\begin{figure}
    \centering

     \includegraphics[width = \textwidth]{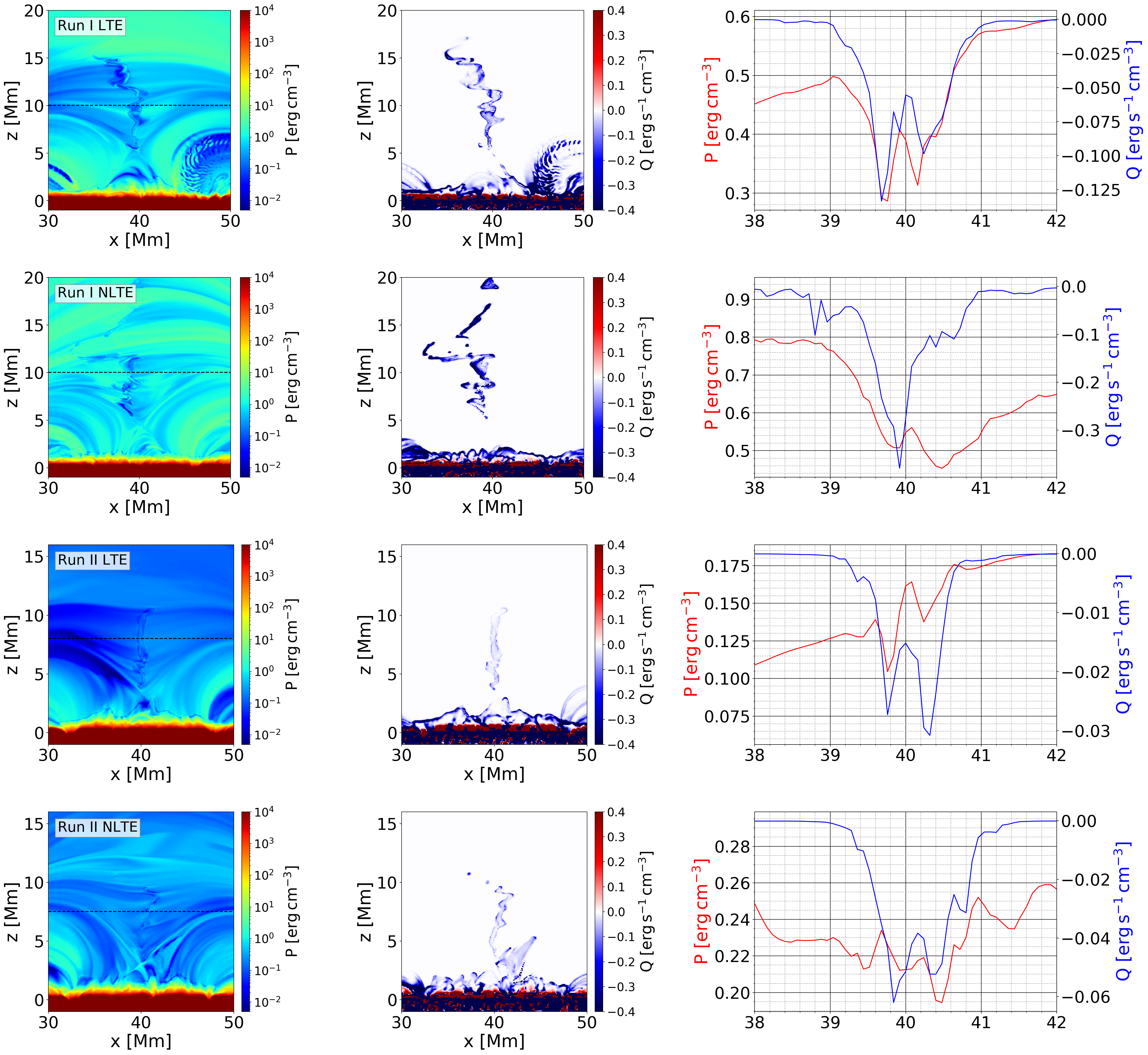}

    \caption{\textbf{Radiative cooling and pressure for one snapshot of Run I and Run II, as well as the corresponding NLTE runs.} The color maps show pressure (left) and radiative cooling (middle) for a zoom-in on the prominence along a vertical slice of the shown snapshots. The plots on the right show the horizontal profiles of averaged pressure and radiative cooling along a horizontal slice. The height of the corresponding horizontal slices is indicated by the dashed black lines in the pressure maps on the left. Due to the shorter runtime of the NLTE simulations, the simulated time frames do not include a prominence that is as vertical as for the LTE simulations. Each row corresponds to one simulation, from top to bottom: Run I LTE, Run I NLTE, Run II LTE and Run II NLTE.} 
    \label{fig:sup-fig3}
\end{figure}

\section{Thermal instabilities}
\label{sec:sup-sec2}

Thermal instabilities have been found to play a role in the formation and dynamics of prominences in numerical simulations. Here we apply the corresponding criterion to snapshots of our simulations to test whether we find similar relations between the prominence material and   unstable regions as in previous work.

\begin{figure}
    \centering
    \includegraphics[width=1\linewidth]{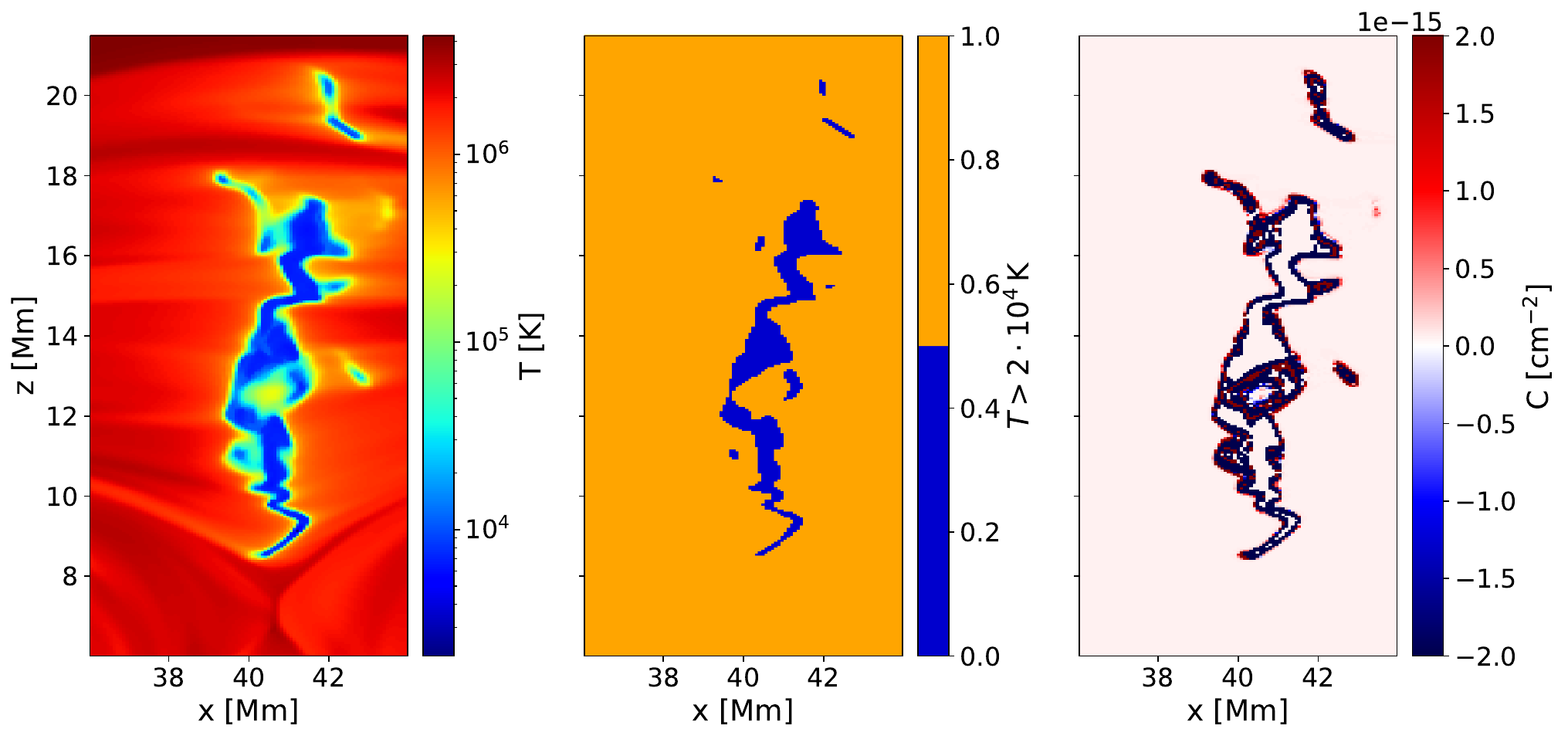}
    \caption{\textbf{Thermal instabilities for a snapshot of the prominence in Run I.} A vertical slice from a snapshot of Run I when the prominence is fully formed, zoomed in to show the prominence material. Left: temperature. Middle: dual colorbar to differentiate between regions with $T>2\cdot 10^4\,\mathrm{K}$ and $T<2\cdot 10^4\,\mathrm{K}$. The blue region roughly shows where the radiative losses from the RT scheme become dominant. The criterion calculated with the optically thin losses is valid for the orange region. Right: isochoric instability criterion $C$.}
    \label{fig:sup-fig4}
\end{figure}

\begin{figure}
    \centering
    \includegraphics[width=1\linewidth]{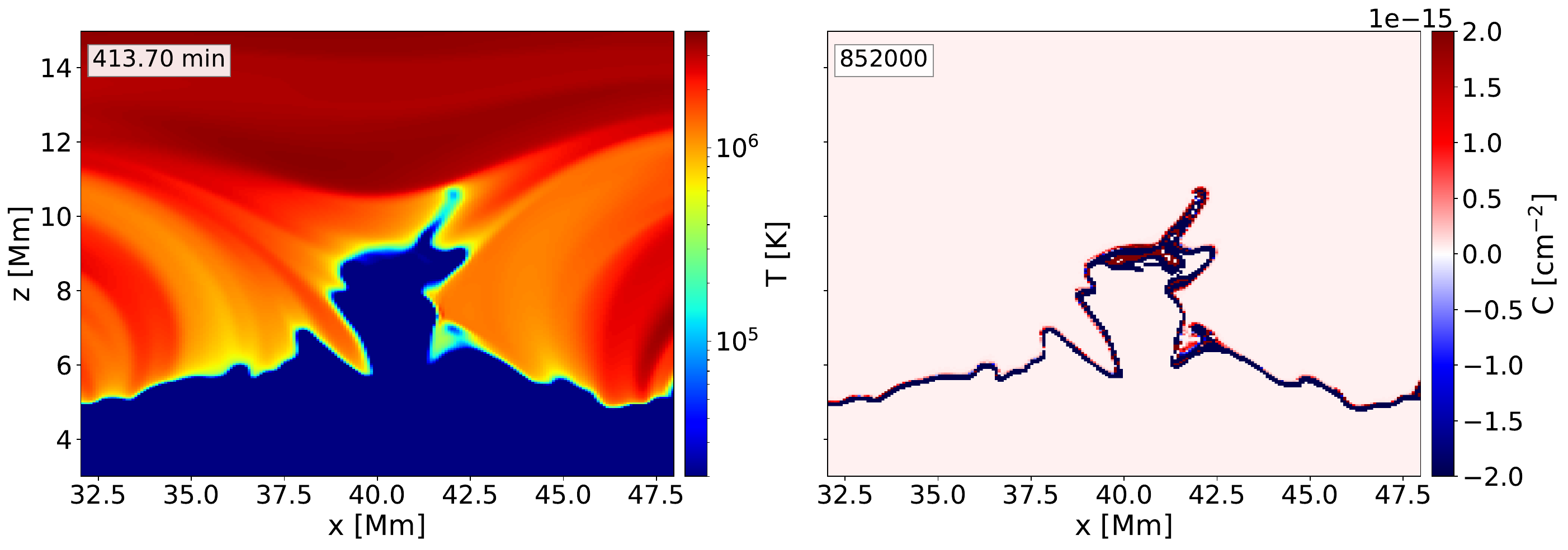}
    \caption{\textbf{Thermal instabilities for a snapshot during the prominence formation process in Run II.} Similar to Supplementary Figure~\ref{fig:sup-fig4}, but for a snapshot during prominence formation in Run II. At the shown time, a dense chromospheric seed is being ejected from the chromosphere into the corona. Left: temperature. Right: isochoric instability criterion $C$. The numbers in the top left corners show the time (left panel) and the iteration number (right panel) of the chosen snapshot.} 
    \label{fig:sup-fig5}
\end{figure}

It has been shown that the isochoric instability criterion

\begin{equation}
    C = {k}^{2}-\displaystyle \frac{1}{\sigma }\left(\displaystyle \frac{\partial {Q}_{H}}{\partial T}-\displaystyle \frac{{n}_{e}{n}_{H}\partial {\rm{\Lambda }}(T)}{\partial T}\right) < 0
    \label{eq:TI}
\end{equation}

\noindent maps well onto the thermally unstable regions in the corona{\cite{moschou2015simulating,lu2024periodic}}. In equation~{\ref{eq:TI}}, the wavenumber $k=2\pi/ \lambda$ depends on the size $\lambda$ of the condensation. $\sigma$ is the Spitzer heat conductivity, $Q_H$ denotes the heating terms in the simulation and the term $\frac{{n}_{e}{n}_{H}\partial {\rm{\Lambda }}(T)}{\partial T}$ corresponds to the tabulated optically thin losses in the simulation (see also the description in the Methods Section). In the energy equation (equation 3 in the Methods Section), the heating terms and the radiative loss term from the radiative transfer scheme do not have a direct dependence on the temperature, therefore only the optically thin losses are considered to calculate $C$, similar to (Lu et al. 2024)\cite{lu2024periodic}. 
The term $\frac{{n}_{e}{n}_{H}\partial {\rm{\Lambda }}(T)}{\partial T}$ is calculated with the overlap interval that is used for the optically thin losses{\cite{rempelEXTENSIONMURAMRADIATIVE2016_2}}. Supplementary Figure~{\ref{fig:sup-fig4}} shows $C$ for a vertical slice through the prominence of Run I. We show here results for $\lambda = 4\,\mathrm{Mm}$, but qualitatively the unstable regions do not change much with varying $\lambda$. Supplementary Figure~{\ref{fig:sup-fig4}} shows that the transition region around the prominence is unstable according to $C < 0$. Similarly, Supplementary Figure~{\ref{fig:sup-fig5}} shows the criterion for a snapshot at the start of prominence formation in Run II. Similar to the already built-up prominence in Supplementary Figure~{\ref{fig:sup-fig4}}, the first dense seed in Supplementary Figure~{\ref{fig:sup-fig5}} is surrounded by an unstable transition region. We do not see unstable regions along coronal loops before prominence formation as seen for e.g. the coronal rain simulated in (Lu et al. 2024)\cite{lu2024periodic}. This suggests that thermal instabilities only start to set in after the formation was started by the first injection, because the transition region of the cool injections are thermally unstable.

\section{Example of a cool injection}
\label{sec:sup-sec3}

This section presents more details on the dynamics of selected injection events. We present two subsequent injection events in Run II that happen at the beginning of the prominence formation process. Injection events like the ones presented below happen rather regularly before and after the prominence formation process starts, whereas the size and velocity of the blobs can vary. At the end of this Section, we show examples of two ejected blobs that do not contribute to the prominence mass.

\begin{figure}
    \centering

    \includegraphics[width=1\linewidth]{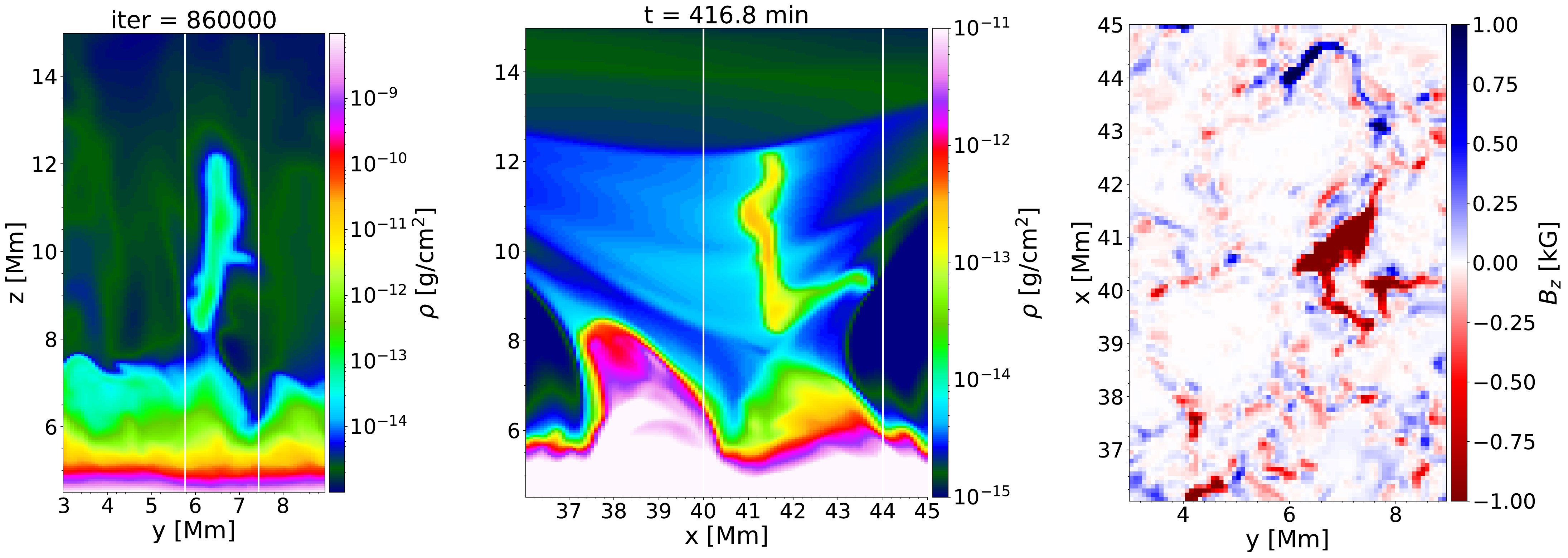}
    \caption{\textbf{An example for an injection event in Run II that is originating from below the Nullpoint.} Shown is a snapshot of part of the domain taken when the dense chromospheric blob has already reached the dipped part of the fieldlines in the corona. Left: Integrated density from the side (the line-of-sight is the x-axis). Middle: Integrated density from the front (the line-of-sight is the y-axis). Right: magnetogram at the surface ($z = 0\,\mathrm{Mm}$). The white vertical lines in the left/middle panel indicate over which range along the other horizontal axis the density in the middle/left panel is integrated. Supplementary Video 10 shows an animation of this Figure.}
    \label{fig:sup-fig6}
\end{figure}

For the selected event, Supplementary Figure~{\ref{fig:sup-fig6}} shows the integrated density from the side (left panel) and the front (middle panel) right after a dense blob has been ejected from the chromosphere into the corona. The right panel shows the corresponding magnetogram at the same time for the region where the injection happens. Supplementary Video 10 follows the dense blob as it is ejected from the chromosphere until it reaches the corona. At the very beginning of the video, a previous ejection is seen at $x\, {\sim}\,40$--$41\,\mathrm{Mm}$, $y\, {\sim}\,3$--$4\,\mathrm{Mm}$, which reaches the corona but disappears shortly afterwards. After that, during $t\, {\sim}\,409$--$415\,\mathrm{min}$, plasma surges up from the chromosphere at $x\, {\sim}\,38$--$41\,\mathrm{Mm}$, $y\, {\sim}\,4$--$7\,\mathrm{Mm}$. When viewed from the front (middle panel in Supplementary Figure~{\ref{fig:sup-fig6}} and Supplementary Video 10), a part of this plasma moves from below the dip towards the top left, while another part moves upwards through the Nullpoint into the corona. The mass of the plasma surge towards the top left is then also partly supplied to the dipped region in the corona while the mass is falling back down towards the Nullpoint. Once the plasma is in the corona, it rises slightly to around $6$--$8\,\mathrm{Mm}$ above the surface, marking the start of prominence formation. Right after this mass has reached the corona, a second surge happens close to the first one during $t\, {\sim}\,414$--$423\,\mathrm{min}$ at $x\, {\sim}\,37$--$40\,\mathrm{Mm}$, $y\, {\sim}\,5$--$8\,\mathrm{Mm}$ that feeds more plasma into the dipped region. In the magnetogram in the right panel, signs of flux cancellation can be clearly seen during the time frame of the video, especially around the stronger red polarity at $x \sim 39$--$42\,\mathrm{Mm}$, $y \sim 6$--$8\,\mathrm{Mm}$ which lies below the ejected plasma blob that is shown in the left and middle panel of Supplementary Figure~\ref{fig:sup-fig6}. This flux cancellation will be further discussed below in Figure~{\ref{fig:sup-fig9}}. 

\begin{figure}
    \centering
    \includegraphics[width=1\linewidth]{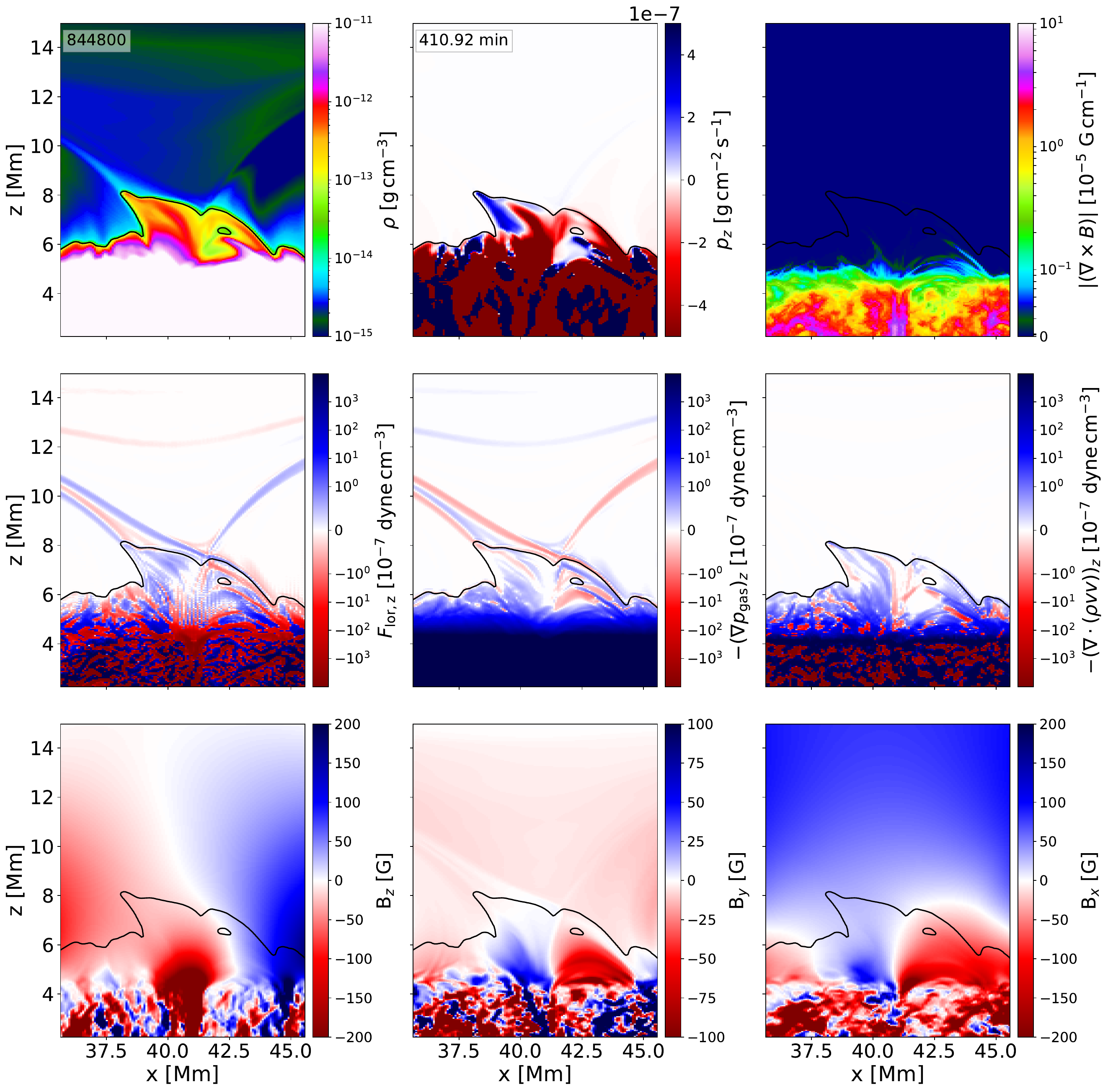}
    \caption{\textbf{Forces in the z-direction and magnetic field components for one snapshot during an injection event}. Top row: density (left), the z-component of the momentum (middle) and the absolute value of the current density $\nabla \times \vec{B}$ (right). Middle row: z-components of the Lorentz force (left), the pressure gradient force (middle) and the advective term (right) (see also equation~2 in the Methods Section). Bottom: z- (left), y- (middle) and x-component (right) of the magnetic field. The shape of the shown injection event is also visible in Supplementary Figure~\ref{fig:sup-fig6}. All quantities in this Figure are averaged over the y-axis in the region $y = 5.8$--$7.4\,\mathrm{Mm}$ around the injection, as indicated by the white vertical lines in the left panel of Supplementary Figure~\ref{fig:sup-fig6}. The black contours are taken at $5\cdot 10^{-13}\,\mathrm{g\,cm^{-3}}$ of the integrated density shown in the top left panel. The solar surface is at $z=4\,\mathrm{Mm}$. Supplementary Video 11 shows an animation of this Figure.}
    \label{fig:sup-fig7}
\end{figure}

Supplementary Figure~{\ref{fig:sup-fig7}} shows the forces in the z-direction (middle row) and the magnetic field components (bottom row) for one snapshot of the first injection event. The top row shows the density, the z-component of the momentum and the absolute value of the current $\nabla \times \vec{B}$. The injection event is shown from the front, in the same way as the integrated density in the middle panel in Supplementary Figure~{\ref{fig:sup-fig6}}. All quantities are averaged over the y-direction as indicated by the region between the two white lines in the left panel of Supplementary Figure~{\ref{fig:sup-fig6}}. Supplementary Video 11 shows an animation of Supplementary Figure~\ref{fig:sup-fig7} for the time frame of the two injection events. For the snapshot shown in Supplementary Figure~{\ref{fig:sup-fig7}}, the plasma below and left of the Null-point has started to rise. At this time, we can see an upward pressure gradient force in the lower part of the surging plasma and an upward directed Lorentz force in the top and left part of the surge. Later, at the beginning of the second injection ($t \sim 414\,\mathrm{min}$), strong upward-directed z-components can be seen in the video at the bottom of the surge at $x \sim 37$--$40\,\mathrm{Mm}$ in the Lorentz and the pressure gradient force. Both forces thus  contribute to the dynamics of the injection. This is also the case for the two horizontal components of the forces, suggesting that the exact dynamics of the ejected chromospheric plasma is complicated. In the regions where the chromospheric plasma is surging upwards, changes in the magnetic field can be seen above the surface in all three components. In the y-component (middle bottom panel), the negative polarity region (blue) at $x \sim 38$--$41\,\mathrm{Mm}$, $z \sim 4$--$8\,\mathrm{Mm}$ strongly changes shape during the first surge. In the z-component (bottom left), small inclusions of positive polarity (blue) can be seen in the negative polarity regions (red) at $x \sim 37$--$39\,\mathrm{Mm}$, $z \sim 4$--$5\,\mathrm{Mm}$ below the location of the second surge, which move and then disappear toward the end of the video. Slight enhancements in the current density (bottom right) are visible at $z \sim 5$--$7\,\mathrm{Mm}$ in the lower part of the surges.

\begin{figure}
    \centering

    \includegraphics[width=1\linewidth]{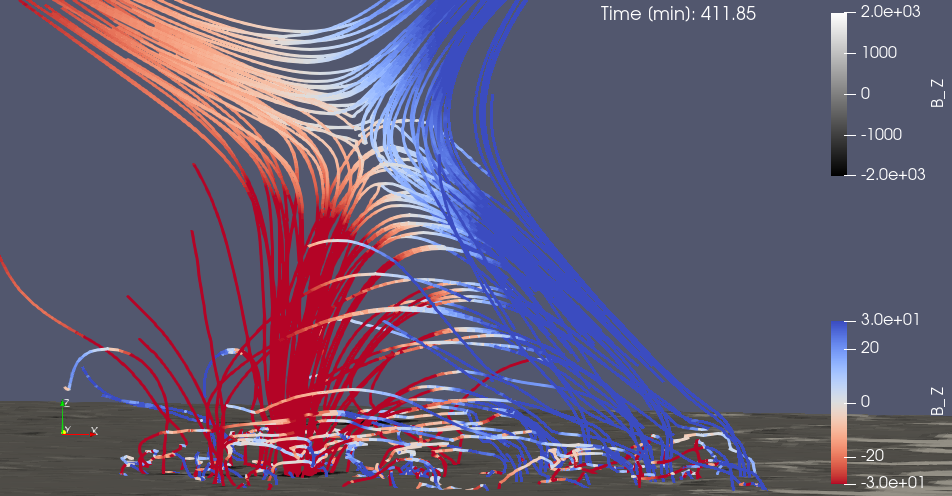}
    \includegraphics[width=1\linewidth]{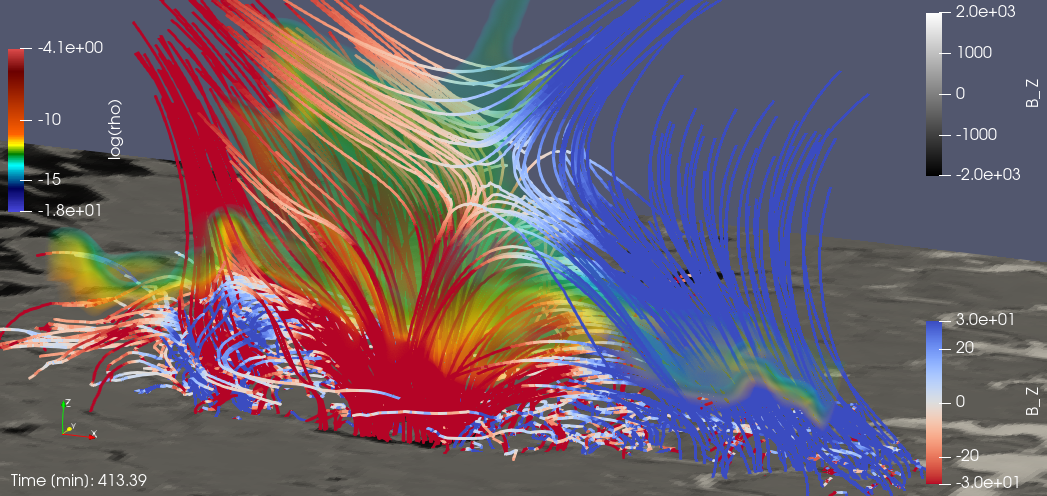}
    \caption{\textbf{Surface magnetogram and 3D rendering of the magnetic field lines around the location of an injection event.} Shown is the region where the dense blobs from Supplementary Figure~\ref{fig:sup-fig6} get injected from the chromosphere into the corona. The bottom panel additionally includes a volume rendering of the plasma in the upper chromosphere. The top and bottom panels show the scene from two slightly different viewing angles. The black and white shading at the solar surface indicates the $B_z$ there. The coloring of the field lines corresponds to $B_z$ in Gauss. The left black polarity at the surface is the same one as the red polarity in Supplementary Figure~\ref{fig:sup-fig6} and Supplementary Figure~\ref{fig:sup-fig9}. The bottom panel includes more seedpoints for the fieldlines compared to the top panel. Supplementary Videos 12 and 13 show an animation of the top and bottom panel of this Figure.}
    \label{fig:sup-fig8}
\end{figure}

Supplementary Figure~{\ref{fig:sup-fig8}} shows a 3D rendering of the magnetic field lines around the location of the two injections. Here, the color of the field lines indicates $B_z$. The Null-point just below the magnetic dips is clearly visible. Supplementary Videos 12 and 13 show animations of the top and bottom panel of this Figure. At and above the surface, the red field lines (negative polarity) and the blue field lines (positive polarity) at the footpoints continuously interact with the surrounding field. The bottom panel of Supplementary Figure~{\ref{fig:sup-fig8}} additionally shows the density in the upper chromosphere to visualize when the injections happen.
As in the video corresponding to Supplementary Figure~{\ref{fig:sup-fig6}} (Supplementary Video 10), the bottom panel of Supplementary Figure~{\ref{fig:sup-fig8}} (Supplementary Video 13) shows that plasma is shooting up from the chromosphere around the location of the left footpoint, which corresponds to the negative polarity at $x \sim 40$--$41\,\mathrm{Mm}$ in Supplementary Figure~{\ref{fig:sup-fig6}}. The right footpoint corresponds to the positive polarity at $x \sim 43$--$45\,\mathrm{Mm}$. Supplementary Figure~{\ref{fig:sup-fig9}} (Supplementary Video 14) shows that the unsigned photospheric flux in the region around the negative footpoint, as marked by the black rectangle in the right panel, is continuously decreasing during the time frame of the injections. Higher up in the chromosphere, changes in the magnetic field configuration around the  footpoints are visible in the 3D field lines. The blue inclusions in $B_z$ from Supplementary Figure~{\ref{fig:sup-fig7}} during the second injection are well visible in the 3D field lines in the bottom panel of Supplementary Figure~{\ref{fig:sup-fig8}} (Supplementary Video 13): at $t  \sim 414\,\mathrm{min}$, twisted blue field lines appear on the left side of the red footpoint, close to a smaller negative polarity at the photosphere. Until the end of the video, they seem to unravel, accompanied by the chromospheric plasma that surges upwards. After the chromospheric plasma is accelerated upwards from the surface, it reaches the corona through the Nullpoint. The video corresponding to the top panel of Supplementary Figure~{\ref{fig:sup-fig8}} (Supplementary Video 12) shows that rearrangements of the field lines around the Null-point happen regularly, allowing the chromospheric mass of the surges to reach the dipped region in the corona above the Nullpoint. Once in the corona, the cool prominence mass is supported against gravity by the Lorentz force, as shown in Supplementary Figure~{\ref{fig:sup-fig10}} for a longer-term average of the forces in the z-direction. This is also visible in the Supplementary Video 11 (corresponding to Supplementary Figure~{\ref{fig:sup-fig7}}) once the chromospheric mass has moved through the Nullpoint into the corona. The Supplementary Videos 6 and 7 (corresponding to Figure \ref{fig:fig3} in the main text) show that these kind of surges happen regularly and at different locations along the x- and y-axis, but they can only reach and stay in the dipped region in the corona when they are magnetically connected to it.

\begin{figure}
    \centering
    \includegraphics[width=1\linewidth]{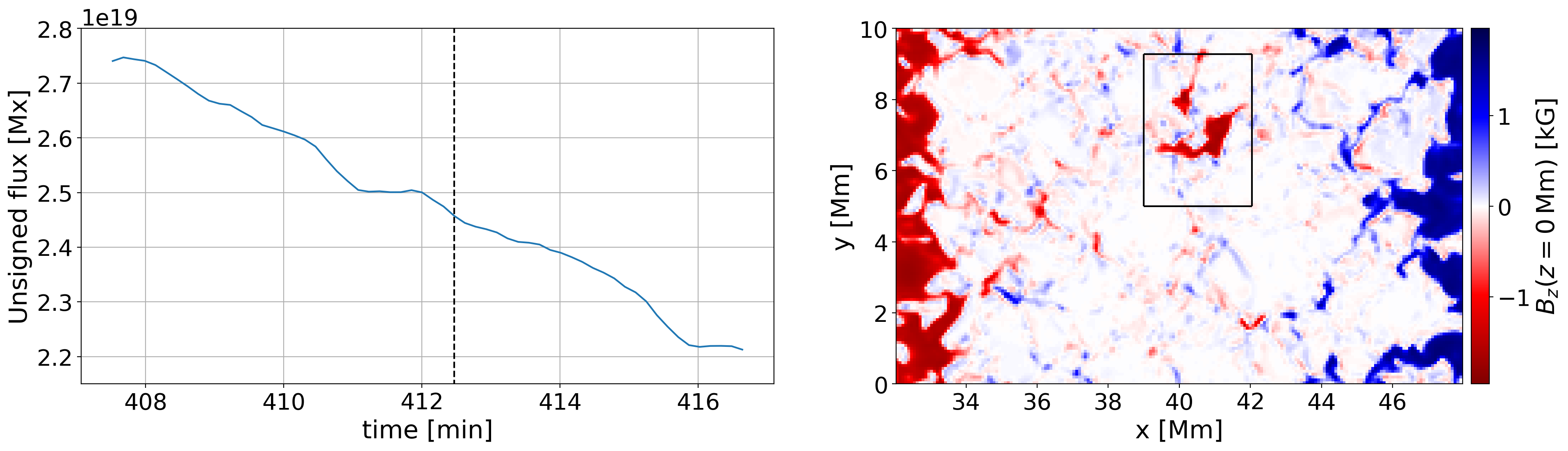}
    \caption{\textbf{Photospheric flux cancellation during an injection event.} Shown is the region where the dense blobs from Supplementary Figure~\ref{fig:sup-fig6} get injected from the chromosphere into the corona. Left: Unsigned magnetic flux at the surface in the region around the negative polarity footpoint at $x\sim 40\,\mathrm{Mm}$, $y\sim 7\,\mathrm{Mm}$, measured within the black rectangle in the right panel. Right: Magnetogram at the surface ($z = 4\,\mathrm{Mm}$ in Supplementary Figure~\ref{fig:sup-fig7}). Supplementary Video 14 shows an animation of this Figure.}
    \label{fig:sup-fig9}
\end{figure}

\begin{figure}
    \centering
    \includegraphics[width=1\linewidth]{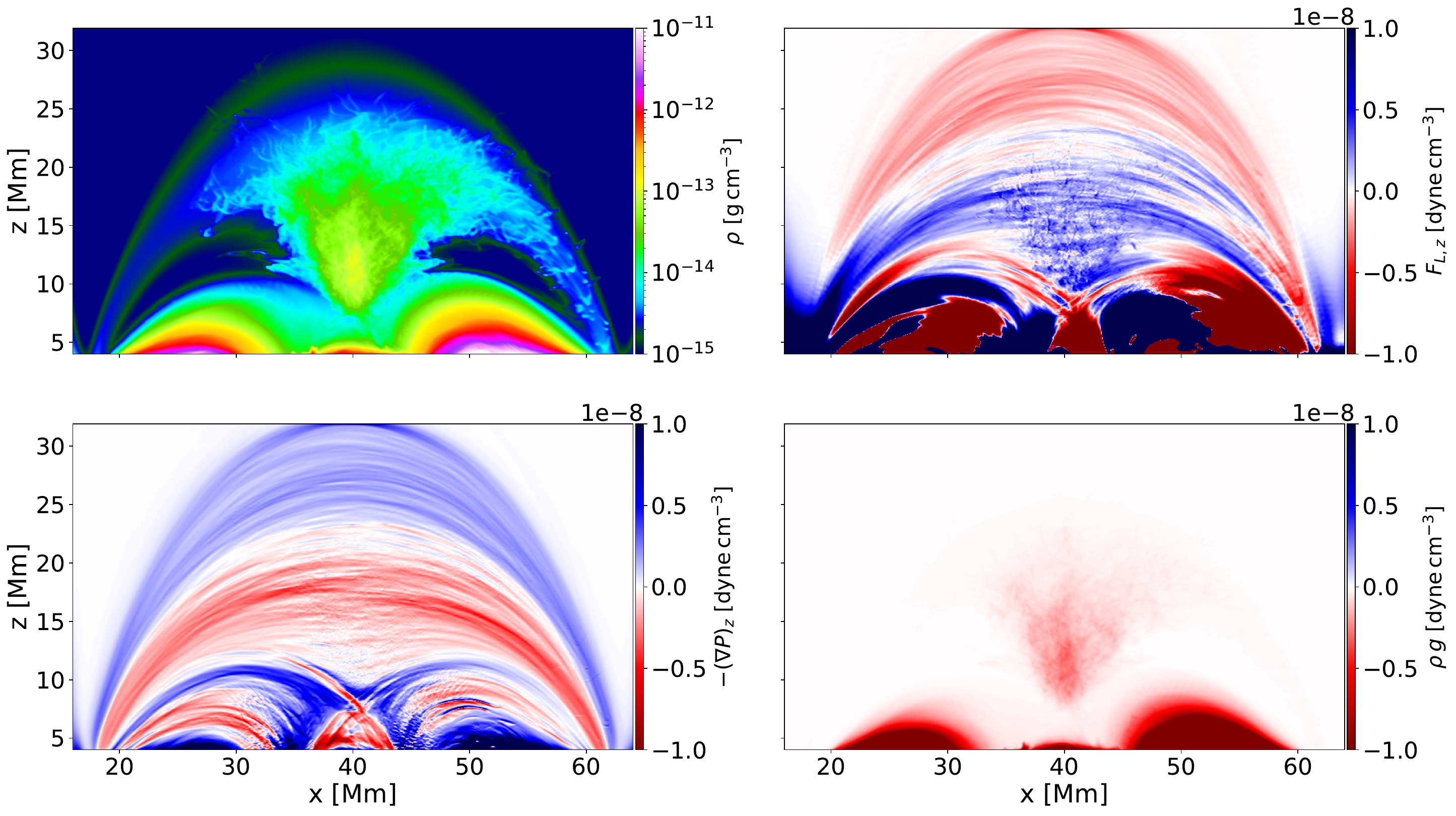}
    \caption{\textbf{Time-averaged density and force balance in the z-direction for Run I.} Time-averages over 200 minutes for the density (top left) and the z-components of the Lorentz force (top right), the pressure gradient force (bottom left) and the gravitational force (bottom right). All panels are averaged over 0.8 Mm along the y-direction, corresponding to 10 vertical y-slices. In the time-averaged density, it is visible that the prominence is moving from side to side during the time frame considered here. Draining to both sides is visible, but draining to the right side dominates in this sample.} 
    \label{fig:sup-fig10}
\end{figure}

\begin{figure}
    \centering
    \includegraphics[width=\linewidth]{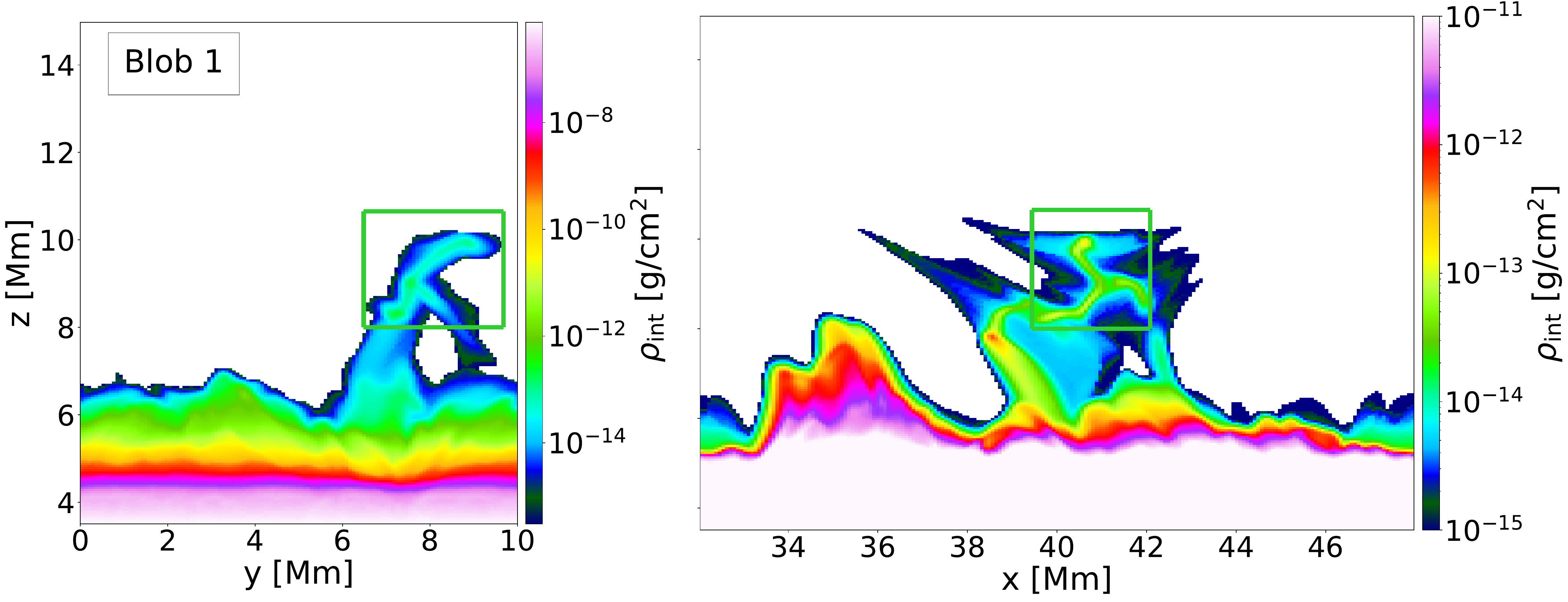}
    \includegraphics[width=\linewidth]{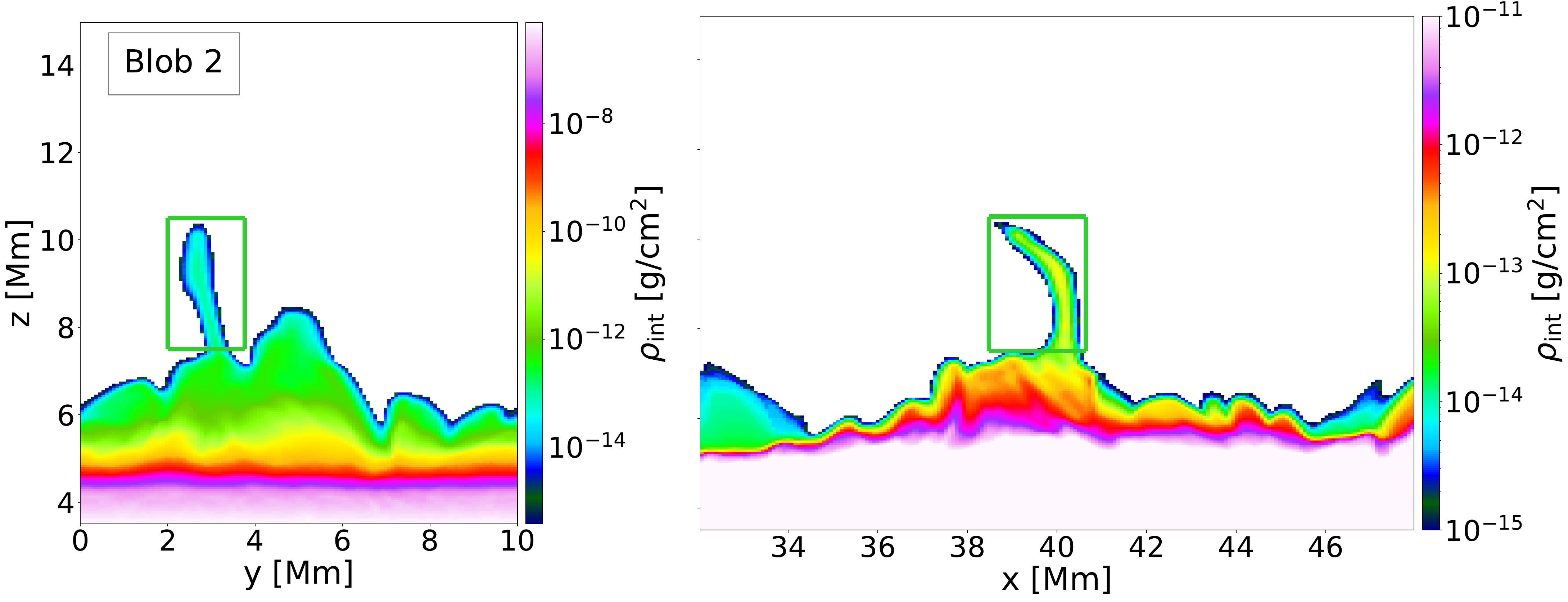}

    \includegraphics[width=\linewidth]{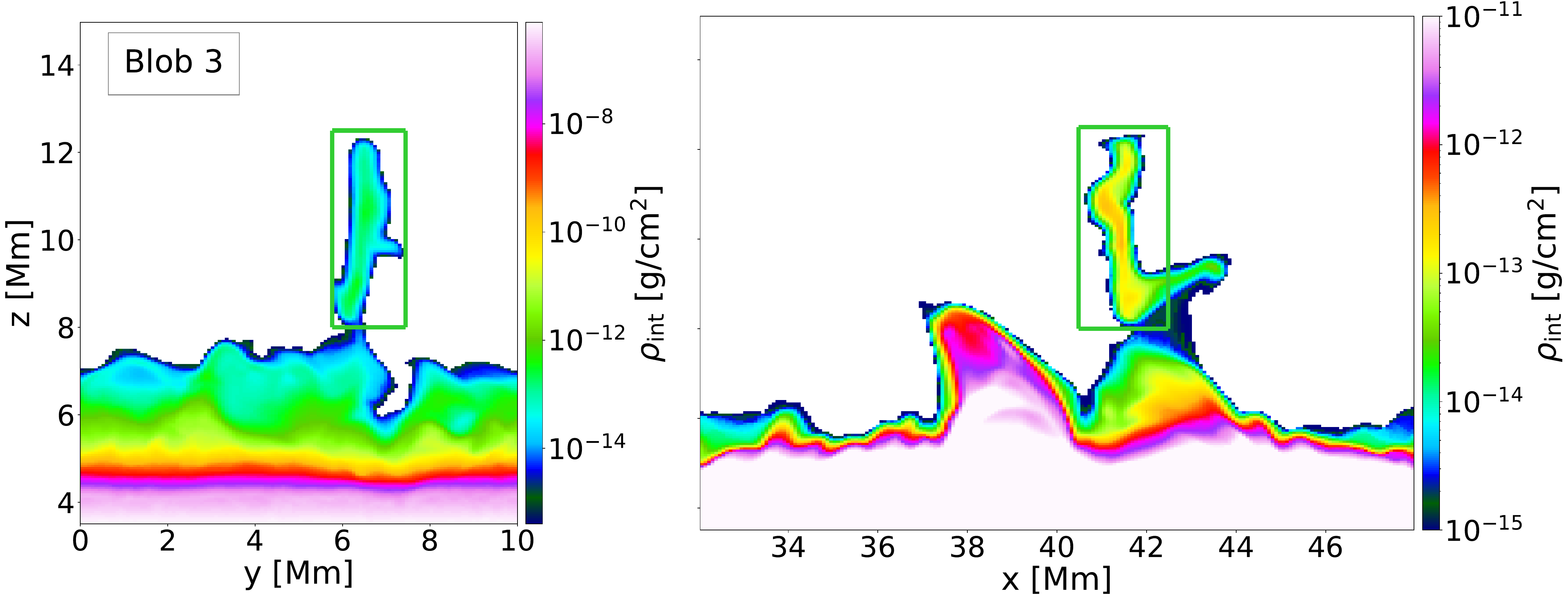}
    \caption{\textbf{Selection of three ejected chromospheric blobs in Run II.} Blobs 1 (top), 2 (middle) and 3 (bottom) in integrated density from the side (left) and the front (right). The blobs are manually selected and marked by the green rectangles. Only plasma with $\rho>10^{-14}\,\mathrm{g\,cm^{-3}}$ is shown and considered for the blob statistics in Supplementary Figure~\ref{fig:sup-fig12}. Blob 3 is the same as shown in Supplementary Figure~\ref{fig:sup-fig6}.}
    \label{fig:sup-fig11}
\end{figure}

\begin{figure}
    \centering
    \includegraphics[width=\linewidth]{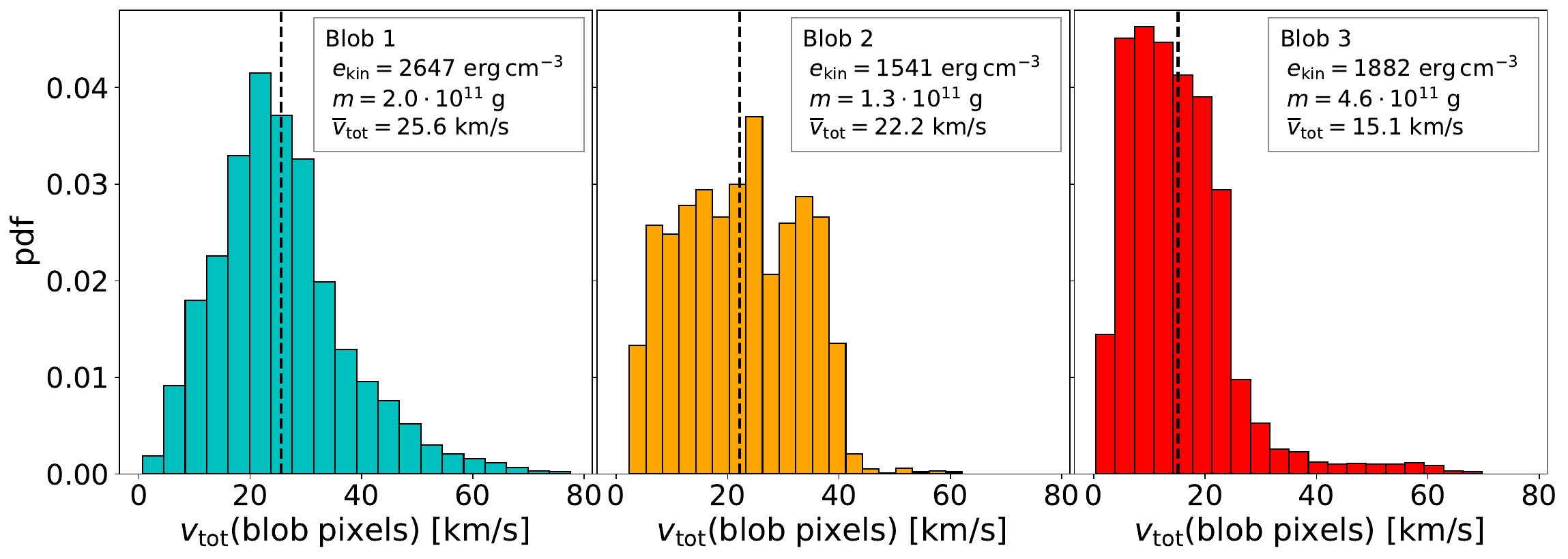}
    \caption{\textbf{Velocity distribution and masses for three ejected chromospheric blobs in Run II.} Histograms of the total velocities of the pixels belonging to blobs 1,2 and 3 as shown in Supplementary Figure~\ref{fig:sup-fig11}. The text boxes show the kinetic energy density, the total mass and the average total velocity of the corresponding blob. The dashed lines indicate the locations of the average velocity in the histogram.}
    \label{fig:sup-fig12}
\end{figure}

In general, some of the ejected blobs feed mass to the prominence, while others do not contribute to the mass build-up. Supplementary Figure~{\ref{fig:sup-fig11}} shows three examples for ejected cool blobs before prominence formation starts in Run II. The first two shown blobs disappear after being injected, whereas the third one stays in the dipped region long enough to start the prominence formation process. This third blob corresponds to the first injection event shown above in Supplementary Figure~{\ref{fig:sup-fig6}}. Supplementary Figure~{\ref{fig:sup-fig12}} shows
 histograms of the total velocity for the pixels belonging to each blob. The text boxes in Supplementary Figure~{\ref{fig:sup-fig12}} show the corresponding kinetic energy density, the total mass and the average total
 velocity of the three blobs. The disappearing blobs 1 and 2 have lower masses, a broader velocity distribution, and higher average velocities. This suggests that an injection should be slow and massive enough to stay in the magnetic
 dips. Generally, it can also happen for some injection events that a part of the prominence material is pushed out of the dipped region by injections with a larger momentum, inducing a draining event. 

\section{Prominence formation and mass supply}
\label{sec:sup-sec4}

In this section, we describe the prominence formation and the mechanisms that contribute to the mass growth of the prominence. Due to the longer runtime of the NLTE version of the code, the results in this section are shown for the LTE runs only. We start by qualitatively describing the formation for both simulation runs in the first paragraph, which is then followed by more quantitative descriptions for each run individually.

\paragraph{Qualitative description}

As described in the main text, the process of prominence formation consists of different parts for both simulation runs. It is started by a random ejection of a dense plasma seed from the chromosphere into the magnetic dips in the corona (see Supplementary Section~\ref{sec:sup-sec3}). This happens on a regular basis during the simulation, but it needs hours of simulated solar time until a large enough seed reaches and stays in the dip of the magnetic field configuration. For Run I, the cool ejections happen along the small loops on both sides of the prominence. The first seed originates from the left side at $t\,{\sim}\,  674\,\mathrm{min}$. For Run II, the cool ejections originate from the region below the magnetic dips: small blobs are ejected upward in discrete events from time to time. The ejection that starts prominence formation in Run II for the first time happens at $t\,{\sim}\,  412\,\mathrm{min}$. 

In both runs, the cool seeds start the formation of the prominence.  Almost as soon as the dense plasma seed settles in the magnetic dip, an inflow of plasma from the corona onto the prominence occurs along magnetic field lines (as described in Section~\ref{sec:sup-sec1}). These inflows consist of hot gas with $T \approx 3 \cdot 10^5$--$10^6\,K$ that cools and condenses onto the edges of the prominence (see Figure~\ref{fig:fig2}F and Supplementary Videos 4 and 5). This process corresponds to the condensation model and it provides approximately $18$--$42$\,\%\ of the mass of the prominence (see next paragraph). However, it is not the only mechanism that supplies the prominence with chromospheric plasma. Blobs of chromospheric plasma, similar to the one that initiated the prominence formation, also get ejected during the further evolution of the prominence. Sometimes, the ejected plasma blobs reach and stay in the magnetic dips, thus adding mass to the existing prominence structure. This is also visible in the Supplementary Videos 6 and 7 which correspond to Figure~\ref{fig:fig3} in the main text. When one of the blobs comes up and stays there, we can see an increase in the integrated density of the prominence (mostly in the left panel which shows the density integrated along the x-axis). This mechanism supplies approximately $58$--$82$\,\%\ of the mass that circulates through the prominence (see next paragraph).

\paragraph{Quantitative estimate}

\begin{figure}
    \centering
    \includegraphics[width = \textwidth]{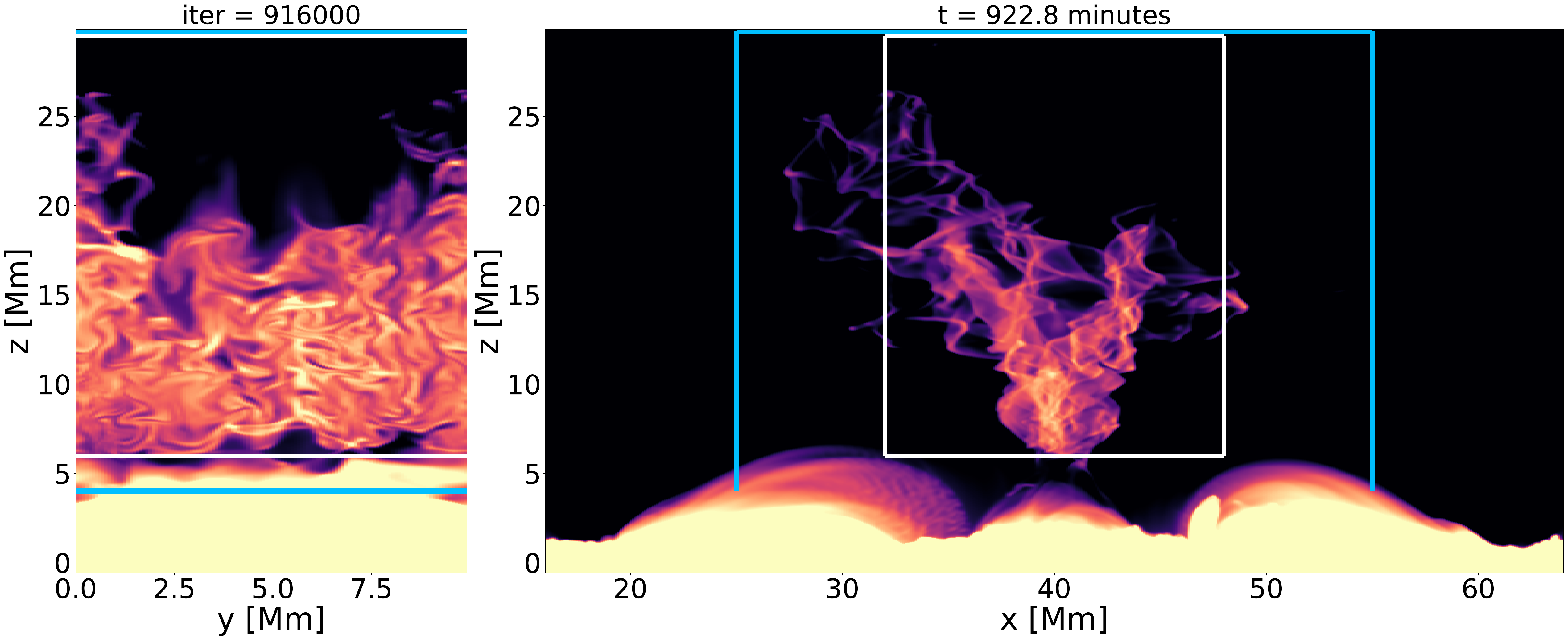}
   
    \caption{\textbf{Analysis of the mass flows into/out of the prominence (I)} Integrated density of the prominence from the side (left panel) and the front (right panel) for one snapshot of Run I. The white rectangle defines the surfaces through which the mass flows for the total mass, the hot flows and cool flows are measured. The blue lines show the surfaces for measuring flows of draining events. The resulting time-integrated mass fluxes are shown in Supplementary Figure~\ref{fig:sup-fig14}.}
    \label{fig:sup-fig13}
\end{figure}

\begin{figure}
    \centering

     \includegraphics[width = \textwidth]{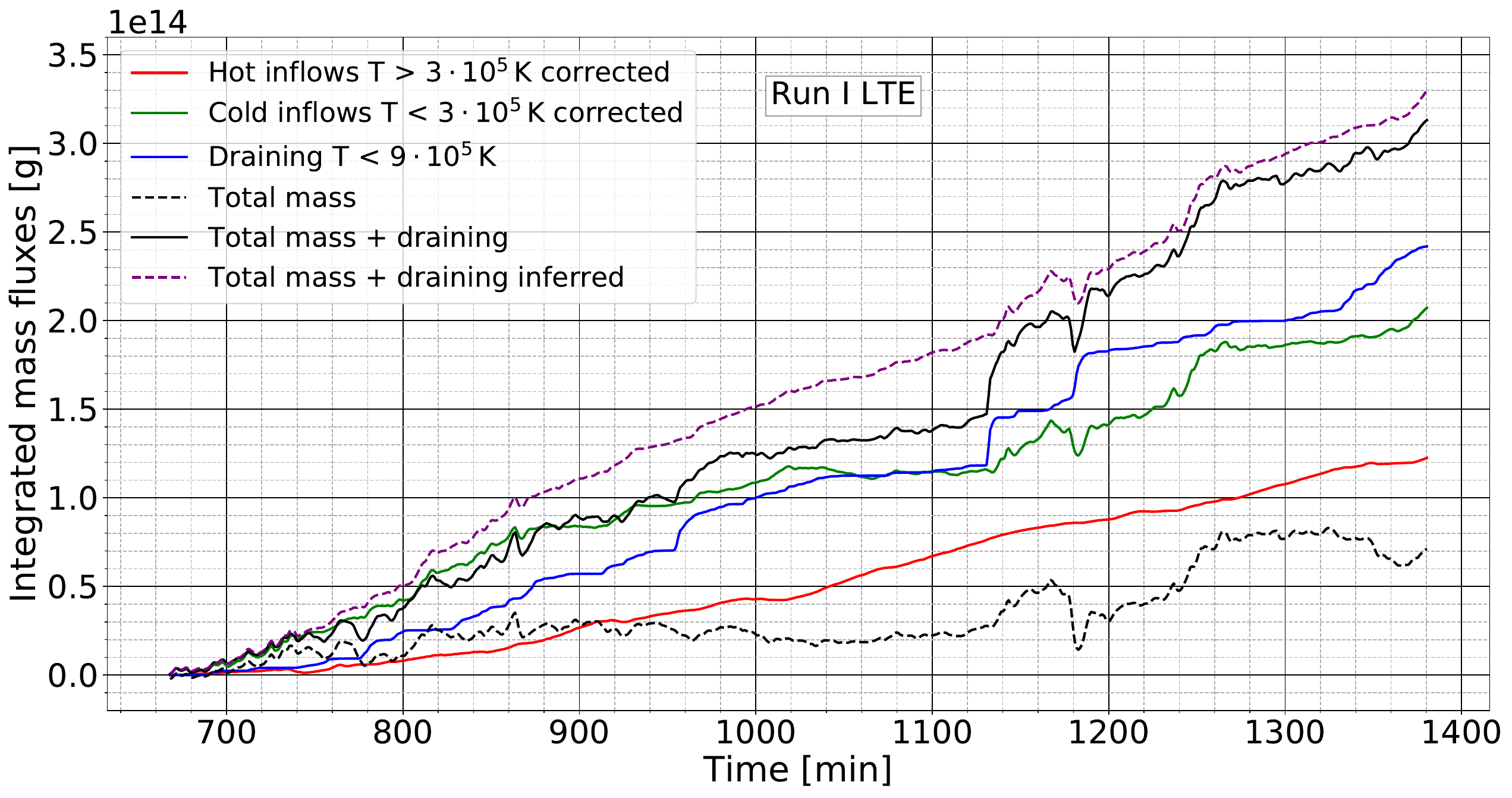}

      \includegraphics[width =\textwidth]{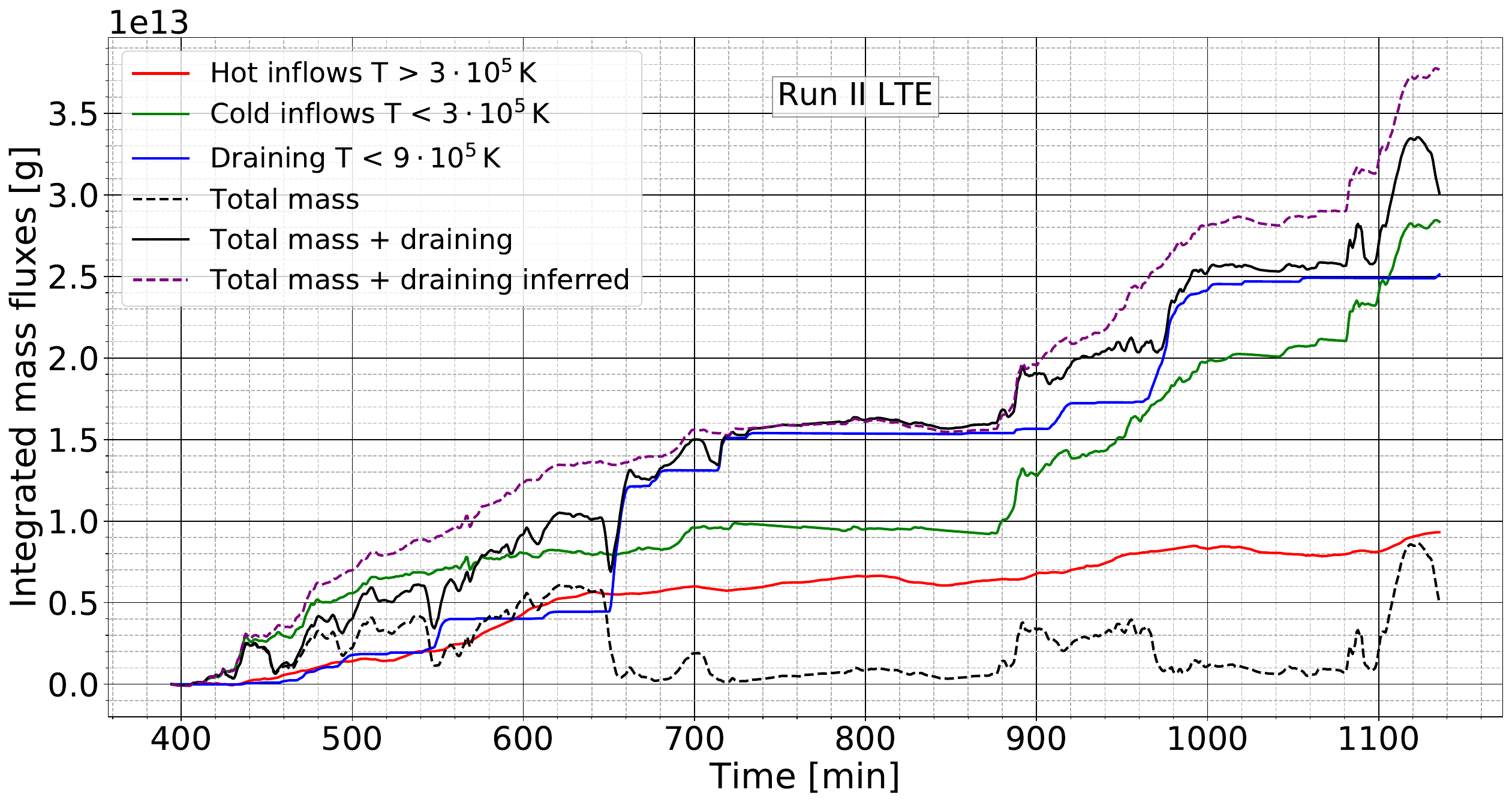}
    \caption{\textbf{Analysis of the mass flows into/out of the prominence (II)} time evolution of the total prominence mass (black dashed line), the total drained mass (blue line), the total mass that circulated through the prominence (solid black and purple dashed line) and the mass supplied by hot (red line) and cool (green line) inflows onto the prominence, as illustrated by Supplementary Figure~\ref{fig:sup-fig13} and described in the text. The top panel shows Run I and the bottom panel shows Run II. The draining (blue line) measures mass going out of the prominence and has a negative contribution. We show the absolute value of the draining here to quantify the amount of mass that circulates through the prominence in total (solid black line).}
    \label{fig:sup-fig14}
\end{figure}

Supplementary Figure~\ref{fig:sup-fig13} and \ref{fig:sup-fig14} quantify the approximate contribution of the two mechanisms, hot inflows, and cool injections, to the mass supply of the prominence, as well as the total mass of the prominence and the total amount of drained mass. Supplementary Figure~\ref{fig:sup-fig13} shows how the relevant mass flows are calculated. We calculate mass fluxes into and out of the prominence through the four surfaces of the shown white rectangular box: left, right, top, and bottom. As the prominence spans the whole box along the y-axis and the horizontal boundaries are periodic, there is no front/back component to the fluxes. 

For flows through the white box surfaces, we distinguish three cases. First: all mass flows going in and out of the box, to get an estimate for the current total mass of the prominence. The measurement is started at prominence formation time, such that the background coronal density is not counted. Second: mass flows for which $T > 3 \cdot 10^5\,K$ holds, to estimate the mass supply by hot siphon inflows. Third: mass flows for which $T < 3 \cdot 10^5\,K$, to estimate the mass supply by cool injections from the side/bottom. For the hot inflows, mass contributions from the left, right, and bottom surfaces are added (the contribution from the top surface is very small and can be neglected). In this way, flows that enter and leave the box without feeding the prominence are not counted for the mass budget. This mainly happens at the bottom edges of the box for Run I: plasma flows along the small loops left and right to the prominence. It enters the white box at the lower part of the side surfaces and leaves the box through the bottom surface. For the mass contribution by cool injections, we want to exclude the top part of the box to cut out the effect of draining events. Therefore, we introduce a vertical cut that divides the side surfaces for the cool flows in a bottom and a top part. This cut lies at 12\,Mm above the surface for Run I and at 6\,Mm above the surface for Run II. For the cool injections contribution, the bottom side parts and the bottom surface are then summed up. 

For Run I, we see that the cool inflows along the side loops get thermalized: plasma that enters the box in the $T < 3 \cdot 10^5\,K$ regime leaves it at a higher temperature. Both the cool and hot flows are therefore corrected for this contribution. The blue rectangular box in Supplementary Figure~\ref{fig:sup-fig13} shows how the amount of drained mass is estimated: all plasma with $T < 9 \cdot 10^5\,K$ that is passing outwards through a blue surface is counted as drained mass (the bottom part of the box is not counted here, as depicted in the map, to exclude the dynamics happening at the bottom below the prominence). The draining temperature cut is set because we want to exclude the hot outflows along the open field lines on both sides of the magnetic arcade. The blue box is much wider than the white one because the draining measurement should not capture the oscillating motions of the prominence structure itself. All mass fluxes are corrected by subtracting background fluxes, which are estimated by doing the same measurement for a time frame of 100 minutes before the prominence starts to form. The size of the white and blue boxes is adjusted to the smaller prominence size for Run II. The measured background fluxes are averaged over time and then subtracted as a constant correction factor. 

Supplementary Figure~\ref{fig:sup-fig14} shows the corresponding mass contribution after integrating the measured fluxes over time. The red and green lines show the contribution of hot and cool inflows. The total amount of draining that has happened until the respective time is shown with the solid blue line. The dashed black line represents the current total mass of the prominence (as estimated by the fluxes through the white box in Supplementary Figure~\ref{fig:sup-fig13}). The limits for the total amount of mass that has circulated through the prominence at the respective time is calculated by taking the sum of total prominence mass and total drained mass (solid black line), and by taking the sum of the hot and cool inflows (dashed purple line).

The shown time frames start a few minutes before the prominences begin to form. The injection of the dense seed from the side/bottom that starts prominence formation happens at 674 minutes/412 minutes for Run I and Run II, respectively. The injection and draining events that are described in the following can also be seen in the integrated density in the Supplementary Videos 6 and 7.

\paragraph{Total mass} Looking at the dashed black line that shows the estimate for the prominence mass, we can see that, from the formation at 674 minutes, the prominence in Run I is growing in mass until it reaches a roughly constant value of $2$--$3 \cdot 10^{13}\,\mathrm{g}$ from around 820 minutes onwards. The full size of the prominence generally extends slightly over the edges of the white box, especially in the top part, where the prominence is widest. Therefore, left and right motions of the oscillating prominence are captured in the total mass measurement. At around 1130 minutes, the prominence mass slightly starts to grow again. The gain in mass during the simulation is also visible in the integrated density from the side (Supplementary Video 6, left panel). Shortly before the end of the simulated time frame, at around 1260-1340 minutes, the total prominence mass reaches a maximum of $8 \cdot 10^{13}\,\mathrm{g}$ before decreasing to $6 \cdot 10^{13}\,\mathrm{g}$. In Run II, the prominence is not stable over time. After the prominence starts to form at 412 minutes, its mass first increases to a maximum of around $6 \cdot 10^{12}\,\mathrm{g}$. At around 640 minutes, a big draining event starts that completely depletes the prominence mass (the dashed black line is not exactly zero afterwards because turbulent motions can still bring chromospheric material into the region enclosed by the white box, such that inflows are measured temporarily). At 890 minutes, the prominence starts to form anew, and stays at a roughly constant mass of $3.5 \cdot 10^{12}\,\mathrm{g}$ before disappearing again at 970 minutes. A new formation starts at 1090 minutes, shortly before the simulated time frame ends.

\paragraph{Draining}
The blue line in Supplementary Figure~\ref{fig:sup-fig14} shows the total amount of draining that has happened before the respective point in time. In both runs, the draining consists of discrete events that can happen to either side of the prominence. Because the height of the prominence in Run I is comparable to the height of the simulated box, Run I also shows draining events at the top blue surface in Supplementary Figure~\ref{fig:sup-fig13}. For Run I, examples of draining events that can be seen in Supplementary Video 6 (corresponding to Figure~\ref{fig:fig3} in the main text), happen on the left side at $t \approx 835$--$860\,\mathrm{min}$ (mid-sized draining) and at $t \approx 930$--$950\,\mathrm{min}$ (big draining), as well as on the right side at $t \approx 750$--$800\, \mathrm{min}$ (big draining) and $t \approx 970$--$1000 \, \mathrm{min}$ (big draining). For Run II (Supplementary Video 7), example draining events can be seen on the left side at $t \approx 460$--$480\,\mathrm{min}$ (small draining), $t \approx 545$--$555\,\mathrm{min}$ (small draining) and $t \approx 700$--$720\,\mathrm{min}$ (big draining), as well as on the right side at $t \approx 490$--$500\,\mathrm{min}$ (small draining) and $t \approx 645$--$660\,\mathrm{min}$ (big draining).

\paragraph{Hot siphon inflows}
The contribution by hot inflows (red line) is continuous for both runs. Hot gas flows from the loop footpoints up to the magnetic dips, where they condense when converging onto the prominence structure. When the prominence is not draining, the siphon inflow is visible in the velocities (see Figure~\ref{fig:fig2}C-F in the main text and the Supplementary Videos 4 and 5). Also in the bottom panel of Supplementary Figure~\ref{fig:sup-fig14}, it is visible for Run II that the siphon flows set in as soon as the prominence starts to build up, but vanish after the prominence disappears (there is hot plasma below the prominence that is flowing all the time along the small magnetic loops on both sides, but the flows that lead higher up into the corona only start after the first dense seed of the prominence appears.). An estimate for how much mass the hot inflows contribute to the prominence follows below.

\paragraph{Cool injections}
 The ejection of cool plasma from the chromosphere into the corona happens in discrete events. Not all ejections contribute mass the the prominence: some ejected plasma blobs fall back to the surface without feeding the prominence (see also Supplementary Figures~\ref{fig:sup-fig11} and \ref{fig:sup-fig12}). For Run II, cold plasma blobs are mostly ejected from below the magnetic Nullpoint, i.e. from directly below the prominence. For Run II (Supplementary Video 7), examples of mass contributions via injection can be seen at  $t \approx 430\,\mathrm{min}$ (shortly after formation, in integrated density visible at $x \approx 40\, Mm$,$y \approx 6$--$8\, Mm$), at $t \approx 560$--$580\,\mathrm{min}$ (shortly after a draining event; in integrated density we see that one part of the ejected material shoots out to the right side and does not supply mass to the prominence). For Run I (Supplementary Video 6), these ejections happen along the small magnetic loops on both sides of the prominence. Little jet-like events happen at the outer footpoints and drive a plasma flow along the magnetic field lines, either back to the surface or into the prominence structure. This can for example be seen in the Supplementary Videos 2 and 6 (corresponding to Figure~\ref{fig:fig2}A and Figure~\ref{fig:fig3} in the main text) at the beginning of the prominence formation process. A description of an injection event in Run II is presented in Supplementary Section~\ref{sec:sup-sec3}.

\paragraph{Total circulated mass}
From the discussion above, we can infer how much mass circulated in total through the prominence structure relative to the mass of the prominence. To calculate the total mass that flowed in and out of the prominence structure, we have two estimates. First, we add the current prominence mass (dashed black line in Figure~\ref{fig:sup-fig14}) and the total amount of draining (blue line) to get the solid black line. This total mass shown by the solid black line could be underestimated because part of the prominence plasma is located in the region between the white and blue surfaces in Supplementary Figure~\ref{fig:sup-fig14}: parts of the prominence are too far out to be counted as prominence mass, but are not leaving the blue surface and are therefore not counted as draining, either. The solid black line thus represents a lower limit for the total circulated mass. To get a second estimate, we calculate the total circulated mass by adding the mass supply contribution from the hot and cool inflows (red and green lines). This is shown in the dashed purple line and represents an upper limit for the mass estimate. For Run I, the total circulated mass is $\sim 2.7$--$2.95 \cdot 10^{14}\,\mathrm{g}$ at 1300 minutes (80 minutes before the end of the simulated time frame). The mass of the prominence at this time is $8 \cdot 10^{13}\,\mathrm{g}$, so the total amount of mass that flowed through the structure is $3.4 - 3.6$ times the prominence mass at this time step. For Run II, the total amount of circulated mass after the first complete drainage of the prominence (at $t \approx 733\,\mathrm{min}$) is $1.4 \cdot 10^{13}\,\mathrm{g}$. The maximum mass of the prominence until this point is $6 \cdot 10^{12}\,\mathrm{g}$, so the ratio between the total drained mass and the maximum prominence mass is 2.3. When we consider also the next two reappearances of the prominence, the total circulated mass at the end of the simulation time (at 1135 minutes) is $3$--$3.7 \cdot 10^{13}\,\mathrm{g}$, which is $6 - 7.4$ times the current mass of $5 \cdot 10^{13}\,\mathrm{g}$ at this time. This strong mass circulation is also noted in observations \cite{liuFIRSTSDOAIA2012_2} and emphasizes how dynamic the prominence is. 

\paragraph{Contributions to mass supply}
With these values of the total circulated mass, we can now estimate how much of this mass is supplied by the hot inflows and cool injections. To calculate the contribution of both mechanisms, we calculate the ratios of the supplied hot and cool mass relative to the total circulated mass as given by the dashed purple curve. For Run I, we average these ratios over the last 400 minutes of the simulation time and take the maximum and minimum value in this time frame to estimate the variation of the ratios. The resulting contribution of hot inflows is $33$--$41$\,\% with an average of $37$\,\%. Consequently, the contribution of cool injections in Run I is  $59$--$67$\,\% with an average of $63$\,\%. In contrast to Run I, the prominence in Run II is unstable. For the ratios, we therefore take all time frames during which the prominence is present, starting from 100 minutes after the first formation. The resulting ratios for the hot siphon contribution is $18$--$42$\,\% with an average of $30$\,\%. The contribution of cool injections in Run II is thus $58$--$82$\,\% with an average of $70$\,\%.For both runs, the cool injections are thus contributing more mass than the hot siphon inflows. The ratios are roughly similar for Run I and Run II, with the tendency that Run II is somewhat more dominated by the cool injections than Run I.

\section{Prominence formation in current numerical simulations}
\label{sec:sup-sec5}

Many numerical prominence simulations use the condensation model as a base for prominence formation \cite{antiochosDynamicFormationProminence1999_2,antiochosThermalNonequilibriumProminences2000a_2,karpenOriginHighSpeedMotions2006b_2,xiaSIMULATIONSPROMINENCEFORMATION2012_2, fanMHDSimulationProminence2018_2,jercic2024prominence_2,donne2024mass_2}. An often used mechanism is evaporation-condensation: Localized static or stochastic heating is manually applied at the loop footpoints, which leads to  evaporation of chromospheric plasma at the loop footpoints and thus to an increase in density at the top of the loop. Due to this increased density, plasma condensations start to form via thermal instabilities \cite{parker1953instability} or thermal non-equilibrium \cite{klimchuk2019distinction} at the top of the loop. The condensed plasma then either falls into pre-existing magnetic dips, or locally drags down the magnetic field lines to form new dips \cite{karpen2001magnetic}. Donné and Keppens (2024) \cite{donne2024mass_2} recently showed that localized footpoint heating is not necessarily needed: in their setup, thermal instabilities are enough to drive siphon inflows that condense onto the prominence.

While we see condensation happening through siphon flows and thermal instabilities (see Supplementary Sections~\ref{sec:sup-sec1} and \ref{sec:sup-sec3}), the formation of our prominence is not started by a condensation mechanism. The formation starts when a cool and dense chromospheric plasma blob is ejected from the surface and gets stuck in the magnetic dips of our configuration. Only when this dense blob stays in the magnetic dips do we see the siphon flows setting in. The onset of the hot flows after the first injection is well visible in the Supplementary Videos 4 and 5 corresponding to Figure~\ref{fig:fig2}C-F in the main text. The formation of our prominence is thus started by injection, and the mass supply to the existing prominence structure is a combination of injection and condensation of the hot inflows.

The models by Kaneko and Yokoyama (2015) \cite{kaneko2015numerical_2} and Jenkins and Keppens (2021) \cite{jenkinsProminenceFormationLevitationcondensation2021_2} form prominences by levitation and condensation: during the formation of the flux rope, denser plasma from the lower corona is lifted up with the magnetic field. Higher up it has a larger density than the surroundings, leading to condensation via thermal instabilities. The second part is somewhat similar to our case, because denser mass from below is first brought higher up in the corona by another mechanism before condensation sets in.

There are also numerical simulations that form/feed prominences by the injection mechanism, with the difference that these simulations do not include a self-consistent treatment of the upper convection zone and photosphere. Huang (2021) \cite{huangUnifiedModelSolar2021_2} and Huang (2025) \cite{huang2025unified} formulated a model that explains both injections and condensation by the same physical mechanism. In Huang (2021) \cite{huangUnifiedModelSolar2021_2}, the authors manually applied different heating profiles in the chromosphere and found that heating in the lower chromosphere leads to injections, while heating in the upper chromosphere leads to evaporation-condensation. In this manner, they found cases where injection and condensation can occur together to feed the prominence, depending on the applied heating profile. In Huang 2025 {\cite{huang2025unified}}, the applied heating was created self-consistently by including magnetic reconnection in the lower and upper chromosphere, confirming the unified model for injections and condensation found in {\cite{huangUnifiedModelSolar2021_2}}.

Li (2025) {\cite{li2025response_2}} performed 2.5D simulations of flux emergence driven eruptions and found a model for flux-emergence-fed injections. The authors found that a filament channel is formed when the angle between the emerging and the pre-existing magnetic flux is 90\textdegree. The continued magnetic reconnection that forms the filament channel leads to the injection of cool plasma during the rise of the newly formed flux rope, feeding cool plasma into the channel. In comparison to our simulation, the injections in this work are driven by a global change of the magnetic field configuration, whereas the injections studied in Supplementary Section {\ref{sec:sup-sec3}} are related to small-scale changes in the magnetic field driven by turbulent convective motions.

\section{Captions for Supplementary Videos}
\label{sec:movies}

The Supplementary Videos can be downloaded at \\ \url{https://doi.org/10.1038/s41550-026-02840-7}.

\vspace{0.5cm}
\noindent {\bf Supplementary Video 1:}
\textbf{Magnetogram for Run I} Video corresponding to Figure \ref{fig:fig1}A in the main text: time evolution of the magnetogram at the $\tau_{500} = 1$ surface for Run I, starting approximately 20 minutes before prominence formation.

\vspace{0.5cm}
\noindent {\bf Supplementary Video 2:}
\textbf{Prominence formation in Run I, part I} Movie corresponding to Figure \ref{fig:fig2}A in the main text: 3D rendering of the prominence density, showing the beginning of prominence formation in Run I. The cool plasma that is injected along the small loop from the left side starts the formation process. The coloring of the plasma shows the logarithmic gas density (in $\mathrm{g\,cm^{-3}}$). For plasma with a density below $10^{-14}\,\mathrm{g\,cm^{-3}}$, the opacity is set to zero, such that the surrounding corona is not visible.

\vspace{0.5cm}
\noindent {\bf Supplementary Video 3:}
\textbf{Prominence formation in Run II, part I} Video corresponding to Figure \ref{fig:fig2}B in the main text: 3D rendering of the prominence density, showing the beginning of prominence formation in Run II. The cool plasma that is injected from below (at the polarity inversion line) starts the formation process. Draining events can also be seen towards the end of the movie. The coloring of the plasma shows the logarithmic gas density (in $\mathrm{g\,cm^{-3}}$). For plasma with a density below $10^{-14}\,\mathrm{g\,cm^{-3}}$, the opacity is set to zero, such that the surrounding corona is not visible.

\vspace{0.5cm}
\noindent {\bf Supplementary Video 4:}
\textbf{Prominence formation in Run I, part II} Video corresponding to Figure \ref{fig:fig2}C-F in the main text: It shows how the prominence in Run I is fed via condensation of hot plasma that is flowing along the magnetic field lines onto the cool prominence structure, driven by a pressure drop at the cool prominence material. For better visibility, the density (top left) and horizontal velocity (top right) are averaged over the current line-of-sight, whereas the temperature (bottom left) and pressure (bottom right) are taken along one vertical slice of the box. The arrows in the top right panel show the line-of-sight-averaged velocity field. Line-of-sight averaged magnetic field lines are added to the bottom left panel.

\vspace{0.5cm}
\noindent {\bf Supplementary Video 5:}
\textbf{Prominence formation in Run II, part II} Video corresponding to Figure \ref{fig:fig2}C-F in the main text: It shows how the prominence in Run II is fed via condensation of hot plasma that is flowing along the magnetic field lines onto the cool prominence structure, driven by a pressure drop at the cool prominence material. For better visibility, the density (top left) and horizontal velocity (top right) are averaged over the current line-of-sight, whereas the temperature (bottom left) and pressure (bottom right) are taken along one vertical slice of the box. The arrows in the top right panel show the line-of-sight-averaged velocity field. Line-of-sight averaged magnetic field lines are added to the bottom left panel.

\vspace{0.5cm}
\noindent {\bf Supplementary Video 6:}
\textbf{Prominence dynamics in Run I} Video corresponding to Figure \ref{fig:fig3} in the main text: Prominence dynamics for Run I, seen from three directions. The movie starts at prominence formation time. Left: Integrated density through the prominence from the side (integration along the x-axis). Top right: Integrated density through the prominence from the front (integration along the y-axis). Bottom right: Density along a horizontal cut through the simulation box, taken at a height of 12 Mm above the surface.

\vspace{0.5cm}
\noindent {\bf Supplementary Video 7:}
\textbf{Prominence dynamics in Run II} Video corresponding to Figure \ref{fig:fig3} in the main text: Prominence dynamics for Run II, seen from two directions. The movie starts shortly before prominence formation. Left: Integrated density through the prominence from the side (integration along the x-axis). Right: Integrated density through the prominence from the front (integration along the y-axis).

\vspace{0.5cm}
\noindent {\bf Supplementary Video 8:}
\textbf{Prominence dynamics in the Shear run} Video corresponding to Figure \ref{fig:fig3} in the main text: Prominence dynamics for the sheared setup of Run I, seen from three directions. Left: Integrated density through the prominence from the side (integration along the x-axis). Top right: Integrated density through the prominence from the front (integration along the y-axis). Bottom right: Density along a horizontal cut through the simulation box, taken at a height of 12 Mm above the surface.

\vspace{0.5cm}
\noindent {\bf Supplementary Video 9:}
\textbf{Magnetogram for Run II} Video corresponding to Extended Data Figure \ref{fig:ed-fig5}A: time evolution of the magnetogram at the $\tau_{500} = 1$ surface for Run II, starting a few minutes before prominence formation.

\vspace{0.5cm}
\noindent {\bf Supplementary Video 10:}
\textbf{Evolution of an injection event in Run II} Video corresponding to Supplementary Figure \ref{fig:sup-fig6} in the Supplementary Text: An example for two subsequent injection events in Run II, happening from below the Nullpoint. Left: Integrated density from the side (line-of-sight is the x-axis). Middle: Integrated density from the front (line-of-sight is the y-axis). Right: magnetogram at the surface (here $z = 0\,\mathrm{Mm}$). The vertical lines in the left/middle panel indicate over which range along the other horizontal axis the density in the middle/left panel is integrated.

\vspace{0.5cm}
\noindent {\bf Supplementary Video 11:}
\textbf{Forces in the z-direction and magnetic field components for an injection event in Run II} Video corresponding to Supplementary Figure \ref{fig:sup-fig7} in the Supplementary Text: Forces in the z-direction and magnetic field components during the injection event shown in Supplementary Figure~\ref{fig:sup-fig6}.  Top row: density (left), the z-component of the momentum (middle) and the absolute value of the current density $\nabla \times \vec{B}$ (right). Middle row: z-components of the Lorentz force (left), the pressure gradient force (middle) and the advective term (right) (see also equation~2 in the Methods Section). Bottom: z- (left), y- (middle) and x-component (right) of the magnetic field. All quantities are averaged over the y-axis in the region $y = 5.8$--$7.4\,\mathrm{Mm}$ around the injection, as shown by the white vertical lines in the left panel of Supplementary Figure~\ref{fig:sup-fig6}. The black contours are taken at $5\cdot 10^{-13}\,\mathrm{g\,cm^{-3}}$ of the integrated density that is shown in the top left panel. The solar surface is here at $z=4\,\mathrm{Mm}$.

\vspace{0.5cm}
\noindent {\bf Supplementary Video 12:}
\textbf{3D rendering of the magnetic field for an injection event in Run II} Video corresponding to the top panel of Supplementary Figure \ref{fig:sup-fig8} in the Supplementary Text: Surface magnetogram and 3D rendering of the magnetic field lines around the location where the dense blobs shown in Supplementary Figure~\ref{fig:sup-fig6} get injected from the chromosphere into the corona. Rearrangements of the field lines around the Null-point are regularly visible. The black and white shading at the solar surface indicates the $B_z$ there. The coloring of the field lines corresponds to $B_z$ in Gauss. The left black polarity at the surface is the same one as the red polarity in Supplementary Figure~\ref{fig:sup-fig6} and Supplementary Figure~\ref{fig:sup-fig9}. 

\vspace{0.5cm}
\noindent {\bf Supplementary Video 13:}
\textbf{3D rendering of the magnetic field and the chromospheric density for an injection event in Run II} Video corresponding to the bottom panel of Supplementary Figure \ref{fig:sup-fig8} in the Supplementary Text: Surface magnetogram, 3D rendering of the magnetic field lines and volume rendering of the chromospheric plasma density around the location where the dense blobs shown in Supplementary Figure~\ref{fig:sup-fig6} get injected from the chromosphere into the corona. While chromospheric plasma is surging upwards, rearrangements of the magnetic field lines above the surface are visible. The coloring of the field lines corresponds to $B_z$ in Gauss. The opacity of the plasma density is adjusted such that only the upper chromosphere is visible. 

\vspace{0.5cm}
\noindent {\bf Supplementary Video 14:}
\textbf{Photospheric flux cancellation for an injection event in Run II} Video corresponding to Supplementary Figure \ref{fig:sup-fig9} in the Supplementary Text: Strong signatures of flux cancellation are visible at the photosphere during the injection events shown in Supplementary Figure~\ref{fig:sup-fig6}. Left: Unsigned flux at the solar surface ($z = 4\,\mathrm{Mm}$ in Supplementary Figure~\ref{fig:sup-fig7}) in the region around the negative polarity footpoint at $x\sim 40\,\mathrm{Mm}$, $y\sim 7\,\mathrm{Mm}$, measured within the black rectangle in the right panel. Right: Magnetogram at the surface ($z = 4\,\mathrm{Mm}$ in Supplementary Figure~\ref{fig:sup-fig7}).

\bibliographystyle{sn-nature}